%% file: main.tex
\pdfoutput=1
\documentclass[pdflatex,sn-nature]{sn-jnl}

\usepackage{graphicx}
\usepackage{amsmath,amssymb,amsfonts}
\usepackage{amsthm}
\usepackage{xcolor}
\usepackage{booktabs}
\usepackage{array,tabularx}
\usepackage{algorithm}
\usepackage{algorithmicx}
\usepackage{algpseudocode}
\usepackage{url}
\usepackage{stylesetting}
\usepackage{microtype}
\usepackage{placeins}

\begin{document}

\title[Closing the loop on memory]{Closing the loop on memory: accurate, shallow and transferable decoding of quantum LDPC codes}

\author*[1]{\fnm{Zhou} \sur{You}}\email{youzhou26@stu.pku.edu.cn}

\author*[2]{\fnm{Ying} \sur{Li}}\email{yli@gscaep.ac.cn}

\author*[1]{\fnm{Xiao} \sur{Yuan}}\email{xiaoyuan@pku.edu.cn}

\author*[1]{\fnm{Yuan} \sur{Yao}}\email{yuan.yao@pku.edu.cn}

\affil[1]{\orgname{Center on Frontiers of Computing Studies, School of Computer Science, Peking University, Beijing 100871, China}}
\affil[2]{\orgname{Graduate School of China Academy of Engineering Physics, Beijing 100193, China}}

\input{sections/abstract.tex}

\keywords{quantum error correction, quantum LDPC codes, belief propagation, real-time decoding, feedback control}

\maketitle

\input{sections/introduction.tex}
\input{sections/results.tex}
\input{sections/discussion.tex}
\input{sections/methods.tex}
\input{sections/backmatter.tex}

\setlength{\bibsep}{0.6em}

\end{document}

\maketitle

\section*{Notation and glossary}

Symbols follow the main text and Methods. Indexed messages $m_{a\to i}$
and the scalar temporal multiplier $m$ are distinct. Ensemble size is $S$;
Relay-BP's requested solution count is $S_{\rm R}$.

\begingroup\SItablefont
\renewcommand{\theHtable}{notation.\arabic{table}}
\begin{longtable}{@{}p{43mm}p{122mm}@{}}
\toprule
Symbol & Meaning \\
\midrule\endfirsthead
\toprule
Symbol & Meaning \\
\midrule\endhead
\bottomrule\endfoot
$H$, $O$ & Detector--fault incidence matrix and logical-observable matrix. \\
$M$, $N$, $K_{\rm L}$ & Numbers of detectors, fault mechanisms and logical observables. \\
$E$, $\mathcal N(a)$, $\mathcal M(i)$ & Tanner edges, variables adjacent to check $a$, and checks adjacent to variable $i$. \\
$s$, $e$, $\hat e$ & Observed syndrome, sampled fault vector and returned correction. \\
$p$, $p_i$, $L_i$ & Physical error rate, fault prior and prior log-likelihood ratio. \\
$m_{a\to i}$, $\mu_{i\to a}$ & Check-to-variable and variable-to-check messages. \\
$R_i$, $\Lambda_i$, $Q_i$ & Variable belief, memory-weighted prior and summed check field. \\
$D_i=R_i-L_i$ & Displacement of the belief from its prior. \\
$\gamma_i$, $\gamma$ & Variable memory strength and a uniform memory strength. \\
$\Gamma=[\gamma_{\min},\gamma_{\max}]$ & Admissible interval for the controller. \\
$[\gamma^-,\gamma^+)$ & Interval for the initial random memory strengths. \\
$\gamma_0$ & Memory strength at which the temporal-channel growth factor is one; not the Mem-BP value $0.15$. \\
$\alpha$ & Min-sum normalization, equal to one throughout. \\
$z=s\oplus H\hat e$, $\rho_i$ & Check residual and local fraction of unsatisfied checks. \\
$c_i$, $T_{\rm conf}$ & Variable confidence and its scale. \\
$\kappa_0$, $\kappa_i$, $\beta$ & Step scale, variable step size and residual-gate weight. \\
$U_i$, $\zeta$, $T_{\rm tgt}$ & Target belief, magnitude bound and target scale. \\
$\epsilon$ & Denominator guard for $D_i$. \\
$K$, $W$, $I_{\max}$ & Update interval, warm-up iterations and total iteration limit. \\
$t$, $\tau$ & Iteration index and convergence time or executed iteration count. \\
$T$ & Number of syndrome-extraction rounds; not an iteration limit. \\
$S$, $B$, $S_{\rm R}$ & Ensemble size, total iteration budget and requested Relay solutions. \\
$m$, $b$, $\lambda$ & Temporal multiplier, channel coefficient and growth factor per iteration. \\
$n$, $P_{\rm L}$, $P$ & Shot count, logical failure rate per shot and statistical test probability. \\
$\lambda_s$, $\mathcal S_s(t)$, $\eta$ & Syndrome escape rate, survival probability and population-tail exponent. \\
$\Gamma_{\rm E}$ & Euler gamma function in Note~4; distinct from the controller interval $\Gamma$. \\
$c_{\rm hw}$, $\tau_{\rm iter}$ & Extra controller clock cycles and time per iteration (Note~5). \\
\end{longtable}
\endgroup

\newpage
\subsection*{Glossary of execution concepts and metrics}

\begingroup\SItablefont
\renewcommand{\theHtable}{glossary.\arabic{table}}
\begin{longtable}{@{}>{\raggedright\arraybackslash}p{43mm}p{122mm}@{}}
\toprule
Term & Definition and usage across the manuscript \\
\midrule\endfirsthead
\toprule
Term & Definition and usage across the manuscript \\
\midrule\endhead
\bottomrule\endfoot
Shot (circuit trial) & One complete $T$-round X-basis trial ($T\in\{6,12,18,20\}$) generating syndrome $s=He$; distinct from a decoder \emph{run}. \\
Run (single run / instance) & One end-to-end decoder call per shot returning $\hat e$; the ensemble runs $S$ concurrent instances (members), whereas Relay-BP runs sequentially. \\
Single-path & Comparison class of decoders in which one trajectory uses one memory field that is never repeatedly redrawn (fixed, or adjusted by feedback): BP, Mem-BP, DMem-BP, DMem-warm, TEMPO-static, TEMPO-BP and BP+OSD+CS-10. \\
Leg / successive legs & One Relay-BP run of up to 60 iterations with a fresh random memory field, started from the previous leg's beliefs $R_i$ (up to 300 legs after warm-up); distinct from independent restarts ($R^{(0)}=L$). \\
Candidate selection & Relay-BP-1 stops at the first valid correction ($H\hat e=s$). Relay-BP-5 requests five valid corrections across successive legs, and the TEMPO ensemble collects valid member corrections; both select the candidate most probable under the priors. \\
Failure count \& $P_{\rm L}$ & Failures or per-shot rate $P_{\rm L}=\text{failures}/\text{shots}$ from non-convergence ($H\hat e\ne s$) or wrong logical class ($O\hat e\ne Oe$); raw counts require matched cohorts. \\
Per-round failure rate ($\epsilon_{\rm L}$) & Normalized convention $\epsilon_{\rm L}=1-(1-P_{\rm L})^{1/T}$ plotted on log axes across differing $T$ (Main Figs.~1e, 2a,b, 3c,d, 4a--d). \\
Iteration count & Sequential BP steps per shot (warm-up and all legs included); means and the 99th percentile span all shots (censored at cap), whereas Main Fig.~1c plots converged medians. \\
Decoding depth & Critical-path iteration count per shot: iteration count for $S=1$, and $\max_{k=1..S}\tau_k$ (capped at $\lfloor B/S\rfloor=\lfloor 18{,}080/S\rfloor$) for the $S$-member ensemble. \\
Summed work \& CPU time & Aggregate ensemble work $\sum_{k=1}^S\tau_k$ (Supplementary Table~7); wall-clock time uses separate single-thread benchmarks (Supplementary Table~4). \\
\end{longtable}
\endgroup
\setcounter{table}{0}

\input{si/note1_properties.tex}

\setcounter{theorem}{0}
\renewcommand{\thetable}{2.\arabic{table}}
\renewcommand{\theHtable}{2.\arabic{table}}
\setcounter{table}{0}
\input{si/note2_theory.tex}
\FloatBarrier
\renewcommand{\thetable}{\arabic{table}}
\renewcommand{\theHtable}{\arabic{table}}
\setcounter{table}{0}

\renewcommand{\thetable}{3.\arabic{table}}
\renewcommand{\theHtable}{3.\arabic{table}}
\setcounter{table}{0}
\input{si/note3_mechanism.tex}
\FloatBarrier
\renewcommand{\thetable}{4.\arabic{table}}
\renewcommand{\theHtable}{4.\arabic{table}}
\renewcommand{\thetheorem}{4.\arabic{theorem}}
\renewcommand{\theHtheorem}{4.\arabic{theorem}}
\setcounter{table}{0}
\input{si/note4_convergence.tex}
\FloatBarrier
\renewcommand{\thetable}{\arabic{table}}
\renewcommand{\theHtable}{\arabic{table}}
\renewcommand{\thetheorem}{\arabic{theorem}}
\renewcommand{\theHtheorem}{\arabic{theorem}}
\setcounter{table}{0}
\input{si/note5_hardware.tex}
\FloatBarrier

\section*{Supplementary Tables}
\addcontentsline{toc}{section}{Supplementary Tables}

\subsection*{Supplementary Table 1: BB-72 failure breakdown}

Supplementary Table~\ref{tab:S1} lists every decoder at every tested noise point of
BB-72.  It is the complete version of the center-point comparison in
the main text, with the failure split into non-convergence and convergence to a
wrong logical class.  This breakdown illustrates the regime on distance-6 BB-72 where valid wrong-class outcomes dominate overall failures rather than non-convergence.  The ``Nonfinite'' column counts returned posteriors containing a
non-finite value; this statistic is reported separately and is not added to the
failure total. At $p=0.4\%$, Relay-BP-1, Relay-BP-5 and the TEMPO ensemble
share a 48,000-shot cohort; the other rows use 20,000 common shots (corresponding to per-shot failure rates of 6.72\% for BP, 6.29\% for BP+OSD+CS-10 and 4.65\% for TEMPO-BP over 20,000 shots, and 5.30\% for Relay-BP-1, 4.39\% for Relay-BP-5 and 3.72\% for the TEMPO ensemble over 48,000 shots).

\input{tables/tableS1.tex}

\subsection*{Supplementary Table 2: paired tests}

Supplementary Table~\ref{tab:S2} collects the exact McNemar tests and their Holm adjustments.
The families are defined as listed in the table: ten center-point comparisons
on the bivariate bicycle codes, eight ensemble-versus-Relay-BP-5 comparisons at
the other error rates, ten TEMPO-BP-versus-BP comparisons, ten
TEMPO-BP-versus-Relay-BP-1 comparisons, twelve new LP20/Kasai-96 comparisons,
nine Mitten-540 comparisons and nine Coset-128 comparisons on 8,000 new shots
per point. Only the LP20/Kasai-96 and Mitten-540 families were declared before
the corresponding samples were collected; the remaining families are post hoc.
The two ensemble-versus-Relay-BP-5 comparisons at $p=0.4\%$ belong only to the
ten-comparison center-point family. Probabilities
below the double-precision range are shown as $<10^{-300}$; the exact integer-tail
logarithms are retained in the CSV companion, so the ordering and the Holm
decision remain exact.  A non-significant entry is a non-rejection, not a
demonstration of equivalence. The reused LP20 and Kasai-96 center rows are
outside every Holm family and are therefore not included in the table.
On 2,000 paired shots at $p=0.175\%$ for LP20, the unadjusted exact $P$ values
against Relay-BP-5 are $2.31\times10^{-33}$ for TEMPO-BP and
$1.79\times10^{-43}$ for the TEMPO ensemble; at $p=0.06875\%$ for Kasai-96,
they are $3.38\times10^{-24}$ and $6.25\times10^{-28}$, respectively.
These reused center-point tests are descriptive, without a Holm adjustment.

\input{tables/tableS2.tex}

\subsection*{Supplementary Table 3: per-seed ensemble results}

Supplementary Table~\ref{tab:S3} gives the three independent 16,000-shot sample sets at
$p=0.4\%$ for the ensemble, Relay-BP-5 and Relay-BP-1.  These replicates verify that the ensemble advantage in both center comparisons reproduces in each independent seed alongside the pooled totals, using the complete member roster; the
per-member limits are 1,205 iterations on the gross code and 1,808 on
BB-72.  The Relay-BP-1 rows use the same seeds but are a descriptive
comparison, not part of the ensemble family.

\input{tables/tableS3.tex}

\subsection*{Supplementary Table 4: isolated timing}

Supplementary Table~\ref{tab:S4} reports every isolated single-worker timing record in the data
layer, on all six codes.  It is the table behind the cost statements in the main
text and Fig.~4e of the main text: the transfer rows use 100 fixed inputs per method, the
bivariate bicycle rows use 5,000 (gross, BB-144) and 10,000 (BB-72) inputs.
The ``Latency?'' column
is important: the ensemble row is the sum of the serial member API calls (10 or
15 members) and is not an ensemble latency.  On the gross code (BB-144), TEMPO-BP is faster in isolated single-worker runtime than BP+OSD+CS-10; on BB-72 the order reverses.  The transfer timing
does not include BP+OSD+CS-10 in the transfer benchmark.

\input{tables/tableS4.tex}

\subsection*{Supplementary Tables 5 and 6: hardware latency estimates}

Supplementary Tables~\ref{tab:S5} and~\ref{tab:S6} give the estimated decoding
time of one window for the gross code (BB-144) and BB-72 that is derived in
Supplementary Note~5.  These entries represent model estimates for a proposed fully parallel FPGA architecture rather than hardware measurements.  The columns are the mean, 99th percentile and
iteration-limit time of one 12-round (gross) or 6-round (BB-72) window,
in $\mu$s, for TEMPO-BP, the TEMPO ensemble and Relay-BP-5; the ensemble is an
ideal parallel depth and the Relay-BP-5 entries use the reference 24~ns
iteration time.  Dividing a window time by the number of rounds gives an
amortized time per round, not a latency, and is not used as a deadline.  The
tables should be read together with the assumptions in Note~5.4, especially the
untested fixed-point controller, the unmodeled sliding window and streaming
operation, the one-decoder-per-member resource assumption and the $c_{\rm hw}=4$
sensitivity.

\input{tables/tableS5.tex}
\input{tables/tableS6.tex}

\IfFileExists{tables/tableS7_work.tex}{%
\subsection*{Supplementary Table 7: work and depth}
Supplementary Table~\ref{tab:S7} reports, for the ten gross-code and
BB-72 operating points, the ensemble size, the per-member iteration cap,
the mean and 99th-percentile ideal-parallel depth of the ensemble, the total
iterations summed over its members, the corresponding Relay-BP-5 iteration
counts and the two ratios.  Across these points, aggregate ensemble work ranges from $0.87\times$ (gross code, $p=0.6\%$) to $2.68\times$ Relay-BP-5 (Supplementary Note~5.3), exposing area and work requirements alongside the bounded depth.  ``Ens. total'' gives the recorded sum over members per shot rather than an $S$-times-depth product; the depth ratio is Relay mean divided by ensemble mean
depth, and the work ratio is ensemble total divided by Relay mean iterations.

\begin{table}[!htbp]
\caption{Ensemble depth and total work against Relay-BP-5 at the ten operating points.}\label{tab:S7}
\input{tables/tableS7_work.tex}
\end{table}
}{}

\IfFileExists{tables/tableS8_parameters.tex}{%
\subsection*{Supplementary Table 8: profiles and ensemble parameters}
Supplementary Table~\ref{tab:S8} lists every TEMPO-BP profile parameter for the
two single-run TEMPO-BP profiles and for every ensemble member, to four significant digits.  It
is the parameter record behind the single-run TEMPO-BP and ensemble results: the
controller interval, residual gate, step scale, target clipping and iteration
cap are part of the algorithm's definition, and the member roster is the one
used in the ensemble.  Rather than reporting a tuning search, the table records the frozen
values used throughout evaluation.

\IfFileExists{tables/tableS8_parameters.tex}{\input{tables/tableS8_parameters.tex}}{}
}{}

\IfFileExists{tables/tableS9_actions.tex}{%
\subsection*{Supplementary Table 9: control actions}
Supplementary Table~\ref{tab:S9} classifies the controller actions on the
gross-code diagnostic sample at $p=0.4\%$: requested memory strengths below $\gamma_{\min}$, above
$\gamma_{\max}$ and inside the interval, with their share of events and of the
total absolute change in $\gamma$.  Matching the main-text breakdown, about 99\% of actions request strengths below $\gamma_{\min}$ (brake), 0.8\% above $\gamma_{\max}$ (hold) and 0.2\% inside $\Gamma$, while the rare hold and interior actions supply about 13\% of the total change in $\gamma$ (evaluated on the 10,000-shot ablation cohort).

\begin{table}[!htbp]
\caption{Control-action classes on the gross code (BB-144) at $p=0.4\%$.}\label{tab:S9}
\input{tables/tableS9_actions.tex}
\end{table}
}{}

\FloatBarrier
\begingroup
\setlength{\parskip}{2pt}
\setlength{\abovecaptionskip}{4pt}
\setlength{\textfloatsep}{6pt plus 2pt minus 2pt}
\setlength{\intextsep}{6pt plus 2pt minus 2pt}
\setlength{\floatsep}{6pt plus 2pt minus 2pt}
\section*{Supplementary Figures}
\addcontentsline{toc}{section}{Supplementary Figures}

\subsection*{Supplementary Figure 1: uniform memory on six codes}

Supplementary Figure~\ref{fig:S1} shows the uniform-memory logical failure rate per shot
on each code, with the TEMPO-BP level on the same cohort. All six codes have
complete fifteen-point grids.
The caps, sample sizes and physical error rates differ between panels;
comparisons are within each panel, not across codes. Supplementary Note~3
reports the counts and failure compositions.

\begin{figure}[!htbp]
\centering
\includegraphics[width=\textwidth]{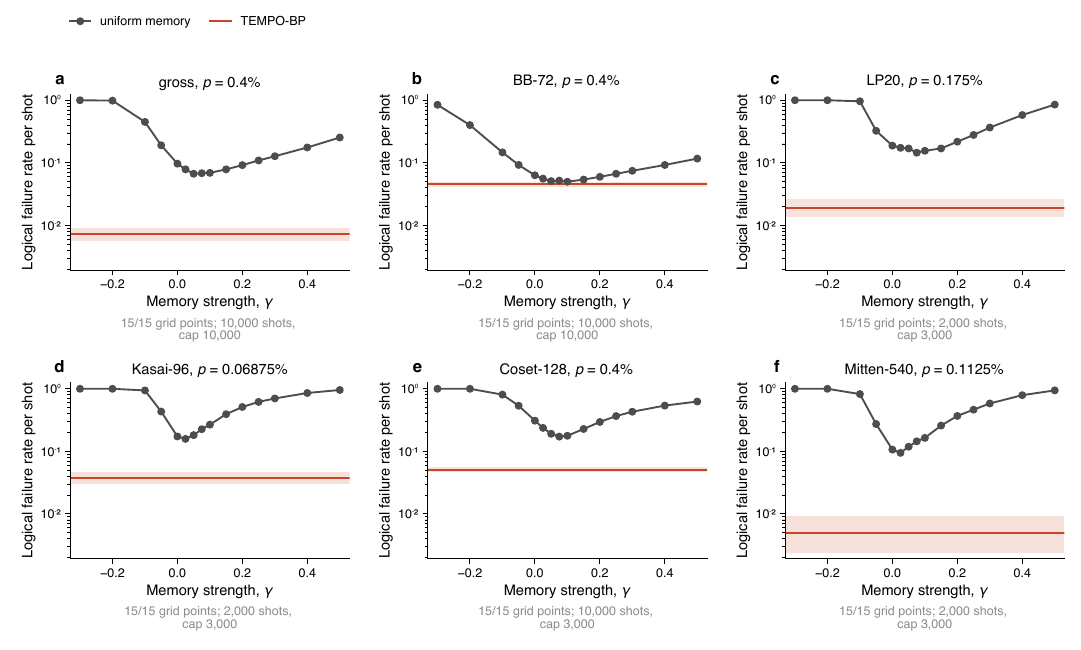}
\caption{Uniform-memory failure curves on the six codes.  Each panel shows the
logical failure rate per shot against a constant memory strength $\gamma$ (gray, exact Clopper--Pearson
95\% intervals) and the TEMPO-BP rate on the same syndrome cohort (accent line
and band).  The panels use the mechanism cohorts at the stated physical error
rate, sample size and iteration cap to probe uniform-memory behavior rather than benchmark accuracy.  The
six codes all show complete fifteen-point grids. Negative memory strengths increase failures to
converge; sufficiently large positive strengths increase logical failures
beyond the observed minimum, with different failure compositions on different
codes.}
\label{fig:S1}
\end{figure}

\subsection*{Supplementary Figure 2: the disordered-field map}

Supplementary Figure~\ref{fig:S2}a is the full $7\times6$ center--width map of
the 42 disordered fields.  Each cell is the number of failures in 4,000 common
gross-code shots; the outlined cells are the three selected on this discovery
sample and re-evaluated on fresh shots in panel b.  The map is the evidence that
the sampled low-failure region includes multiple intervals: the best discovery cell has $c=0.10,w=0.20$, but the neighboring
$c=0.10,w=0$ cell is almost as good, and wider intervals at the same center are
worse.  Panel~b confirms a shared 6.6\%--8.2\% failure floor across these best static intervals (664, 689 and 820 failures in 10,000 shots), nearly an order of magnitude above TEMPO-BP (71 failures on the same shots).

\begin{figure}[!htbp]
\centering
\includegraphics[width=0.92\textwidth]{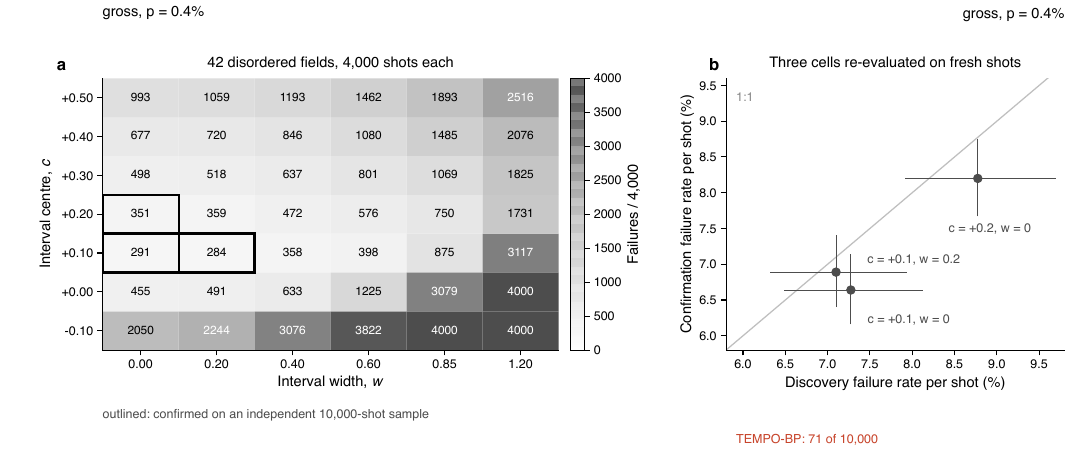}
\caption{Disordered-memory scan on the gross code.  \textbf{a}, Failure count in
4,000 common shots for each interval center $c$ and width $w$; the color scale
is the count out of 4,000.  The three outlined cells were selected on this
discovery sample.  \textbf{b}, Discovery logical failure rate per shot against confirmation
logical failure rate per shot on an independent 10,000-shot sample for the three selected cells
(errors, exact Clopper--Pearson 95\% intervals); the diagonal is equality.  The TEMPO-BP
confirmation rate is below the displayed axis range, so its count is shown as a
direct label, 71 of 10,000, rather than as a point or line.  }
\label{fig:S2}
\end{figure}
\FloatBarrier

\subsection*{Supplementary Figure 3: per-syndrome time scales}

Supplementary Figure~\ref{fig:S3}a is the split-half check behind the rank
correlation quoted in the main text: each point is one hard syndrome, with the
fitted convergence rate from the first 64 seeds against the rate from the last 64.
The points lie close to the diagonal, so the difficulty ordering is reproducible
even though the absolute rates are not tightly estimated for the slowest rows.
Panel b shows the distribution of the effective time scale $1/\hat\lambda_s$ over the 400
hard syndromes; the 1st and 99th percentiles differ by a factor of about 627.
The histogram characterizes the selected hard stratum, with the inverse fitted rate $1/\hat\lambda_s$ serving as an effective convergence time scale.

\begin{figure}[!htbp]
\centering
\includegraphics[width=0.92\textwidth]{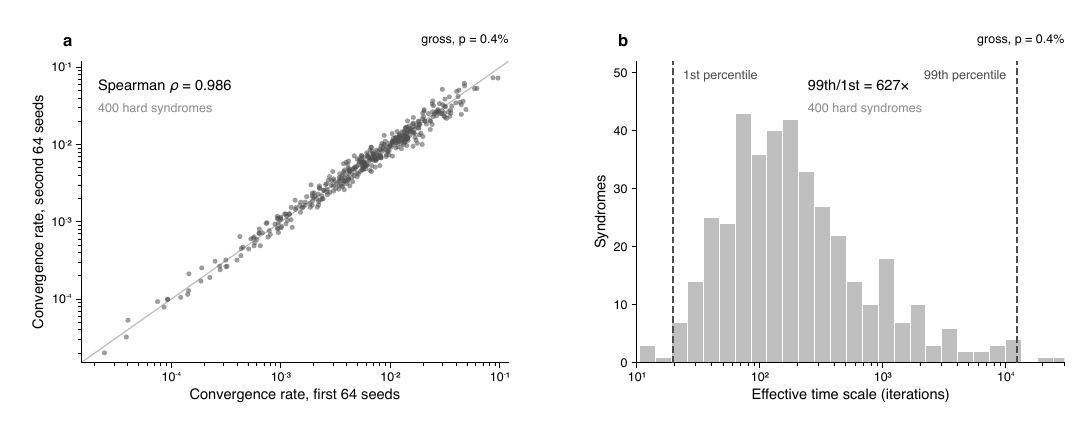}
\caption{Per-syndrome convergence-time statistics on the gross code at
$p=0.4\%$.  \textbf{a}, Convergence rate fitted on the first 64 seeds against the rate
fitted on the last 64, for the 400 hard syndromes; the line is equality and the
Spearman correlation is 0.986.  \textbf{b}, Distribution of the effective time scale $1/\hat\lambda_s$ over the same syndromes; dashed lines mark the 1st
and 99th percentiles (about 20 and 12,300 iterations, ratio 627).  The rates are
censored shifted-exponential estimates characterizing late-time convergence behavior (Supplementary Note~4).}
\label{fig:S3}
\end{figure}
\FloatBarrier

\subsection*{Supplementary Figure 4: all decoders on BB-72}

Supplementary Figure~\ref{fig:S4} completes the single-path, repeated-leg and ensemble
comparisons on BB-72. Accuracy panels show BP, Mem-BP, DMem-BP and TEMPO-BP (a);
DMem-warm, TEMPO-static and TEMPO-BP (b); and BP+OSD+CS-10, Relay-BP-1,
Relay-BP-5, the TEMPO ensemble and TEMPO-BP (c). Mean and 99th-percentile
iteration panels show BP and Mem-BP (d), DMem-warm, Relay-BP-1 and TEMPO-BP (e), and
Relay-BP-5 and the TEMPO ensemble (f). The single-path group comprises BP,
Mem-BP, BP+OSD+CS-10, DMem-BP, DMem-warm and TEMPO-BP; Relay-BP-1 and
Relay-BP-5 search over successive legs, and the TEMPO ensemble searches in parallel.
On BB-72, the single-path decoders separate much more narrowly, with TEMPO-BP
having the lowest failure rate among them at the center. Against Relay-BP-1, TEMPO-BP achieves fewer failures at all five noise rates, whereas Relay-BP-1 requires fewer iterations and less single-core time. DMem-BP and TEMPO-static are shown
for completeness in panels a and b, respectively; they are ablations, and their
iteration counts are not plotted in panels d--f.

\begin{figure}[!htbp]
\centering
\includegraphics[width=\textwidth]{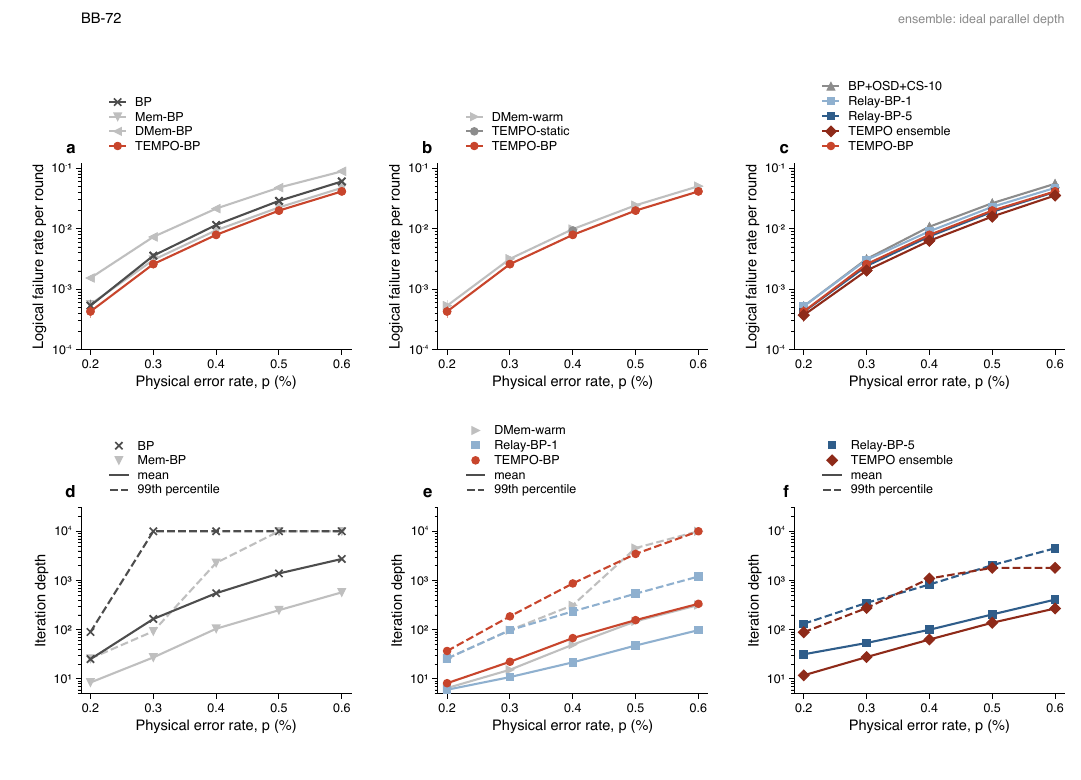}
\caption{Accuracy and iteration depth of all decoders on BB-72.
\textbf{a}--\textbf{c}, Logical failure rate per round against physical error
rate for BP, Mem-BP, DMem-BP and TEMPO-BP (\textbf{a}); DMem-warm,
TEMPO-static and TEMPO-BP (\textbf{b}); and BP+OSD+CS-10, Relay-BP-1,
Relay-BP-5, the TEMPO ensemble and TEMPO-BP (\textbf{c}). TEMPO-static was
measured only at $p=0.4\%$. \textbf{d}--\textbf{f}, Mean and
99th-percentile iterations for BP and Mem-BP (\textbf{d}), DMem-warm and
Relay-BP-1 and TEMPO-BP (\textbf{e}), and Relay-BP-5 and the TEMPO ensemble (\textbf{f});
the ensemble uses ideal parallel depth. Error bars are exact
two-sided 95\% Clopper--Pearson intervals transformed per round.  Sample sizes
are those of Supplementary Table~1; Relay-BP-1, Relay-BP-5 and the ensemble
at $p=0.4\%$ use the same 48,000-shot benchmark cohort.}
\label{fig:S4}
\end{figure}

\IfFileExists{figures/figS5.pdf}{%
\subsection*{Supplementary Figure 5: cross-code growth rates}

Supplementary Figure~\ref{fig:S5} collects the growth-branch diagnostics on
one scale.  Panel a shows the median rate of the growing trajectories against
memory strength for the six codes, with interquartile bars and the exact law
$\lambda^2-(1+\gamma)\lambda-\gamma=0$; points with fewer than five growing
trajectories are omitted.  Panel b shows the fraction of unconverged
trajectories that are growing across all 19 memory strengths.  The figure is
descriptive; the amplitude is the largest absolute variable belief.

\begin{figure}[!htbp]
\centering
\includegraphics[width=\textwidth]{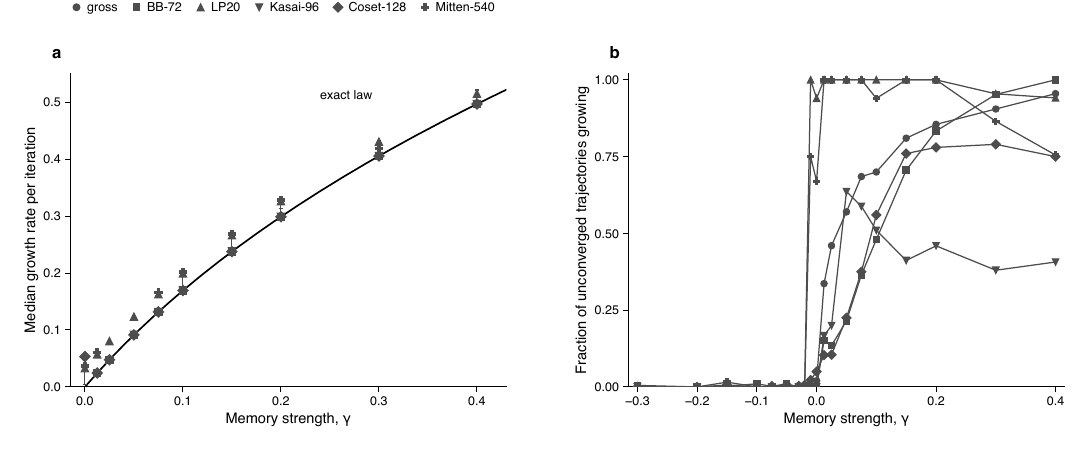}
\caption{Cross-code growth rates of unconverged trajectories.  \textbf{a},
Median late-time growth rate of the growing trajectories (rate $>10^{-3}$;
amplitude: largest absolute variable belief) against memory strength $\gamma\ge0$ for the
six codes, with interquartile bars and only points with at least five growing
trajectories; the black curve is $\ln\lambda_+$ from
$\lambda^2-(1+\gamma)\lambda-\gamma=0$.  \textbf{b}, Fraction of unconverged
trajectories that are growing against $\gamma$ for all 19 values and the six
codes.  Marker shapes follow the shared code map.  }
\label{fig:S5}
\end{figure}
}{}

\IfFileExists{figures/figS6.pdf}{%
\FloatBarrier
\subsection*{Supplementary Figure 6: temporal doubling time}
\setcounter{figure}{5}
\begin{figure}[!htbp]
\centering
\includegraphics[width=0.52\textwidth]{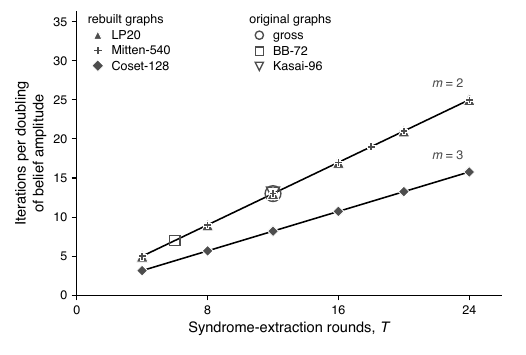}
\caption{Asymptotic doubling time of the unsigned, prior-free temporal channel
against the number of syndrome-extraction rounds $T$. Lines, the theorem gives
$T+1$ for $m=2$ and $(T+1)\ln2/\ln3$ for $m=3$. Filled symbols, exact
certificates on detector graphs rebuilt with different $T$ (triangles, LP20;
crosses, Mitten-540; diamonds, Coset-128). Open symbols, detector graphs of the
gross code, BB-72 and Kasai-96, which satisfy the hypotheses of
the theorem but on which no sampled trajectory grows at this rate.}
\label{fig:S6}
\end{figure}
}{}

\IfFileExists{figures/figS7.pdf}{%
\clearpage
\subsection*{Supplementary Figure 7: one certified control action}
\setcounter{figure}{6}
\begin{figure}[!htbp]
\centering
\includegraphics[width=0.52\textwidth]{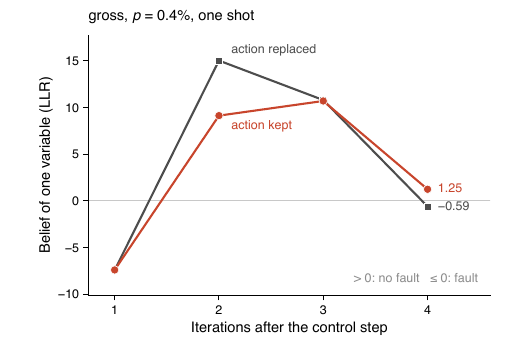}
\caption{Belief (log-likelihood ratio) of one fault variable after a control
step in two branches that differ in a single action. In this stored trajectory
the decoder reaches a syndrome-valid correction in the correct logical class
at the fourth iteration after the control step (iteration 16) only in the branch
retaining the action. This single-trajectory comparison illustrates how retaining one bounded hold action alters belief evolution over four iterations.}
\label{fig:S7}
\end{figure}
}{}

\FloatBarrier
\endgroup

%% file: sections/abstract.tex
\abstract{Quantum low-density parity-check (qLDPC) codes have attracted interest for fault-tolerant quantum computing because their high-rate constructions offer a route to reduced qubit overhead. Realizing this potential requires reliable decoding. Belief propagation (BP) forms the basis of many qLDPC decoders, but can fail to converge under circuit-level noise. Memory-based variants can improve convergence, yet selecting memory strengths that work reliably across different syndromes remains challenging. Here we introduce TEMPO-BP, which adapts each variable’s memory strength to the evolving decoding state through local feedback. On the gross code at a physical error rate of 0.4\%, single-run TEMPO-BP achieves a per-shot logical failure rate of 0.73\%, compared with 2.51\% for BP+OSD+CS-10. Ensembles of 10–15 concurrent instances match or lower failure counts relative to Relay-BP-5 at all ten tested operating points, with per-instance iteration caps at least tenfold lower than its serial limit. Without retuning, the gross-code parameters yield fewer observed failures than Relay-BP-5 in its published gross-code configuration across four additional code families. Ablations support state-dependent memory control, while a loop-gain analysis characterizes specific amplification modes in the decoding dynamics.}

%% file: sections/introduction.tex
\section{Introduction}\label{sec:intro}

Quantum error correction (QEC) is essential for protecting quantum information against noise and enabling fault-tolerant computation~\cite{terhal2015qec}. Quantum low-density parity-check (qLDPC) codes are promising candidates for this purpose: their high-rate constructions encode many logical qubits within a single block while retaining sparse stabilizer checks, offering a route to reduced qubit overhead~\cite{breuckmann2021qldpc,bravyi2024high,xu2024constant}. Favorable code parameters alone, however, do not ensure reliable operation. Realizing their benefits also requires algorithms that accurately and efficiently decode syndrome information from noisy circuits. A practical decoder should achieve low logical failure rates under circuit-level noise, scale to larger decoding problems, and sustain sufficient throughput to avoid a backlog of syndrome data~\cite{tang2026hierarchical}. When subsequent logical operations depend on decoded outcomes, both mean decoding latency and its tail must also be controlled~\cite{skoric2023parallel,tang2026hierarchical}.

\begin{figure}[!t]
\centering
\includegraphics[width=\textwidth]{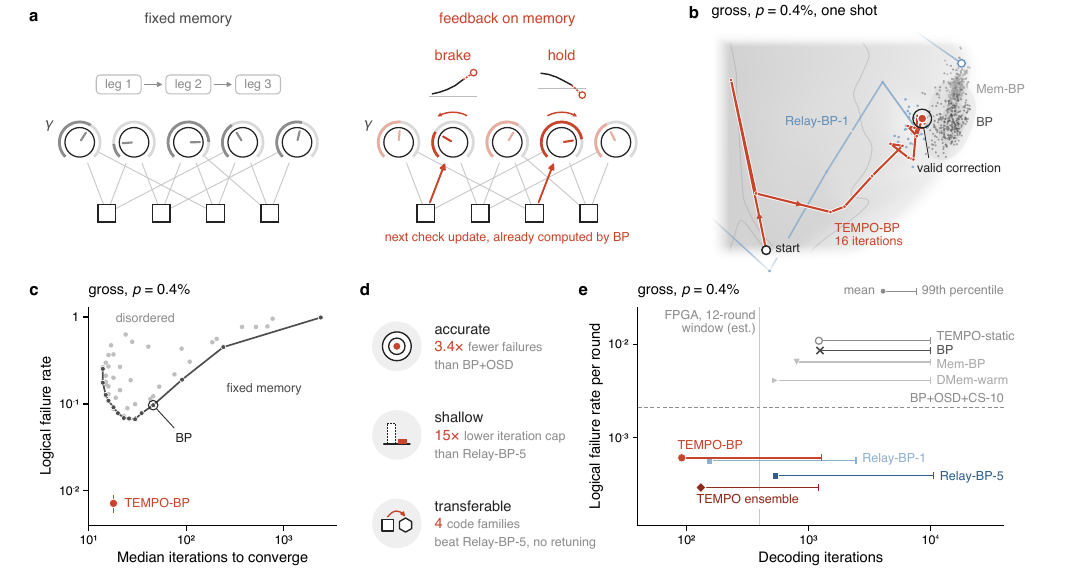}
\caption{\textbf{TEMPO-BP closes the loop on memory.} \textbf{a}, Relay-BP fixes every variable's $\gamma$ before each leg and draws a fresh field for the next leg (left); TEMPO-BP adjusts $\gamma$ online from BP's next check update, lowering $\gamma$ (brake) for target strengths below $\Gamma$ and raising it (hold) above $\Gamma$ (right). \textbf{b}, Paths for an illustrative gross-code hard shot ($p=0.4\%$) on the first two principal components of $\tanh(R/2)$ (dots, iterates). TEMPO-BP converges in 16 iterations and Relay-BP-1 in 126 after starting a new leg past an 80-iteration warm-up (ring), whereas BP and Mem-BP keep circling (iterations 10--300, clouds). Smoothed shading interpolates unsatisfied checks; tickless monotone asinh axes zoom around the valid correction. \textbf{c}, Per-shot failure rate versus converged median iterations (gross code, $p=0.4\%$) for uniform fixed memory (joined by $\gamma$; circle, plain BP; 10,000 shots, at most 10,000 iterations), disordered fixed fields (light; 4,000 shots, at most 3,000 iterations) and TEMPO-BP (same 10,000 uniform-scan shots, at most 10,000 iterations). \textbf{d}, Overview of accuracy, depth and transfer: one TEMPO-BP run has 3.4 times fewer failures than BP+OSD+CS-10 on the same 100,000 gross-code shots at $p=0.4\%$ (730 against 2,510). The per-member iteration limit of the 15-member ensemble is 15 times below Relay-BP-5 on the gross code (1,205 against 18,080), with no more failures at any of ten points. Across four other code families, one TEMPO-BP run has fewer failures than Relay-BP-5 at all 12 tested points without retuning. \textbf{e}, Per-round failure rate versus iterations (gross code, $p=0.4\%$; marker, mean; bar, 99th percentile; ensemble, mean and 99th-percentile decoding depth over concurrent members). Dashed line, BP+OSD+CS-10 (iterations not comparable). Shots: 100,000 (48,000 for Relay-BP-1, Relay-BP-5 and ensemble). Vertical line, FPGA 12-round model estimate (400 iterations at 30~ns per iteration, 1~$\mu$s period; modeling assumptions in Supplementary Note~5).}
\label{fig:concept}
\end{figure}

Belief propagation (BP) provides a common starting point for constructing qLDPC decoders~\cite{breuckmann2021qldpc,poulin2008iterative}. Its local message updates on sparse factor graphs have a regular structure amenable to parallel implementation. Standard BP, however, can stall or oscillate without producing an error estimate consistent with the measured syndrome, limiting its decoding accuracy~\cite{poulin2008iterative}. In quantum LDPC codes, short cycles in the decoding factor graph and quantum syndrome degeneracy can correlate circulating messages and trap iterative beliefs. While neural decoders and graph neural networks can improve accuracy~\cite{bausch2024learning,gong2024graphneural}, they introduce offline training costs and inference overhead whose generalization across code structures and noise conditions requires evaluation, motivating training-free modifications to message-passing dynamics. Addressing this difficulty has led to methods that modify the message-passing dynamics or supplement BP with additional decoding steps.

One approach augments BP with ordered-statistics decoding (OSD), which uses BP-derived reliabilities to construct syndrome-consistent corrections through linear-algebraic post-processing, improving accuracy at the cost of cubic-scaling Gaussian elimination on each unconverged shot~\cite{panteleev2021degenerate,roffe2020decoding}. Another approach modifies the message-passing dynamics by incorporating algorithmic memory of previous beliefs~\cite{kuo2022exploiting}. Relay-BP extends this approach by drawing variable-specific memory strengths and passing beliefs between successive decoding legs, with the strengths resampled between legs but held fixed within each leg~\cite{muller2025relay}. This strategy demonstrates the value of memory, yet holding strengths fixed within each leg requires chaining many sequential legs to escape trapped states, resulting in a heavy-tailed latency distribution. 

Adaptive feedback has already been explored in iterative decoding through adjustments to step sizes and message gains~\cite{chen2025improved,cordova2026syndromegain}, stabilizer feedback~\cite{wang2012feedback}, and restart strategies~\cite{taghipour2026rewinding}. In memory BP and Relay-BP, however, memory strengths are held fixed within each decoding run or relay leg~\cite{kuo2022exploiting,muller2025relay}. This motivates treating each variable's memory strength as an adaptive control parameter. The question is how to update these strengths from the evolving beliefs and local syndrome residuals, improving convergence within a decoding trajectory while retaining the local computational structure of BP.

Here we close the loop on memory. TEMPO-BP (TEmpered Memory via Predictive Online control) adjusts the influence of previous beliefs on subsequent updates. It uses an exact local relation: given the current decoder state and the next incoming check messages, a variable's next belief is an affine function of its memory strength. Every few iterations, the controller uses this relation to make a gradual, bounded adjustment toward a strength that would steer the next belief toward a small target of opposite sign, gated by the local fraction of unsatisfied syndrome checks and variable confidence. These updates typically act as a local brake that lowers memory where a belief is being reinforced and hold or raise it where the incoming messages would reverse its sign. The required check messages are computed one step ahead and reused in the next iteration, preserving local, parallel message passing without additional check-node updates.

In memory experiments under circuit-level noise with idling errors, a single TEMPO-BP run on the $[[144,12,12]]$ bivariate bicycle (gross) code at a 0.4\% physical error rate fails on 0.73\% of shots (circuit trials), against 9.84\% for BP, 4.82\% for warm-started disordered memory and 2.51\% for BP with ordered-statistics post-processing. Because separate instances are independent, ten to fifteen executed in parallel match or reduce failure counts relative to Relay-BP-5 at all ten operating points, under an iteration limit ten to fifteen times lower (1,205 or 1,808 iterations) and for as many copies of the decoder. Parameters tuned on the gross code carry over unchanged to four codes from other families, where a single run produces fewer failures than Relay-BP-5 in its published configuration at all twelve points tested and runs up to 118 times faster on one core.

A loop-gain analysis provides a dynamical interpretation. A variable's memory strength acts as the gain of a feedback loop. On a certified zero-prior family, plain BP is marginal and positive memory amplifies beliefs ($\lambda_+\approx 1+2\gamma$), while repeated fault mechanisms across $T$ syndrome-extraction rounds form a second, temporal amplification channel with asymptotic growth factor $m^{1/(T+1)}$. For $-1<\gamma<0$, both roots of the associated two-dimensional coefficient block lie inside the unit circle. Any memory field fixed beforehand must strike one balance between speed and convergence across all variables and syndromes, whereas feedback sets that balance variable by variable. On the gross code, the best fixed field across our uniform and 42-interval scans fails about nine times more often than TEMPO-BP.

%% file: sections/results.tex
\section{Results}\label{sec:results}

\subsection{Memory control by predictive feedback}\label{sec:control}

Under circuit-level noise, a circuit defines a detector error model: a binary matrix $H$ whose rows are detectors (parities of measurement outcomes) and whose columns are independent fault mechanisms with prior probabilities $p_i$. Given a syndrome $s$, min-sum BP iterates check-to-variable messages $m_{a\to i}$ and variable beliefs $R_i$ from prior log-likelihood ratios $L_i=\ln[(1-p_i)/p_i]$, stopping when the signs of the beliefs reproduce the syndrome (Methods). Memory BP~\cite{kuo2022exploiting,muller2025relay} replaces each variable's prior with a mixture of the prior and its previous belief,
\begin{equation}\label{eq:memory}
R_i^{(t+1)}=(1-\gamma_i)\,L_i+\gamma_i\,R_i^{(t)}+Q_i^{(t+1)},\qquad Q_i^{(t+1)}=\sum_{a\ni i}m_{a\to i}^{(t+1)},
\end{equation}
where $\gamma_i$ is the memory strength. Setting $\gamma_i=0$ recovers BP. Mem-BP assigns one positive strength to all variables, whereas Relay-BP draws the $\gamma_i$ at random from an interval extending below zero and holds them fixed during each leg: a run of up to 60 iterations with a fresh random memory field, started from the previous leg's beliefs~\cite{muller2025relay}. Terms such as shot, leg and depth are defined in the Notation and glossary of the Supplementary Information.

In our scans, a fixed uniform memory strength trades decoding speed against convergence (Fig.~\ref{fig:concept}c). Across 10,000 common shots of the gross code at a physical error rate $p=0.4\%$ (allowing 10,000 iterations), plain BP fails on 9.7\%. Positive memory accelerates converged shots---cutting their median iteration count from 46 for BP to 14 at $\gamma=0.4$---yet the per-shot failure rate dips only shallowly to 6.7\% at $\gamma=0.05$ before climbing to 25\% at $\gamma=0.5$. Negative memory produces the opposite trade-off, raising failures to 45\% at $\gamma=-0.1$ and preventing convergence altogether at $\gamma=-0.3$. Positive and negative strengths fail in different ways: almost all failures at high $\gamma$ end with belief magnitudes above $10^{12}$ or non-finite beliefs, whereas uniform negative memory keeps amplitudes bounded and slows decoding in this scan.

Fixed random memory fields remain limited by this speed--convergence trade-off across the scanned interval grid (Fig.~\ref{fig:concept}c). We scanned 42 random intervals---drawing variable strengths independently---selected the best three on 4,000 gross-code shots at $p=0.4\%$, and evaluated them on 10,000 fresh shots. The lowest observed failure rate in confirmation is 6.6\% for a zero-width interval with no negative strengths, compared with 6.7\% for the best uniform strength. On the same 10,000 shots, a single TEMPO-BP run fails on 0.71\%, a 9.4-fold reduction (95\% confidence interval 7.6 to 12.0, without adjustment for selecting the lowest confirmation rate; Supplementary Note~3). The gain over the best uniform strength likewise reaches a factor of 3.4 on Coset-128, versus a 9\% failure reduction on distance-6 BB-72 where valid wrong-class corrections dominate (Supplementary Note~3).

Equation~\eqref{eq:memory} points directly to a local remedy because the check field $Q_i^{(t+1)}$ depends only on variable messages already sent, not on the new memory strengths. Once $Q_i^{(t+1)}$ is known, variable $i$'s next belief is an exact affine function of its own memory strength (Fig.~\ref{fig:law}d),
\begin{equation}\label{eq:response}
R_i^{(t+1)}(\gamma_i)=\bigl[L_i+Q_i^{(t+1)}\bigr]+\gamma_i\,\bigl[R_i^{(t)}-L_i\bigr],
\end{equation}
whose slope is the variable's current displacement from its prior and whose intercept is the memoryless BP next-step belief. Computing $Q_i^{(t+1)}$ requires the next iteration's check-node update, which the decoder must perform anyway; TEMPO-BP computes it one step early and reuses it, adding no check-node updates.

To retain positive memory's speed while locally countering belief reinforcement, TEMPO-BP steers each free variable's one-step prediction toward a small target of opposite sign (Fig.~\ref{fig:concept}a). Every $K$ iterations (the control interval; $K=4$ on the gross code and transfer codes, $K=8$ on BB-72), each free variable applies equation~\eqref{eq:response}, solves for the memory strength $\gamma_i^\star$ that would bring its next belief to that target (zero if the current belief is zero) and projects the result onto an admissible interval $\Gamma=[\gamma_{\min},\gamma_{\max}]$ with $\gamma_{\min}<0<\gamma_{\max}$. It then moves partway toward that projection, taking larger steps for confident variables and variables adjacent to unsatisfied checks (Methods). When $\gamma_i^\star<\gamma_{\min}$, the controller lowers $\gamma_i$ toward $\gamma_{\min}$ (brake). When $\gamma_i^\star>\gamma_{\max}$, it raises $\gamma_i$ toward $\gamma_{\max}$ (hold). Increasing the memory strength shifts the next-belief prediction toward the current sign when $R_i^{(t)}[R_i^{(t)}-L_i]>0$. Starting from moderately positive memory for initial speed, the controller selectively lowers strength where reinforcement persists, escaping the cyclic trajectories of fixed-memory decoding (Fig.~\ref{fig:concept}b) and yielding the accuracy, bounded depth and transfer summarized in Fig.~\ref{fig:concept}d.

\subsection{Single-path decoding on bivariate bicycle codes}\label{sec:single}

\begin{figure}[!htbp]
\centering
\includegraphics[width=\textwidth]{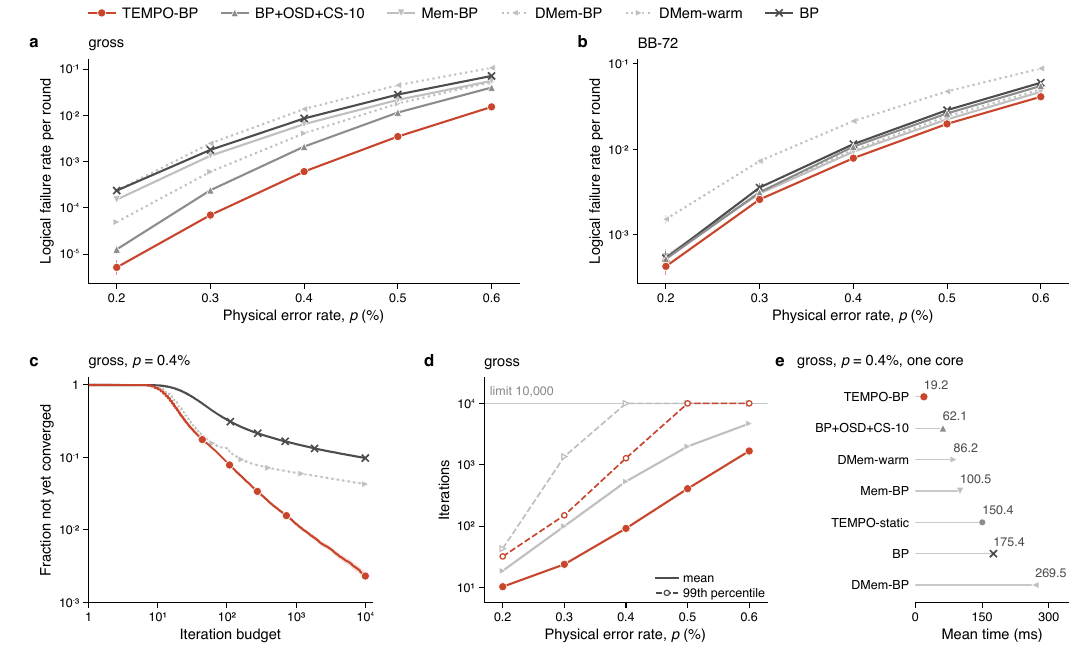}
\caption{\textbf{Single-path decoding on bivariate bicycle codes.} \textbf{a},\textbf{b}, Logical failure rate per round against physical error rate $p$ for the six single-path decoders shown in the legend on the gross code (\textbf{a}) and BB-72 (\textbf{b}) in X-basis memory experiments of 12 and 6 rounds; the two panels have separate vertical ranges. A failure is a shot that does not converge or converges to the wrong logical class; rates per round are obtained from rates per shot as described in Methods. Shots per point, in order of increasing $p$: 500,000, 100,000, 100,000, 15,000 and 10,000 (\textbf{a}) and 30,000, 15,000, 20,000, 15,000 and 10,000 (\textbf{b}). \textbf{c}, Fraction of shots not yet converged against the iteration budget on the gross code at $p=0.4\%$ for BP, DMem-warm and TEMPO-BP. \textbf{d}, Mean (solid) and 99th-percentile (dashed) number of iterations on the gross code for DMem-warm and TEMPO-BP; the horizontal line marks their common limit of 10,000 iterations. \textbf{e}, Mean time per shot on one processor core at $p=0.4\%$ for the seven single-path decoders with isolated timings (5,000 shots each). Error bars and bands, exact 95\% confidence intervals.}
\label{fig:single}
\end{figure}

Evaluating online feedback independently of multi-leg resampling requires comparing single-path decoders on X-basis memory circuits of the public Relay-BP benchmark for BB-72 (6 rounds) and the gross code (BB-144, 12 rounds)~\cite{relaybprepo} from $p=0.2\%$ to $0.6\%$ (Methods and Supplementary Information). In single-path decoding, each trajectory uses one memory field---either fixed or adjusted online by feedback---rather than resampling across successive legs. Unless stated otherwise, decoders share common shot cohorts and any unconverged shot counts as a logical failure. To isolate online feedback from Relay-BP's search over successive legs (examined in Section~\ref{sec:ensemble}), this single-path comparison comprises plain BP, Mem-BP, disordered fixed memory (DMem-BP), warm-started DMem-warm (80 Mem-BP iterations followed by one DMem-BP run, up to 10,000 total iterations), ordered-statistics post-processing (BP+OSD+CS-10) and TEMPO-BP. DMem-warm is the main open-loop baseline. The text quotes failure rates per shot and failure counts; Fig.~\ref{fig:single}a,b plots rates per round (Methods).

Across all five error rates on both bivariate bicycle codes, a single TEMPO-BP run achieves the lowest logical failure rate among all tested single-path decoders (Fig.~\ref{fig:single}a,b). On the gross code from $p=0.2\%$ to $0.6\%$, TEMPO-BP reduces failures by factors of 3.5 to 46 relative to BP, 2.8 to 9.5 relative to DMem-warm and 2.3 to 3.4 relative to BP+OSD+CS-10. At the center point $p=0.4\%$ (100,000 common shots), TEMPO-BP fails on 0.73\% of shots (730 failures; $6.10\times 10^{-4}$ per round), against 2.51\% for BP+OSD+CS-10 ($2.12\times 10^{-3}$ per round), 4.82\% for DMem-warm ($4.11\times 10^{-3}$), 7.49\% for Mem-BP ($6.46\times 10^{-3}$), 9.84\% for BP ($8.59\times 10^{-3}$) and 15.3\% for un-warm-started DMem-BP. On BB-72, TEMPO-BP again leads every baseline at all five noise rates: at $p=0.4\%$ (20,000 common shots), it fails on 4.65\% of shots (929 failures; $7.90\times 10^{-3}$ per round), against 5.51\% for Mem-BP, 5.75\% for DMem-warm and 6.29\% for BP+OSD+CS-10.

These gains remain statistically significant on both codes after Holm correction at $p=0.4\%$ against every baseline with a paired test (Holm-adjusted exact McNemar $P<10^{-300}$ on the gross code and $P\le 8.8\times 10^{-13}$ on BB-72; Supplementary Table~2). Error-mode splits further show different dominant failure modes on the two codes. On the gross code, fixed-memory baselines fail predominantly by not converging within the iteration cap (9.82\% unconverged for BP and 4.28\% for DMem-warm at $p=0.4\%$), whereas TEMPO-BP cuts non-convergence to 231 shots (0.23\%) alongside 499 valid wrong-class corrections (0.50\%). On distance-6 BB-72, by contrast, convergence is rarely limiting: TEMPO-BP already converges on all but 38 of 20,000 shots (0.19\%), leaving valid corrections in the wrong logical class as the dominant barrier (Supplementary Table~1).

Online feedback also cuts both the sequential iteration counts pacing parallel hardware and single-core CPU runtime on the gross code (Fig.~\ref{fig:single}c--e). At $p=0.4\%$ (Fig.~\ref{fig:single}c), TEMPO-BP and DMem-warm track similarly at low iteration budgets before open-loop DMem-warm stalls near 4.28\% unconverged shots, whereas adaptive control continues resolving hard syndromes down to 0.23\%---18.5 times lower than DMem-warm and 42.5 times lower than BP (9.82\%). Under the 10,000-iteration cap (Fig.~\ref{fig:single}d), TEMPO-BP averages 92 iterations at $p=0.4\%$ with a 99th percentile of 1,278, compared with 533 mean iterations and a saturated 10,000-iteration tail for DMem-warm, keeping a lower mean across all five noise rates and avoiding tail saturation until $p\ge0.5\%$. In isolated single-core timings over 5,000 gross-code shots at $p=0.4\%$ (Fig.~\ref{fig:single}e), TEMPO-BP averages 19.2~ms per shot---3.2 times faster than BP+OSD+CS-10 (62.1~ms) and 4.5 times faster than DMem-warm (86.2~ms), ahead of Mem-BP (100.5~ms), TEMPO-static (150.4~ms), BP (175.4~ms) and DMem-BP (269.5~ms; Supplementary Table~4). Plotted iteration counts omit the small controller cycle overhead every $K$ steps, which enters the hardware estimate in Supplementary Note~5.

\subsection{Successive legs and parallel ensembles}\label{sec:ensemble}

Relay-BP searches random memory fields sequentially within one run: after an 80-iteration Mem-BP warm start, it chains up to 300 legs until collecting one valid correction (Relay-BP-1, single-solution) or five (Relay-BP-5, multi-solution), returning the candidate most probable under the priors~\cite{muller2025relay}. The TEMPO ensemble instead launches $S$ independent TEMPO-BP runs in parallel.

Even as a single trajectory without repeated legs, TEMPO-BP achieves statistically significant failure reductions over single-solution Relay-BP-1 ($S_{\rm R}=1$) at six of the ten tested operating points---by 19\% and 21\% on the gross code at $p=0.5\%$ and $0.6\%$, and across $p=0.3\%$--$0.6\%$ on BB-72 (Fig.~\ref{fig:concept}e, Supplementary Fig.~4c,e and Supplementary Tables~1 and~2). At the gross-code center point, where Relay-BP-1 uses a separate 48,000-shot cohort, the paired test uses 5,000 matched shots, with 26 failures for TEMPO-BP and 32 for Relay-BP-1. On the gross code at $p=0.4\%$, TEMPO-BP also uses fewer iterations (mean 92 and 99th percentile 1,278 against 154 and 2,458; Fig.~\ref{fig:concept}e) and less single-core time (19.2~ms against 26.4~ms; Supplementary Table~4). Across all five gross-code points, TEMPO-BP has a lower mean iteration count than Relay-BP-1 and a lower 99th percentile at all three points where neither decoder reaches its cap. On BB-72, TEMPO-BP records fewer failures than Relay-BP-1 at all five noise rates (significantly at $p=0.3\%$--$0.6\%$; Supplementary Tables~1 and~2), whereas Relay-BP-1 requires fewer iterations and less single-core time (mean 22 against 68 iterations and 0.93~ms against 3.53~ms at $p=0.4\%$; Supplementary Fig.~4e).

\begin{figure}[!htbp]
\centering
\includegraphics[width=\textwidth]{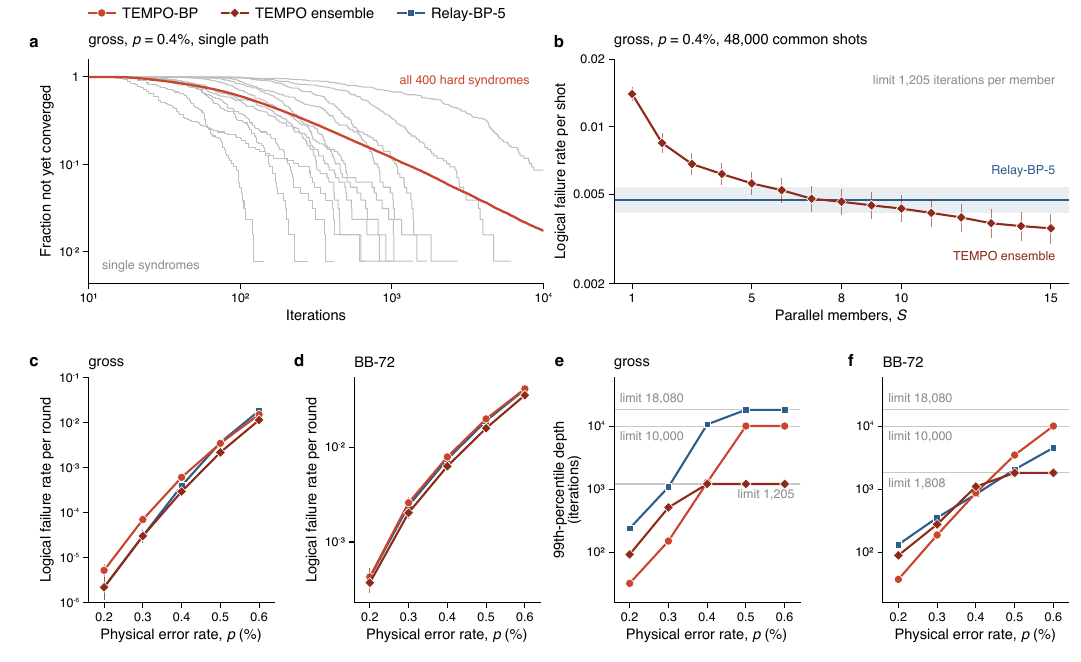}
\caption{\textbf{Convergence-time tails and parallel ensembles.} \textbf{a}, Fraction of runs not yet converged against iterations for 400 hard syndromes of the gross code at $p=0.4\%$, each decoded with 128 seeds: all runs pooled (red) and twelve single syndromes spanning the 5th to 95th percentiles of median convergence time (gray). \textbf{b}, Logical failure rate per shot against the number of parallel members $S$, for the first $S$ members of the 15-member TEMPO ensemble on 48,000 common shots of the gross code at $p=0.4\%$; each member is limited to 1,205 iterations. Line and band, Relay-BP-5 on the same shots and its exact 95\% confidence interval; the ensemble has fewer failures from $S=8$ (222 against 227). \textbf{c},\textbf{d}, Logical failure rate per round against $p$ for single-run TEMPO-BP, the TEMPO ensemble and Relay-BP-5 on the gross code (\textbf{c}; 15 members) and BB-72 (\textbf{d}; 10 members). \textbf{e},\textbf{f}, 99th percentile of the decoding depth against $p$ on the same shots; depth equals sequential iterations for individual decoder runs and the largest iteration count among concurrent members for the ensemble. Light lines mark the iteration limits (1,205 or 1,808 per member; 10,000 for TEMPO-BP; 18,080 for Relay-BP-5). Shots per point as in Fig.~\ref{fig:single}, except the TEMPO ensemble and Relay-BP-5 use 48,000 at $p=0.4\%$. Error bars and band, exact 95\% confidence intervals; \textbf{c} and \textbf{d} have separate vertical ranges.}
\label{fig:ensemble}
\end{figure}

Residual non-convergence on the gross code is associated with reproducible differences in convergence-time scales across syndromes, motivating independent parallel runs to truncate hard tails (Fig.~\ref{fig:ensemble}a). At $p=0.4\%$, roughly one third of single-run TEMPO-BP failures (231 of 730) still fail to converge within 10,000 iterations. Decoding 400 hard syndromes (requiring more than 100 initial iterations) with 128 random initial memory fields each reveals fitted convergence scales that are highly reproducible across disjoint seed halves (rank correlation 0.986). Across syndromes, those time scales span a factor of about 630 between the 1st and 99th percentiles, consistent with the observed population power-law tail with exponent below one over the sampled iteration window. Consistent with randomized search theory~\cite{luby1993restart,gomes2000heavytails}, running independent 1,205-iteration replicas of one TEMPO-BP configuration cuts failures on 100,000 common gross-code shots from 1,360 (one replica) to 927 (two) and 419 (fifteen; Supplementary Note~4).

Because concurrent TEMPO-BP instances exchange no state, organizing them into a parallel $S$-member ensemble keeps the combined iteration budget within Relay-BP's limit while lowering the bound on decoding depth tenfold to fifteenfold. Setting each member's iteration cap to $\lfloor 18{,}080/S\rfloor$ yields $S=15$ members (1,205 iterations each) on the gross code and $S=10$ (1,808 iterations each) on BB-72, returning the valid correction most probable under the priors. While parallel width with independent random seeds captures much of the tail reduction, parameter diversity across the ensemble roster provides a further statistically significant failure reduction over identical-parameter replicas (376 against 419 failures on 100,000 shots; exact McNemar $P=0.009$).

Across all ten bivariate-bicycle operating points, the full TEMPO ensemble yields equal or lower logical failure rates than five-solution Relay-BP-5, with the largest margins at high physical error rates (Fig.~\ref{fig:ensemble}b--d). On 48,000 common shots at $p=0.4\%$, adding members sequentially puts the ensemble ahead of Relay-BP-5 from $S=8$ on the gross code (Fig.~\ref{fig:ensemble}b) and from $S=2$ on BB-72. At full width, the ensemble matches or reduces Relay-BP-5's failure count at all ten points, achieving fewer failures at eight and identical counts at the two lowest gross-code rates ($13$ in 500,000 at $p=0.2\%$ and $36$ in 100,000 at $p=0.3\%$). At $p=0.4\%$, per-shot failure rates fall from 0.47\% ($3.95\times 10^{-4}$ per round) to 0.35\% ($2.94\times 10^{-4}$ per round) on the gross code and from 4.39\% ($7.45\times 10^{-3}$) to 3.72\% ($6.30\times 10^{-3}$) on BB-72. At $p=0.5\%$ and $0.6\%$ on the gross code, failures drop by 39\% and 36\%, a regime in which even single-run TEMPO-BP has fewer failures than Relay-BP-5 (Fig.~\ref{fig:ensemble}c). After Holm correction, all seven differences at higher noise ($p\ge 0.3\%$ on BB-72 and $p\ge 0.4\%$ on the gross code) are statistically significant (Supplementary Table~2), and both $p=0.4\%$ advantages replicate across all three independent 16,000-shot sample sets (Supplementary Table~3). Both error types are less frequent for the ensemble on the gross code at $p=0.4\%$: on their respective 100,000- and 48,000-shot cohorts, a single TEMPO-BP run and the TEMPO ensemble record non-convergence rates of 0.23\% and 0.08\%, and wrong-class rates of 0.50\% and 0.27\%.

The ensemble achieves this accuracy at a bounded decoding depth, using $S$ concurrent members to reduce both mean and tail iteration depth (Fig.~\ref{fig:ensemble}e,f). Mean ensemble depth falls below Relay-BP-5's mean iteration count at all ten points by factors of 1.5 to 10, and 99th-percentile depth is lower at nine of the ten points (1,097 against 828 on BB-72 at $p=0.4\%$). Per-member caps also prevent tail blow-up: on the gross code at $p=0.6\%$, Relay-BP-5 saturates its 18,080-iteration limit while ensemble depth never exceeds 1,205 (Fig.~\ref{fig:ensemble}e). On BB-72 at $p=0.5\%$ and $0.6\%$, 99th-percentile ensemble depth reaches its cap of 1,808 below both Relay-BP-5 (2,047 and 4,520) and single-run TEMPO-BP (3,469 and 10,000; Fig.~\ref{fig:ensemble}f and Supplementary Table~1). Achieving those depths requires $S$ concurrent decoder copies executing fewer summed iterations than Relay-BP-5 on the gross code at $p=0.6\%$ ($0.87\times$) and 1.04 to 2.68 times as many on average across the remaining nine points (Supplementary Table~7).

\subsection{Transfer across code families without retuning}\label{sec:transfer}

A decoder whose parameters must be tuned for every code is of limited use while codes are still being designed. To test transferability, we applied the frozen gross-code configurations of TEMPO-BP and the TEMPO ensemble to four codes built on distinct principles: a $[[1122,148,\le20]]$ lifted-product code (LP20)~\cite{cain2026shor}, a $[[1152,580,\le12]]$ code of rate close to one half (Kasai-96)~\cite{kasai2026orthogonality,zhao2026ultrahigh}, a $[[128,16,12]]$ coset code (Coset-128)~\cite{aydin2026coset} and a $[[540,108,18]]$ non-Abelian lifted-product code (Mitten-540)~\cite{bhardwaj2026highrate} (Table~\ref{tab:transfer}). We simulated each code at three noise rates under the bivariate bicycle noise model (idling errors included) against BP and Relay-BP-5 in its published gross-code configuration. This protocol therefore evaluates out-of-the-box parameter carryover relative to Relay-BP-5's published gross-code baseline, without per-code retuning of either decoder.

\input{tables/table1.tex}

\begin{figure}[!htbp]
\centering
\includegraphics[width=\textwidth]{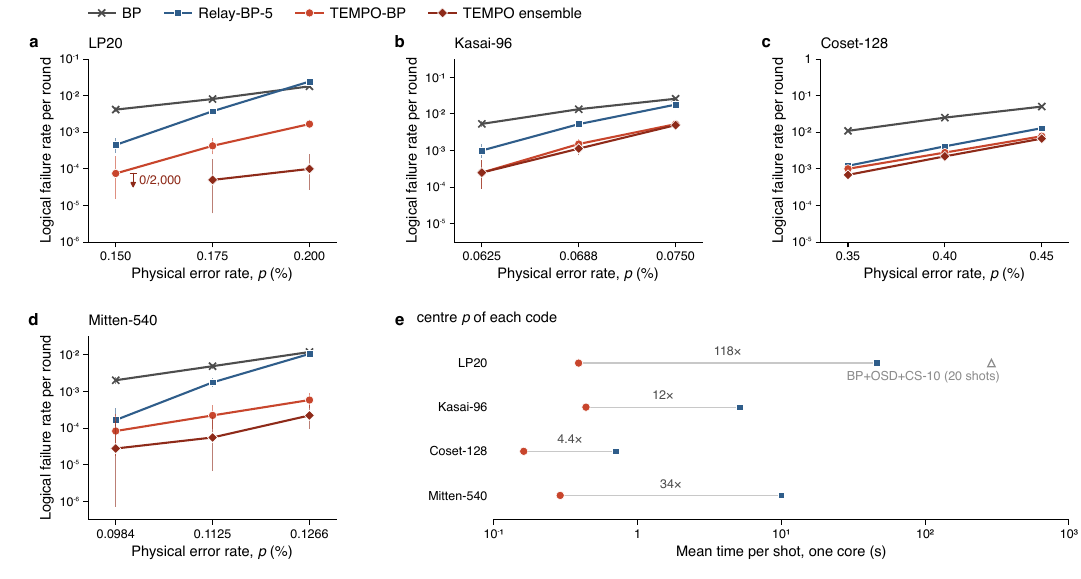}
\caption{\textbf{One parameter set across code families.} \textbf{a}--\textbf{d}, Logical failure rate per round against physical error rate for LP20 (\textbf{a}, 20 rounds), Kasai-96 (\textbf{b}, 12 rounds), Coset-128 (\textbf{c}, 12 rounds) and Mitten-540 (\textbf{d}, 18 rounds). Panels compare BP, Relay-BP-5 in the configuration published for the gross code, and TEMPO-BP and the TEMPO ensemble with the parameters selected on the gross code. Each point comprises 2,000 shots (10,000 in \textbf{c}). Error bars, exact 95\% confidence intervals; the cap with a downward arrow in \textbf{a} marks the one-sided 95\% upper limit for zero failures in 2,000 shots. \textbf{e}, Mean time per shot on one processor core at the center error rate of each code for Relay-BP-5 and TEMPO-BP (100 shots each, timed in isolation), with their ratio. Open triangle, BP+OSD+CS-10 on LP20 in a separate test on 20 shots.}
\label{fig:transfer}
\end{figure}

Across all four transfer code families without retuning, a single TEMPO-BP run fails less often than five-solution Relay-BP-5 at every one of the twelve tested operating points, and the TEMPO ensemble lowers failures further at eleven points while matching single-run TEMPO-BP at one (Fig.~\ref{fig:transfer}a--d and Table~\ref{tab:transfer}). At the center operating points summarized in Table~\ref{tab:transfer}, single-run TEMPO-BP reduces failures relative to Relay-BP-5 by factors of 8.5 on LP20 (17 versus 145 failures in 2,000 shots; 0.85\% versus 7.25\%), 3.4 on Kasai-96 (36 versus 123; 1.80\% versus 6.15\%), 1.5 on Coset-128 (330 versus 485 in 10,000 shots; 3.30\% versus 4.85\%) and 7.9 on Mitten-540 (8 versus 63 in 2,000 shots; 0.40\% versus 3.15\%). Under higher noise, these margins widen further: on LP20 at $p=0.2\%$ (2,000 shots), BP fails 609 times and Relay-BP-5 fails 778 times---more than BP---whereas TEMPO-BP fails 66 times and the ensemble 4 times. At the highest noise rate of Mitten-540 (2,000 shots), BP fails 391 times and Relay-BP-5 fails 351 times, against 21 for TEMPO-BP and 8 for the ensemble. Corresponding counts are 543 and 392 against 123 and 116 on Kasai-96 (2,000 shots), and 4,620 and 1,449 against 897 and 778 on Coset-128 (10,000 shots). Paired comparisons against Relay-BP-5 confirm statistically significant reductions at every operating point of LP20, Kasai-96 and Coset-128 and at the two higher noise rates of Mitten-540 (unadjusted at the reused LP20 and Kasai-96 centers; Methods and Supplementary Table~2), while lowering Mitten-540's lowest-noise failure count from 6 to 3 (TEMPO-BP) and 1 (ensemble) in 2,000 shots.

Cross-family speedups in iteration count and wall-clock time exceed the gains in accuracy (Fig.~\ref{fig:transfer}e). On LP20---whose detector error model is the largest tested (10,395 detectors, 139,722 fault mechanisms)---Relay-BP-5 averages 5,551 to 17,550 iterations versus 75 to 620 for TEMPO-BP. On one isolated CPU core at the center noise rates, TEMPO-BP achieves speedups over Relay-BP-5 of 118-fold on LP20 (0.39~s against 46~s at $p=0.175\%$), 34-fold on Mitten-540, 12-fold on Kasai-96 and 4.4-fold on Coset-128 (Supplementary Table~4). Because of its computational cost, BP+OSD+CS-10 was omitted from the transfer sweeps after averaging 289~s per shot in a separate 20-shot LP20 timing check.

\subsection{Memory strength as a loop gain}\label{sec:law}

\begin{figure}[!htbp]
\centering
\includegraphics[width=\textwidth]{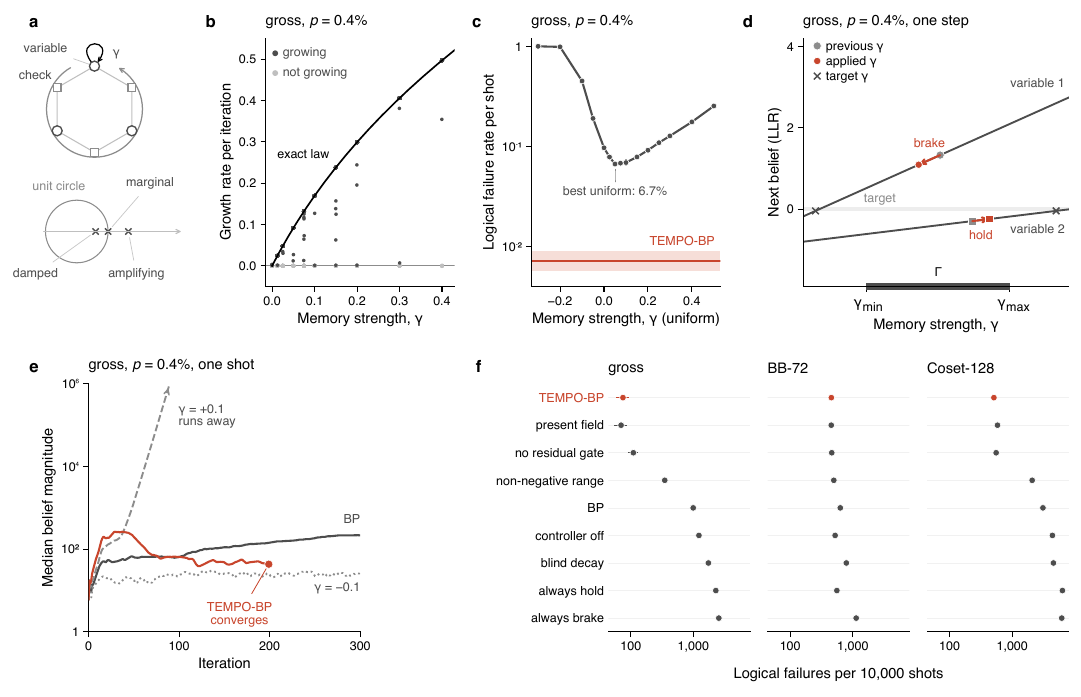}
\caption{\textbf{Memory strength as a loop gain and controller ablations.} \textbf{a}, A belief returns to its variable around a cycle of the Tanner graph, and the memory strength $\gamma$ feeds it back directly (top). For equation~\eqref{eq:law} with $b=1$, both roots lie inside the unit circle for $-1<\gamma<0$; the larger root lies on it for plain BP ($\gamma=0$) and outside it for $\gamma>0$ (bottom; positions illustrative). \textbf{b}, Growth rate per iteration of the largest belief magnitude under uniform memory on the gross code at $p=0.4\%$, one point per trajectory that had not converged after 3,000 iterations (about 200 per memory strength; Methods). Dark points, growing trajectories (rate above $10^{-3}$); light points, trajectories below this growth-rate threshold in the measured window. Line, $\ln\lambda_+$ from equation~\eqref{eq:law} with $b=1$. \textbf{c}, Logical failure rate per shot of uniform-memory decoding on 10,000 common shots of the gross code at $p=0.4\%$ with at most 10,000 iterations. Line and band, TEMPO-BP on the same shots and its exact 95\% confidence interval. \textbf{d}, The next belief of a variable is exactly affine in its memory strength (equation~\eqref{eq:response}). Two variables of one trajectory are shown with the target band, the memory strength that would reach it (crosses), the admissible interval $\Gamma$ and the move made by the controller (gray to red marker). \textbf{e}, Median belief magnitude over variables during the decoding of one shot (illustrative example selected from a bounded screen) by plain BP, by uniform memory of either sign and by TEMPO-BP. \textbf{f}, Failures per 10,000 shots for TEMPO-BP and for controller ablations (Methods), on diagnostic samples of the gross code, BB-72 and Coset-128, with at most 10,000, 10,000 and 3,000 iterations.}
\label{fig:law}
\end{figure}

A variable's memory strength acts as the gain of a feedback loop, and on an exactly solvable family of states plain BP sits at unit gain (Fig.~\ref{fig:law}a). In a Tanner graph with cycles, outgoing messages return around loops, while memory feeds $\gamma_i$ times each belief back once more. To isolate this interplay exactly, we consider zero-prior states with uniform $\gamma\ge0$ where every check holds a tie for its minimum incoming magnitude, each variable receives from its checks $c$ times its own magnitude, and the signs agree with the syndrome on checks with nonzero minima. On this invariant family, nonlinear min-sum reduces to a two-dimensional linear map whose growth factors $\lambda$ satisfy
\begin{equation}\label{eq:law}
\lambda^2-(b+\gamma)\,\lambda-\gamma=0,\qquad b=c-1,
\end{equation}
where $b$ is a non-negative integer fixed by the family (Supplementary Note~2). We constructed such a state with $b=1$ explicitly on the complete gross-code detector error model (936 detectors, 8,784 fault mechanisms and 30,672 edges) and verified invariance in integer arithmetic. At $b=1$, plain BP ($\gamma=0$) yields roots 1 and 0 (marginal); any $\gamma>0$ pushes the dominant root outside the unit circle ($\lambda_+\approx 1+2\gamma$), driving exponential growth of the dominant mode. Conversely, $-1<\gamma<0$ moves both roots of this $2\times2$ coefficient block inside the unit circle (giving Schur stability of the coefficient block, though the tied family need not remain invariant for $\gamma<0$ and uniform negative memory slows decoding in the scan; Fig.~\ref{fig:law}c,e).

The growing branch of unconverged gross-code trajectories under uniform positive memory has a median growth rate matching the eigenvalue prediction of equation~\eqref{eq:law} (Fig.~\ref{fig:law}b). We tracked trajectories that remain unconverged at 3,000 iterations on separate gross-code diagnostic samples ($p=0.4\%$, roughly 200 trajectories per $\gamma$ across nine strengths in $[0.0125,0.4]$). Across this range, the growing fraction rises from 34\% to 96\%, and over half of growing trajectories match $\ln\lambda_+(\gamma)$ to four significant digits (a pattern shared on BB-72, Kasai-96 and Coset-128; Supplementary Fig.~5). On LP20 and Mitten-540, however, most growing trajectories outpace equation~\eqref{eq:law}---reaching a growth factor of 1.22 instead of 1.18 at $\gamma=0.1$, close to separate zero-prior positive-cone certificates (Supplementary Notes~2 and~3)---pointing to amplification along a second loop.

Repeated syndrome-extraction rounds create this second amplification channel along temporal chains, with a per-iteration growth factor that approaches unity as the round window $T$ increases. Faults flipping the same check in consecutive rounds form time chains closed at the two time boundaries. We prove that in a $T$-round memory experiment satisfying sufficient detector-graph conditions met by all six codes, unsigned zero-prior min-sum multiplies an explicit strictly positive state by an integer $m\ge2$ every $T+1$ iterations (Supplementary Note~2). Under the same hypotheses, every strictly positive initial message state has asymptotic growth factor $m^{1/(T+1)}$. In agreement with this proof, most sampled unconverged plain-BP trajectories on LP20 and Mitten-540 grow at this temporal rate, whereas the other four codes show no match to their strictly positive-cone rates, although some Coset-128 trajectories match a distinct signed temporal rate (Supplementary Note~3), while predicting a priori which loop captures an individual signed trajectory remains open.

On the gross code, controller ablations show that negative memory, ineffective when applied uniformly in the scan (Fig.~\ref{fig:law}c,e), helps when feedback applies it selectively along the affine response line (Fig.~\ref{fig:law}d) together with rare hold actions (Fig.~\ref{fig:law}f). Across 10,000 gross-code diagnostic shots at $p=0.4\%$, 99.0\% of the $2\times10^{9}$ controller actions request strengths below $\gamma_{\min}$, yet the 0.8\% requesting above $\gamma_{\max}$ and 0.2\% inside $\Gamma$ supply 13\% of total memory-strength change (Supplementary Table~9). Removing that state selectivity causes large regressions: against 75 TEMPO-BP failures on the 10,000-shot diagnostic sample, the \emph{always brake} ablation fails 2,552 times (worse than 994 for plain BP, 1,227 for controller off and 1,738 for blind decay), forbidding negative memory ($\gamma_{\min}=0$) leaves 349 failures and dropping unsatisfied-check gating leaves 110. Tests across codes confirm the same qualitative pattern: every memory-range, action-selection or gating ablation raises failures on Coset-128, LP20 and BB-72, although their relative ordering shifts by code and unsatisfied-check gating has no statistically significant effect on BB-72 (454 against 461, Holm $P=0.63$, versus $P=1.3\times 10^{-4}$ on the gross code and Holm $P\le 0.037$ on Coset-128 and LP20); replacing $Q_i^{(t+1)}$ with $Q_i^{(t)}$ (\emph{present field}) increases failures on Coset-128 from 500 to 581 (exact McNemar $P=2\times10^{-6}$, Holm $P=8.7\times10^{-6}$), with no statistically significant difference detected on BB-144, BB-72 or LP20 (Holm $P\ge 0.68$; Supplementary Note~3).

Open-loop diagnostics show that the best fixed memory strength in the tested grid varies across codes, whereas feedback adapts memory from the evolving decoder state. On separate diagnostic samples (capped at 10,000 iterations on the gross code and 3,000 on the transfer codes), drawing a single open-loop field from Relay-BP's published gross-code interval (without warm start, successive legs or candidate selection) fails on 15.6\% of gross-code shots. Across the transfer codes, the same open-loop sampling fails, almost always without converging, on 52\% of LP20 shots (against 1.9\% for TEMPO-BP), 81\% of Kasai-96 (against 3.8\%), 38\% of Mitten-540 (against 0.5\%) and 39\% of Coset-128 (against 5.1\%; Supplementary Note~3). Even after scanning uniform strengths per code, the best fixed $\gamma$ shifts from 0.025 to 0.075 across the transfer codes and still leaves 3.4 to 7.6 times as many observed failures as TEMPO-BP on LP20, Kasai-96 and Coset-128, and 189 versus 10 failures on Mitten-540, without retuning TEMPO-BP.

%% file: tables/table1.tex
\begin{table}[!htbp]
\centering
\caption{Codes and center-point decoding results. Each decoder entry is failures / shots over the complete X-memory experiment.}
\label{tab:transfer}
\begingroup
\fontsize{7.5}{9}\selectfont
\setlength{\tabcolsep}{1.7pt}
\renewcommand{\arraystretch}{1.25}
\begin{tabularx}{\textwidth}{@{}>{\raggedright\arraybackslash}p{13.8mm}>{\raggedright\arraybackslash}X>{\raggedright\arraybackslash}p{22.8mm}>{\raggedright\arraybackslash}p{18.6mm}>{\raggedright\arraybackslash}p{15.0mm}*{4}{>{\Centering\arraybackslash}p{10.6mm}}>{\raggedright\arraybackslash}p{13.8mm}@{}}
\toprule
Code & Construction / reference & $[[n,k,d]]$ & Detectors $\times$\newline fault mechanisms & $p$ (\%);\newline center in bold & BP & Relay-\newline BP-5 & TEMPO-\newline BP & TEMPO\newline ensemble & Parameters \\
\midrule
{BB-144}\textsuperscript{$\dagger$} & Bivariate bicycle~\cite{bravyi2024high} & $[[144,12,12]]$ & 936 $\times$\newline 8,784 & 0.2, 0.3\newline \textbf{0.4}, 0.5\newline 0.6 & 9,836 /\newline 100,000 & 227 /\newline 48,000 & 730 /\newline 100,000 & 169 /\newline 48,000 & tuned \\
{BB-72}\textsuperscript{$\dagger$} & Bivariate bicycle~\cite{bravyi2024high} & $[[72,12,6]]$ & 252 $\times$\newline 2,232 & 0.2, 0.3\newline \textbf{0.4}, 0.5\newline 0.6 & 1,343 /\newline 20,000 & 2,105 /\newline 48,000 & 929 /\newline 20,000 & 1,786 /\newline 48,000 & tuned \\
{LP20} & Lifted product~\cite{cain2026shor} & $[[1122,148,\leq 20]]$ & 10,395 $\times$\newline 139,722 & 0.15\newline \textbf{0.175}\newline 0.2 & 300 /\newline 2,000 & 145 /\newline 2,000 & 17 /\newline 2,000 & 2 /\newline 2,000 & from BB-144 \\
{Kasai-96} & Affine permutations~\cite{kasai2026orthogonality,zhao2026ultrahigh} & $[[1152,580,\leq 12]]$ & 3,744 $\times$\newline 77,184 & 0.0625\newline \textbf{0.0688}\newline 0.075 & 301 /\newline 2,000 & 123 /\newline 2,000 & 36 /\newline 2,000 & 27 /\newline 2,000 & from BB-144 \\
{Coset-128} & Coset based~\cite{aydin2026coset} & $[[128,16,12]]$ & 832 $\times$\newline 10,880 & 0.35\newline \textbf{0.4}\newline 0.45 & 2,631 /\newline 10,000 & 485 /\newline 10,000 & 330 /\newline 10,000 & 261 /\newline 10,000 & from BB-144 \\
{Mitten-540} & Non-Abelian\newline lifted product~\cite{bhardwaj2026highrate} & $[[540,108,18]]$ & 4,104 $\times$\newline 62,748 & 0.0984\newline \textbf{0.1125}\newline 0.1266 & 170 /\newline 2,000 & 63 /\newline 2,000 & 8 /\newline 2,000 & 2 /\newline 2,000 & from BB-144 \\
\bottomrule
\end{tabularx}
\par\vspace{3pt}
\fontsize{7.5}{9}\selectfont Failures refer to the complete X-memory experiment (any logical observable), not a per-round rate (syndrome rounds $T=12$ for BB-144, Kasai-96 and Coset-128; $T=6$ for BB-72; $T=20$ for LP20; $T=18$ for Mitten-540). The four transfer codes reuse the BB-144 single-run TEMPO-BP and ensemble parameters. BB-72 uses its own size-specific profiles; BB-144 is the tuning reference. Distance bounds and reported static distances are reproduced from the data metadata, not certified circuit distances. The detector matrix dimensions are rebuilt from the archived center-point circuits using the experiment's undecomposed DEM settings and verified elementwise against its saved decoder input matrices.\par\smallskip \textsuperscript{$\dagger$} Different center-point cohorts: BB-144: BP and TEMPO-BP, 100,000 shots (9.84\% and 0.73\%); Relay-BP-5 and TEMPO ensemble, 48,000 shots (0.47\% and 0.35\%). BB-72: BP and TEMPO-BP, 20,000 shots (6.72\% and 4.65\%); Relay-BP-5 and TEMPO ensemble, 48,000 shots (4.39\% and 3.72\%).
\endgroup
\end{table}

%% file: sections/discussion.tex
\section{Discussion}\label{sec:discussion}

We have introduced TEMPO-BP, which adjusts each variable's memory strength during decoding using local feedback. Whereas fixed memory fields face an open-loop trade-off between convergence speed and cycle-induced runaway, closing the loop at each variable mitigates these instabilities within local message passing itself. In the X-basis memory experiments studied here, this approach improves performance over the fixed-memory and post-processed baselines tested under uniform circuit-level noise, outperforming cubic-scaling BP+OSD post-processing on the bivariate bicycle codes while retaining linear $O(|E|)$ per-iteration message passing. Independent trajectories also allow decoding depth to be reduced through parallel execution, while parameters selected on the gross code remain effective without retuning on four codes from other families. These results demonstrate how online memory adaptation can support accurate decoding, flexible allocation of decoding effort and parameter transfer.

Our analysis gives memory a dynamical interpretation as a gain acting on feedback paths in the decoding graph. For the two-dimensional coefficient blocks obtained by exact coefficient reduction on the analyzed zero-prior state families, this gain determines whether particular modes are amplified ($\lambda_+\approx 1+2\gamma$ for $\gamma>0$, where the nonnegative cone is invariant) or damped (giving Schur stability of the coefficient block for $-1<\gamma<0$, though the nonnegative cone need not remain invariant under negative memory), although these relations do not establish convergence of the full adaptive decoder. The numerical diagnostics provide a complementary insight: the value of memory depends on where and when it is applied. Uniformly negative memory can stall decoding, yet negative strengths become useful when applied selectively via the one-step affine response and unsatisfied-check gating. Likewise, although the controller predominantly favors lower strengths, always lowering them performs poorly. The improvement therefore depends on the state-dependent choice of updates, rather than on a general preference for stronger or weaker memory.

The differences between codes clarify the scope of this improvement. Where non-convergence dominates---as on the $[[144,12,12]]$ gross code and LP20---controlling the message dynamics substantially reduces logical failures (cutting unconverged gross-code shots 42.5-fold at $p=0.4\%$, from 9.82\% for BP to 0.23\%, or 231 shots). Where most failures already correspond to valid corrections in the wrong logical class---as on distance-6 BB-72 ($[[72,12,6]]$), where TEMPO-BP converges on 99.81\% of shots and 95.9\% of remaining failures (891 of 929) are wrong-class corrections (Supplementary Table~1)---improving single-trajectory convergence alone offers less room for progress, whereas comparing multiple candidate corrections across the parallel ensemble lowers wrong-class failures further. This distinction motivates studying how adaptive memory influences both the time required to find a correction and the logical class ultimately reached. The successful transfer of a common parameter set further suggests that a rule for responding to decoder state can remain useful across different graph structures, motivating tests under changes in noise and syndrome-extraction schedules.

Independent trajectories also allow decoding depth to be exchanged for parallel resources. Translating this algorithmic trade-off into practical latency gains will require implementations that account for controller arithmetic, numerical precision and available hardware (roughly eight AMD XCVU19P FPGAs or dedicated ASIC logic for 15 concurrent gross-code X decoders; Supplementary Note~5). On the gross code, 99.8\% of raw targets $\gamma_i^\star$ are projected onto the interval endpoints $\gamma_{\min}$ or $\gamma_{\max}$ ($\bar\gamma_i\in\{\gamma_{\min},\gamma_{\max}\}$, with 98.97\% brake and 0.83\% hold before partial stepping; Supplementary Table~9), although low-bitwidth fixed-point robustness will also depend on whether quantization alters the denominator guard, target sign or projection branch. For continuous decoding, a further question concerns the number of syndrome rounds processed together~\cite{skoric2023parallel}. While the estimated mean ensemble decoding time on the gross code at $p=0.4\%$ ($3.94~\mu$s) is three times below the $12~\mu$s accumulation span of a 12-round window, in the temporal-loop models we can solve, the amplification per iteration ($m^{1/(T+1)}$) approaches unity as this number grows. Window length may therefore affect the dynamics of decoding as well as its computational cost, motivating the joint study of memory control, window selection and parallel execution.

More broadly, TEMPO-BP formulates memory selection as a feedback-control problem within message passing. The decoder's evolving beliefs and syndrome residuals provide information for adjusting its own update rules. Exploring how such local information can guide other decoder parameters---such as check normalization factors, clipping thresholds or soft ensemble arbitration (where oracle selection lowers gross-code diversity-cohort failures from 376 to 193; Supplementary Note~4)---may offer further ways to improve iterative decoding without abandoning its computational structure.

%% file: sections/methods.tex
\section{Methods}\label{sec:methods}

\subsection{Overview}

TEMPO-BP has three layers, which we describe in turn: a memory min-sum BP iteration that it shares with Mem-BP, DMem-BP and Relay-BP; a controller that updates the memory strengths every $K$ iterations; and an optional ensemble that runs several controlled trajectories in parallel and keeps the valid candidate most probable under the priors. Algorithms~\ref{alg:tempo} and~\ref{alg:ensemble} summarize the procedure. Proofs of the properties stated below, implementation details that affect numerical results bit for bit, and the provenance of every experiment are given in the Supplementary Information and the code repository.

\subsection{Decoding problem}

A circuit with circuit-level noise is converted into a detector error model with Stim~\cite{gidney2021stim}. It consists of a binary check matrix $H\in\mathbb F_2^{M\times N}$, whose rows are detectors and whose columns are independent fault mechanisms, an observable matrix $O\in\mathbb F_2^{K_{\rm L}\times N}$ and priors $p_i\in(0,1)$. A shot produces a syndrome $s=He\in\mathbb F_2^M$, and the decoder returns $\hat e\in\mathbb F_2^N$. The shot \emph{converges} if $H\hat e=s$, and is a \emph{logical failure} if it does not converge or if $O\hat e\ne Oe$. We report logical failure rates throughout, so unconverged shots always count as failures, because an output with $H\hat e\ne s$ fails to return a syndrome-consistent Pauli correction online even if its final hard decision happens to satisfy $O\hat e=Oe$. Where failures are split, they are split into non-converged shots and converged shots in the wrong logical class. We write $\mathcal N(a)$ for the variables in check $a$, $\mathcal M(i)$ for the checks containing variable $i$, $d_i=|\mathcal M(i)|$ and $L_i=\ln[(1-p_i)/p_i]$. Prior to iterative decoding, degree-one checks ($|\mathcal N(a)|=1$, incident to a single variable $i$) are peeled by fixing $\hat e_i=s_a$ and updating $s\leftarrow s\oplus s_a H_{\cdot,i}$; message passing and controller updates act on the remaining free variables.

\subsection{Memory min-sum iteration}

Each iteration $t=1,2,\dots$ performs a normalized min-sum check-node update,
\begin{equation}
m_{a\to i}^{(t)}=(-1)^{s_a}\,\alpha\!\!\prod_{j\in\mathcal N(a)\setminus i}\!\!\operatorname{sgn}\mu_{j\to a}^{(t-1)}\cdot\!\!\min_{j\in\mathcal N(a)\setminus i}\!\bigl|\mu_{j\to a}^{(t-1)}\bigr|,
\end{equation}
forms the memory-weighted prior $\Lambda_i^{(t)}=(1-\gamma_i)L_i+\gamma_i R_i^{(t-1)}$ of equation~\eqref{eq:memory}, and updates the variables,
\begin{equation}
R_i^{(t)}=\Lambda_i^{(t)}+Q_i^{(t)},\quad Q_i^{(t)}=\!\!\sum_{a\in\mathcal M(i)}\!\! m_{a\to i}^{(t)},\quad
\mu_{i\to a}^{(t)}=R_i^{(t)}-m_{a\to i}^{(t)},\quad \hat e_i^{(t)}=\mathbb 1\bigl[R_i^{(t)}\le0\bigr].
\end{equation}
Decoding stops as soon as $H\hat e^{(t)}=s$. Messages start from $m^{(0)}=0$ and $\mu^{(0)}_{i\to a}=R_i^{(0)}=L_i$; we use $\alpha=1$ throughout. The decoders differ only in how $\gamma_i$ is chosen: $\gamma_i=0$ gives BP, a constant gives Mem-BP, a per-shot random draw $\gamma_i\sim\mathcal U[\gamma^-,\gamma^+)$ that is then kept fixed gives DMem-BP. By contrast, TEMPO-BP draws its initial strengths in the same way from an interval of its own and then updates them online.

\subsection{The controller}

After every $K$-th iteration that has not converged, and if at least one more iteration is allowed, the controller performs four steps for every free variable $i$, independently and in parallel.

\paragraph{Look-ahead.} From the current variable-to-check messages $\mu^{(t)}$ it computes the next check-to-variable messages $m^{(t+1)}$ and stores them; iteration $t+1$ uses them instead of recomputing them. This yields $Q_i^{(t+1)}$ and, by equation~\eqref{eq:response}, the exact next belief as a function of $\gamma_i$. Because the stored messages replace the check-node update of the next iteration, a run of $\tau$ iterations performs exactly $\tau$ check-node updates for any $K$ (Supplementary Note~1, Proposition~3).

\paragraph{Where to act.} The fraction of unsatisfied checks around $i$, $\rho_i=d_i^{-1}\sum_{a\in\mathcal M(i)}z_a$ with $z=s\oplus H\hat e^{(t)}$, and the variable's confidence $c_i=\tanh(|R_i^{(t)}|/T_{\rm conf})$ set the step size
\begin{equation}\label{eq:methods-step}
\kappa_i=\operatorname{clip}_{[0,1]}\!\bigl(\kappa_0\,c_i\,[(1-\beta)+\beta\rho_i]\bigr).
\end{equation}
Here $\kappa_0$ is the base step, $T_{\rm conf}$ a confidence scale and $\beta$ the weight given to unsatisfied checks. The step is largest for confident variables next to violated checks. With $\beta=1$ the controller acts only in the first neighborhood of unsatisfied checks (Supplementary Note~1, Proposition~6); the configurations used here have $\beta<1$, so that other variables take small steps.

\paragraph{Where to go.} The target belief has the opposite sign of the current one and a small magnitude when $\zeta>0$ and the current belief is nonzero; otherwise the target is zero,
\begin{equation}\label{eq:methods-target}
U_i=-\zeta\tanh\bigl(R_i^{(t)}/T_{\rm tgt}\bigr),\qquad |U_i|\le\zeta,
\end{equation}
where $\zeta$ is the target magnitude and $T_{\rm tgt}$ its scale; setting equation~\eqref{eq:response} equal to it gives the unconstrained memory strength
\begin{equation}\label{eq:methods-ustar}
\gamma_i^\star=\frac{U_i-L_i-Q_i^{(t+1)}}{D_i^{(t)}}\qquad\text{if } |D_i^{(t)}|\ge\epsilon,
\end{equation}
where $D_i^{(t)}=R_i^{(t)}-L_i$ is the displacement of the belief from its prior, and $\gamma_i^\star=\gamma_i$ otherwise. For $|D_i^{(t)}|\ge\epsilon$, its projection $\bar\gamma_i=\Pi_\Gamma(\gamma_i^\star)$ onto the admissible range $\Gamma=[\gamma_{\min},\gamma_{\max}]$ is the unique minimizer of the squared distance between the predicted and the target belief over $\Gamma$ (Supplementary Note~1, Proposition~2).

\paragraph{How far to go.} The memory strength moves part of the way,
\begin{equation}
\gamma_i\leftarrow\Pi_\Gamma\bigl((1-\kappa_i)\,\gamma_i+\kappa_i\,\bar\gamma_i\bigr).
\end{equation}
Between control steps $\gamma$ is held fixed. With $\kappa_i<1$, as for all configurations used here, the partial update brings the next-belief prediction no farther from the target than keeping the old memory strength, with the current state and next check field held fixed.

The extra work of a control step is $O(|E|)$ to accumulate $\rho_i$ and $Q_i$ along the edges plus $O(1)$ scalar operations per variable, the same order as one BP iteration; it needs no extra syndrome computation because $H\hat e$ is already evaluated for the stopping test (Supplementary Note~1, Proposition~4). If no control step is executed, TEMPO-BP reproduces DMem-BP with the same memory field and main-stage initial state and iteration budget iteration by iteration (Supplementary Note~1, Proposition~5); this is the TEMPO-static ablation.

\subsection{Algorithms}

\begin{algorithm}[t]
\caption{TEMPO-BP for one shot}\label{alg:tempo}
\begin{algorithmic}[1]
\Require $H$, $L$, syndrome $s$; parameters $\theta=(I_{\max},W,K,\Gamma,[\gamma^-,\gamma^+),\kappa_0,T_{\rm conf},\beta,\zeta,T_{\rm tgt},\epsilon)$; shot index $j$
\State derive the random stream of shot $j$; peel weight-one checks
\State $m\gets0$, $\mu\gets L$, $R\gets L$
\If{$W>0$} run up to $W$ iterations of BP ($\gamma\equiv0$); \Return if converged; keep $R$, set $m\gets 0$, $\mu\gets L$ \EndIf
\State draw $\gamma_i\sim\mathcal U[\gamma^-,\gamma^+)$ for all $i$
\For{$t=1,\dots,I_{\max}-W$}
  \State use stored $(m,Q)$ if present (and clear cache), else compute $m$ from $\mu$ and $Q_i\gets\sum_{a\in\mathcal M(i)}m_{a\to i}$ \Comment{check-node update}
  \State $\Lambda\gets(1-\gamma)\odot L+\gamma\odot R$; \ $R\gets\Lambda+Q$; \ $\mu\gets R-m$; \ $\hat e\gets\mathbb 1[R\le0]$
  \If{$H\hat e=s$} \Return $\hat e$, converged \EndIf
  \If{$t \bmod K=0$ \textbf{and} $t<I_{\max}-W$}
     \State compute and store $m^{(t+1)}$ from $\mu$; form $Q^{(t+1)}$ \Comment{look-ahead}
     \For{each free variable $i$ \textbf{in parallel}}
        \State compute $\rho_i,\ c_i,\ \kappa_i,\ U_i,\ \gamma_i^\star,\ \bar\gamma_i$ via equations~\eqref{eq:methods-step}--\eqref{eq:methods-ustar}; \ $\gamma_i\gets\Pi_\Gamma((1-\kappa_i)\gamma_i+\kappa_i\bar\gamma_i)$
     \EndFor
  \EndIf
\EndFor
\State \Return last $\hat e$, not converged
\end{algorithmic}
\end{algorithm}

\begin{algorithm}[t]
\caption{TEMPO ensemble}\label{alg:ensemble}
\begin{algorithmic}[1]
\Require ordered member parameters $\theta^{(1)},\dots,\theta^{(S)}$; total budget $B$
\For{$k=1,\dots,S$ \textbf{in parallel}}
   \State $(\hat e^{(k)},\text{conv}_k,\tau_k)\gets$ TEMPO-BP$(\theta^{(k)})$ with $I_{\max}=\lfloor B/S\rfloor$
\EndFor
\If{no member converged} \Return $\hat e^{(1)}$, not converged \EndIf
\State \Return the converged $\hat e^{(k)}$ of minimum weight $\sum_{i:\hat e^{(k)}_i=1}L_i$ (ties: smallest $k$) \Comment{depth $\max_k\tau_k\le\lfloor B/S\rfloor$}
\end{algorithmic}
\end{algorithm}

Shots are decoded independently. Every shot uses its own random stream, derived from the global seed and the shot index by a bijective mixing function, so results do not depend on batching, thread count or execution order. Weight-one checks are resolved by peeling before message passing. The optional warm start $W$ is zero for all single-run TEMPO-BP configurations and is used by one member of the BB-72 ensemble.

\subsection{Parameter selection}

For each bivariate bicycle code, one single-run TEMPO-BP configuration and one ensemble were selected on tuning samples at $p=0.4\%$ that are disjoint from all evaluation samples, and then frozen. The gross-code configuration uses $K=4$ and the BB-72 configuration $K=8$; the 15 gross-code members use $K\in\{4,8\}$ and the 10 members of the BB-72 ensemble $K\in\{1,4,8,16,80\}$. The first gross-code member has the same parameters as the single-run TEMPO-BP configuration, apart from its iteration limit. The transfer experiments reuse the gross-code configuration and ensemble unchanged. All parameter values of the single-run TEMPO-BP configurations and of every ensemble member are listed in Supplementary Table~8 and distributed with the code~\cite{tempocode}.

\subsection{Codes, circuits and noise}

All experiments are X-basis memory experiments decoded with all logical X observables. For the bivariate bicycle codes we use the circuits of the public Relay-BP benchmark~\cite{relaybprepo}, with as many rounds of syndrome extraction as the code distance (6 for BB-72 and 12 for the gross code, BB-144). Both use its uniform circuit-level noise model: every two-qubit gate is followed by two-qubit depolarizing noise of strength $p$, every idle qubit by single-qubit depolarizing noise of strength $p$, and preparations and measurements fail with probability $p$; the final data-qubit readout is noiseless. The four transfer codes use exactly this noise model, at different strengths, and a number of rounds equal to the distance or to its best known upper bound. LP20 is the $[[1122,148,\le20]]$ lifted-product code of ref.~\cite{cain2026shor} (20 rounds), Kasai-96 the $[[1152,580,\le12]]$ code with permutation matrices of size 96 of ref.~\cite{zhao2026ultrahigh}, a construction introduced in ref.~\cite{kasai2026orthogonality} (12 rounds), Coset-128 the $[[128,16,12]]$ coset code of ref.~\cite{aydin2026coset} (12 rounds) and Mitten-540 the $[[540,108,18]]$ non-Abelian lifted-product code of ref.~\cite{bhardwaj2026highrate} (18 rounds). For Coset-128 and Mitten-540 we use the check matrices and gate schedules released by the authors, with Mitten-540 represented as a CSS CNOT circuit preserving the published entangling-gate order; for LP20 and Kasai-96, we built deterministic grouped X/Z edge-color schedules following the reported measurement order. The noise model applied to Mitten-540 is the uniform model above and not the hardware-specific model of ref.~\cite{bhardwaj2026highrate}.

Idling errors are included on every code at the same strength as gate errors. Idle errors can be a significant source of error in circuit-level studies of qLDPC memories~\cite{berthusen2025local,sun2026layer}, yet simulations of large high-rate codes often neglect them~\cite{pecorari2025highrate,bhardwaj2026highrate} or report a baseline without them~\cite{xu2024constant}, which makes operating points look more favorable than a schedule with many idle locations warrants. Because benchmarking across heterogeneous noise models confounds decoder comparisons~\cite{tang2026hierarchical}, we keep this single idle-included circuit model fixed across all codes; the resulting transfer operating points therefore differ from the original code studies and should be interpreted under this idle-noise model and each code's schedule, which leaves a large fraction of qubits idle in every layer.

\subsection{Baselines}

All baselines use the same detector error models as TEMPO-BP, the same samples except where stated and, on bivariate bicycle codes, the configurations of the public Relay-BP benchmark~\cite{relaybprepo}. BP is normalized min-sum with $\alpha=1$; Mem-BP uses the constant 0.15; DMem-BP draws memory strengths from the benchmark's interval $[-0.2263,0.6216)$. DMem-warm runs 80 iterations of Mem-BP and, if they do not converge, one DMem-BP run of at most 9,920 iterations, starting from the resulting beliefs; this is the single-path counterpart of Relay-BP without successive legs. Relay-BP-$S_{\rm R}$ runs the same 80-iteration Mem-BP stage followed by up to 300 legs as defined in Results; it stops after collecting $S_{\rm R}$ valid corrections and returns the one of minimum weight $\sum_{i:\hat e_i=1}L_i$, the most probable under the priors~\cite{muller2025relay}. BP+OSD+CS-10 runs 80 iterations of min-sum BP followed by ordered-statistics decoding with combination sweep of order 10, using the \texttt{ldpc} package~\cite{roffe2020decoding}. TEMPO-static is DMem-BP with the initial memory interval of the selected TEMPO-BP configuration, that is, TEMPO-BP with the controller switched off. BP, Mem-BP, DMem-BP, DMem-warm, TEMPO-static and TEMPO-BP are limited to 10,000 iterations; Relay-BP-1 keeps the limit of Relay-BP, 18,080 iterations. Where the public benchmark and the Relay-BP paper differ (for example in the number of BP iterations before OSD), we follow the public benchmark, whose configurations are documented in our repository.

\subsection{Mechanism diagnostics}

The experiments of Fig.~\ref{fig:concept}c and Fig.~\ref{fig:law} are diagnostics on separately drawn samples and do not enter the benchmark results. They use a research copy of the decoder that adds switches for the ablations and records belief amplitudes. With all switches off, it reproduces the released decoder bit for bit in corrections, convergence, iteration counts and solution weights on 3,000 shots of each bivariate bicycle code with the controller on and on 600 with it off. Within an experiment every arm decodes the same syndromes and, where applicable, starts from the same memory field. Uniform-memory arms use a constant $\gamma$ and no controller. Disordered arms draw $\gamma_i$ independently and uniformly from an interval and keep it fixed. The ablations alter the following controller elements: \emph{always brake} and \emph{always hold} keep the timing and step sizes and replace the projected target by $\gamma_{\min}$ or $\gamma_{\max}$; \emph{blind decay} moves $\gamma_i$ toward $\gamma_{\min}$ with a constant step, retaining the denominator and raw-target finiteness guards. Two structural ablations alter the interval or gating: \emph{non-negative range} sets $\gamma_{\min}=0$ and, on BB-72, also raises the initialization lower bound to zero; \emph{no residual gate} sets $\beta=0$ and rescales $\kappa_0$ so that the step far from unsatisfied checks is unchanged. The diagnostic label \emph{runaway} denotes a maximum belief magnitude over free variables above $10^{12}$ at the last executed iteration, with non-finite terminal values represented by infinity; it does not certify asymptotic divergence. The \emph{present field} ablation keeps the controller unchanged except for the field used in the prediction, replacing $Q_i^{(t+1)}$ in equation~\eqref{eq:response} by $Q_i^{(t)}$, the field that produced the present belief. For the growth rates of Fig.~\ref{fig:law}b and Supplementary Note~3, we decoded with a uniform memory strength and a limit of 3,000 iterations, selecting about 200 unconverged trajectories per code and strength where available (the numbers examined are given in Supplementary Note~3). For each selected trajectory, we tracked the largest belief magnitude over all variables (stopping when it exceeds $10^{200}$) and defined the growth rate as the mean slope of the logarithm of this magnitude over the last quarter of the record. A trajectory counts as growing if its rate exceeds $10^{-3}$ per iteration; the fractions quoted change by at most four percentage points when this threshold is halved or doubled. The exact certificates of the reinforcing channels are described in Supplementary Note~2.

\subsection{Evaluation protocol}

Rates are reported per shot with exact (Clopper--Pearson) 95\% intervals; the diagnostic tables of Supplementary Note~3 use Wilson intervals. Rates per round are $1-(1-P_{\rm L})^{1/T}$, computed via this geometric conversion as a comparative reporting convention across circuit depths $T$ rather than as a measured hazard per round (Supplementary Information). Decoders compared at a given operating point decode the same samples, with two exceptions at $p=0.4\%$ on the bivariate bicycle codes: the comparison of the ensemble with Relay-BP-5 used its own samples of 48,000 shots, and TEMPO-BP and Relay-BP-1 were compared on fixed matched subsets of 10,000 (BB-72) and 5,000 (gross code) shots. Paired comparisons use the exact two-sided McNemar test on discordant shots, with Holm correction within families that Supplementary Table~2 lists with their sizes and status. Only the comparisons on LP20, Kasai-96 and Mitten-540 were fixed in a dated plan before the corresponding shots were collected. The other families were defined after data had been seen and are exploratory: the ten comparisons at $p=0.4\%$ on the bivariate bicycle codes, the comparisons of TEMPO-BP with Relay-BP-1 and with BP, those of the ensemble with Relay-BP-5 at the other error rates, and those on Coset-128. A test that does not reject is not taken as evidence of equivalence. Three cohort-construction conventions apply to the transfer tests: the center points of the LP20 and Kasai-96 scans reuse an earlier run and lie outside the Holm families; the center point of Mitten-540 was chosen from a pilot scan with Relay-BP-5 alone; and for Coset-128 the counts quoted are those of 10,000 pooled shots per point, whereas paired tests use the 8,000 that were collected after a first sample of 2,000 had been examined.

Decoding effort is counted in executed BP iterations, including warm-up iterations and every Relay-BP leg. In a fully parallel hardware decoder a message-passing iteration takes a fixed number of clock cycles~\cite{muller2025relay}. The hardware model in Supplementary Note~5 assigns additional controller cycles every $K$ iterations (modeled analytically rather than measured on hardware, and omitted from raw BP iteration counts~\cite{tang2026hierarchical}). For an ensemble we report its depth, the maximum over members of their iteration counts, which measures completion time in units of a common iteration cost under ideal concurrent execution and is bounded by the per-member limit. The total number of iterations summed over members is reported separately (Supplementary Table~7). Wall-clock times were measured separately from the accuracy runs with a single worker process and one thread on an AMD Ryzen Threadripper PRO 9995WX (Ubuntu 24.04, Python 3.13, Rust decoder core in double precision), on 100 to 10,000 inputs per decoder and code as given in Supplementary Table~4. These timings compare our implementation of TEMPO-BP with the public implementations of the baselines.

%% file: sections/backmatter.tex
\backmatter

\bmhead{Data availability}
The benchmark datasets, detector error models and full simulation archives are deposited on Zenodo at \url{https://doi.org/10.5281/zenodo.24280168}~\cite{tempodata}, while per-point accuracy and latency tables, paired-test tables, timing records and experiment provenance are also provided in the repository~\cite{tempocode}. The circuits for the bivariate bicycle codes are those of the public Relay-BP benchmark~\cite{relaybprepo}.

\bmhead{Code availability}
The decoder (Rust core with Python bindings), all parameter sets, the scripts that reproduce the experiments and figures, and the exact-arithmetic certificates of Supplementary Note~2 are available at \url{https://github.com/Yomure/tempo-bp}~\cite{tempocode}.

%% file: si/note1_properties.tex
\section*{Supplementary Note 1\quad Properties of the TEMPO-BP update}
\addcontentsline{toc}{section}{Supplementary Note 1: Properties of the TEMPO-BP update}

This note states and proves the properties of the update that Methods uses.  The
statements concern the decoder map in exact arithmetic rather than establishing logical failure rates, global convergence or general detector-graph stability.
The loop-gain law and its certificates on named detector graphs are the subject
of Supplementary Note~2.

\subsection*{1.1 Notation and one iteration}

The iteration is the memory min-sum update shared with Mem-BP and
Relay-BP~\cite{kuo2022exploiting,muller2025relay}.  Let $H\in\mathbb F_2^{M\times N}$ have rows indexed by checks $a$ and columns by
variables $i$, let $\mathcal N(a)$ be the variables adjacent to $a$, let
$\mathcal M(i)$ be the checks adjacent to $i$, and let $d_i=|\mathcal M(i)|$.
The prior log-likelihood ratio of variable $i$ is
$L_i=\ln[(1-p_i)/p_i]$.  For a free variable $i$ and iteration $t\ge1$, the
normalized min-sum check update is
\begin{equation}\label{eq:si-check}
m_{a\to i}^{(t)}
=(-1)^{s_a}\prod_{j\in\mathcal N(a)\setminus i}\operatorname{sgn}\mu_{j\to a}^{(t-1)}
\cdot\min_{j\in\mathcal N(a)\setminus i}\bigl|\mu_{j\to a}^{(t-1)}\bigr|,
\end{equation}
with normalization $\alpha=1$ throughout.  The memory-weighted prior, posterior,
extrinsic message and hard decision are
\begin{align}
\Lambda_i^{(t)}&=(1-\gamma_i^{(t)})L_i+\gamma_i^{(t)}R_i^{(t-1)},\label{eq:si-prior}\\
R_i^{(t)}&=\Lambda_i^{(t)}+Q_i^{(t)},\qquad
Q_i^{(t)}=\sum_{a\in\mathcal M(i)}m_{a\to i}^{(t)},\label{eq:si-posterior}\\
\mu_{i\to a}^{(t)}&=R_i^{(t)}-m_{a\to i}^{(t)},\qquad
\hat e_i^{(t)}=\mathbb 1\bigl[R_i^{(t)}\le0\bigr].\label{eq:si-message}
\end{align}
We write $D_i^{(t)}=R_i^{(t)}-L_i$ for the displacement of the belief from its
prior.  The initial state is $m^{(0)}=0$, $\mu_{i\to a}^{(0)}=R_i^{(0)}=L_i$ (or, if $W>0$ warm-up iterations of plain BP with $\gamma\equiv0$ are run first without converging, the main stage retains the warm-up beliefs as $R_i^{(0)}$, resets $m^{(0)}=0$ and $\mu_{i\to a}^{(0)}=L_i$, and draws the main-stage memory strengths $\gamma_i^{(1)}\sim\mathcal U[\gamma^-,\gamma^+)\subseteq\Gamma$).
A weight-one check is peeled before message passing; the propositions below
concern free variables, and fixed variables are never updated by the controller.
The decoder stops at the first $t$ with $H\hat e^{(t)}=s$.

The controller runs after main-stage iteration $t$ ($1\le t\le I_{\max}-W$) only if the iteration did not converge,
at least one further iteration is allowed, and $t$ is a multiple of the update
interval $K$ (warm-up iterations are not counted); on all other main-stage iterations, $\gamma_i^{(t+1)}=\gamma_i^{(t)}$.  It computes the next check
field
\begin{equation}\label{eq:si-field}
Q_i^{(t+1)}=\sum_{a\in\mathcal M(i)}m_{a\to i}^{(t+1)}
\end{equation}
from the current $\mu^{(t)}$, and stores $m^{(t+1)}$ so that iteration $t+1$
does not recompute it.  With
\begin{equation}\label{eq:si-kappa}
\rho_i^{(t)}=\frac1{d_i}\sum_{a\in\mathcal M(i)}z_a^{(t)},\qquad
c_i^{(t)}=\tanh\!\frac{|R_i^{(t)}|}{T_{\rm conf}},\qquad
g_i^{(t)}=(1-\beta)+\beta\rho_i^{(t)},
\end{equation}
where $z^{(t)}=s\oplus H\hat e^{(t)}$ and $\rho_i=0$ if $d_i=0$, the step size is
\begin{equation}\label{eq:si-step}
\kappa_i^{(t)}=\operatorname{clip}_{[0,1]}\!\bigl(\kappa_0\,c_i^{(t)}g_i^{(t)}\bigr).
\end{equation}
The target belief and the unconstrained memory strength are
\begin{equation}\label{eq:si-target}
U_i^{(t)}=-\zeta\tanh\!\frac{R_i^{(t)}}{T_{\rm tgt}},\qquad
\gamma_i^\star=
\begin{cases}
\gamma_i^{(t)}, & |D_i^{(t)}|<\epsilon,\\[2pt]
\dfrac{U_i^{(t)}-L_i-Q_i^{(t+1)}}{D_i^{(t)}}, & \text{otherwise},
\end{cases}
\end{equation}
and the update is
\begin{equation}\label{eq:si-update}
\bar\gamma_i=\Pi_\Gamma(\gamma_i^\star),\qquad
\gamma_i^{(t+1)}=\Pi_\Gamma\!\Bigl((1-\kappa_i^{(t)})\gamma_i^{(t)}
+\kappa_i^{(t)}\bar\gamma_i\Bigr),
\end{equation}
where $\Gamma=[\gamma_{\min},\gamma_{\max}]$ and
$\Pi_\Gamma(x)=\min(\max(x,\gamma_{\min}),\gamma_{\max})$.  All variables are
updated from the state after iteration $t$, independently of one another.

\subsection*{1.2 The propositions}

\begin{proposition}[Exact one-step response]\label{prop:response}
Suppose the controller has updated the memory strengths after iteration $t$ and
iteration $t+1$ uses the stored messages $m^{(t+1)}$.  Then, in exact arithmetic,
\begin{equation}\label{eq:si-response}
R_i^{(t+1)}=L_i+Q_i^{(t+1)}+\gamma_i^{(t+1)}D_i^{(t)}
\end{equation}
for every free variable $i$.
\end{proposition}

\begin{proof}
Iteration $t+1$ uses the stored check messages, so its message sum is exactly
$Q_i^{(t+1)}$ of \eqref{eq:si-field}.  By \eqref{eq:si-prior},
$\Lambda_i^{(t+1)}=(1-\gamma_i^{(t+1)})L_i+\gamma_i^{(t+1)}R_i^{(t)}
=L_i+\gamma_i^{(t+1)}(R_i^{(t)}-L_i)$.  Adding $Q_i^{(t+1)}$ gives
\eqref{eq:si-response}.
\end{proof}

The response is affine in $\gamma_i^{(t+1)}$ with slope $D_i^{(t)}$ and
intercept $L_i+Q_i^{(t+1)}$.  This relation is an exact algebraic identity rather than a local linearization: the
stored check field is the field that the decoder would have computed at
iteration $t+1$ anyway.  In floating point the implementation accumulates $Q_i$
and the posterior in the same edge order, but the two sides of
\eqref{eq:si-response} can still differ by rounding; the identity is exact in the
arithmetic sense used by the proofs and is the relation the controller solves.

\begin{proposition}[The projected target is the constrained optimum]\label{prop:projection}
If $|D_i^{(t)}|\ge\epsilon$, then $\bar\gamma_i=\Pi_\Gamma(\gamma_i^\star)$ is
the unique minimizer over $\gamma\in\Gamma$ of
\begin{equation}\label{eq:si-objective}
\bigl(L_i+Q_i^{(t+1)}+\gamma D_i^{(t)}-U_i^{(t)}\bigr)^2 .
\end{equation}
\end{proposition}

\begin{proof}
The objective is a strictly convex quadratic in $\gamma$.  Its unconstrained
minimizer sets the residual to zero, which is $\gamma_i^\star$ because
$D_i^{(t)}\ne0$.  The minimizer of a strictly convex one-dimensional function over
a closed interval is the projection of the unconstrained minimizer onto that
interval.
\end{proof}

\paragraph{Corollary (landing).}
If $|D_i^{(t)}|\ge\epsilon$, $\kappa_i^{(t)}=1$ and $\gamma_i^\star\in\Gamma$, then
$R_i^{(t+1)}=U_i^{(t)}$: the next belief lands exactly on the target, with magnitude at most $\zeta$ and, when $\zeta>0$ and $R_i^{(t)}\ne0$, the
opposite sign; it is zero when $\zeta=0$ or $R_i^{(t)}=0$.  With the deployed
parameters $\kappa_0<1$ and $c_i<1$, so $\kappa_i<1$ and this limiting case does
not occur; it only fixes the direction of the update.

\begin{proposition}[No additional check-node updates]\label{prop:noextra}
For any update interval $K\ge1$, if the main stage executes $\tau$ iterations,
it performs exactly $\tau$ check-node updates.
\end{proposition}

\begin{proof}
After an iteration at which the controller runs, the check messages
$m^{(t+1)}$ are computed once and stored; iteration $t+1$ consumes them and does
not call the check pass.  After every other iteration, iteration $t+1$ computes
its own check messages.  No look-ahead is made after a converged iteration or
after the last allowed iteration.  Hence each executed iteration accounts for
exactly one check pass, whether or not it follows a control step.
\end{proof}

The look-ahead moves the check pass of iteration $t+1$ to the end of iteration
$t$.  It does not add a check pass to the run; it changes when the pass is done,
so that the controller can use its output.

\begin{proposition}[Cost and parallelism]\label{prop:cost}
Let $E$ be the edge set of the Tanner graph and $N$ and $M$ the numbers of
variables and checks.  A check pass and a variable pass cost $O(|E|)$; the hard
decision and the syndrome test cost $O(N)$ and $O(|E|)$, respectively; one
controller update costs $O(|E|+N)$.  The look-ahead check pass is $O(|E|)$ but
replaces the next iteration's pass.  Hence a main stage of $\tau$ iterations costs
$O\bigl(\tau(|E|+N+M)\bigr)$ time and $O(|E|+N+M)$ memory, and each check, variable or
controller pass is parallelizable across its respective nodes.
\end{proposition}

\begin{proof}
The check pass visits each edge once, and the variable pass visits each edge
once through the prefix and suffix sums.  The controller accumulates
$Q_i^{(t+1)}$ and the unsatisfied-check count along the same edges, which is
$O(|E|)$, and performs a constant number of scalar operations per free variable,
which is $O(N)$.  The syndrome $H\hat e$ is already computed for the stopping
test, so the controller reuses it rather than computing a second syndrome.  The
look-ahead check pass is the pass that iteration $t+1$ would otherwise perform.
Summing over iterations gives the time bound; the stored messages, beliefs and
memory field give the memory bound.  Check updates are independent across
checks, variable updates across variables, and controller updates across
variables once $Q^{(t+1)}$ and $z^{(t)}$ are available.
\end{proof}

The controller uses $\tanh$, clipping and one scalar division (guarded by
$|D_i^{(t)}|\ge\epsilon$), so it introduces no message-passing dependency beyond
the edge sweep.  The controller's edge sweep and scalar operations are costed explicitly in Supplementary Note~5.

\begin{proposition}[Reduction to disordered memory BP]\label{prop:dmem}
If no controller update is executed in the main stage (for example when
$K\ge I_{\max}-W$, or when the controller is switched off), then, for the same initial
interval, random stream, seed, shot index, peeling plan,
warm start, main-stage initial state, and iteration limit,
TEMPO-BP and DMem-BP follow the same trajectory iteration by iteration.
\end{proposition}

\begin{proof}
Both decoders draw the initial memory field from the same stream and the same
interval.  With no update, $\gamma_i$ keeps that value.  The check pass,
variable pass, hard decision and stopping rule are the same functions of the
same state, so induction on the iteration gives the claim.
\end{proof}

The implementation has a unit test comparing the two modes, and the finite-sample
identity check described in Supplementary Note~3 compares the research copy with
the production decoder on 3,000 adaptive and 600 static shots of each bivariate
bicycle code.

\begin{proposition}[Locality for $\beta=1$]\label{prop:locality}
If $\beta=1$ and every check adjacent to variable $i$ is satisfied after
iteration $t$, then $\kappa_i^{(t)}=0$ and $\gamma_i$ is unchanged.  The
controller then acts only on the first neighborhood of the unsatisfied checks.
\end{proposition}

\begin{proof}
With $\beta=1$, $g_i^{(t)}=\rho_i^{(t)}$.  If all adjacent checks are satisfied,
$\rho_i^{(t)}=0$, hence $g_i=0$ and $\kappa_i=0$.  The implementation skips the
variable when $\kappa_i=0$, so $\gamma_i$ is unchanged.  A variable with
$\kappa_i>0$ must have at least one unsatisfied adjacent check.
\end{proof}

Both single-run configurations selected for the paper have $\beta<1$ (0.801 for the gross
code and 0.941 for BB-72), so in the deployed decoder variables outside
the first neighborhood take the small step $\kappa_0c_i(1-\beta)$ rather than
exactly zero.  Proposition~\ref{prop:locality} is the $\beta=1$ limit, not a
description of the deployed locality.

\subsection*{1.3 Scope and implementation audit}

Propositions~\ref{prop:response}--\ref{prop:locality} are exact statements about
one update or one run of the implemented map.  Graph-specific dynamics, convergence-time scaling, multi-code empirical performance and hardware costs are treated in Supplementary Notes~2--5.

The implementation has the following qualifications.
\begin{itemize}
\item \emph{Target magnitude.}  The code forms
$U_i=-\zeta\tanh(R_i/T_{\rm tgt})$, so $|U_i|\le\zeta$, with equality possible
when the floating-point $\tanh$ saturates. For finite inputs and $\zeta>0$, the strict
inequality holds in exact arithmetic; the non-strict bound also covers
floating-point saturation.
\item \emph{Outer projection.}  The code applies
$\operatorname{clip}_{[\gamma_{\min},\gamma_{\max}]}$ to the blended value in
\eqref{eq:si-update}.  In exact arithmetic this second projection is inactive,
because a convex combination of two points of $\Gamma$ lies in $\Gamma$; the
implementation keeps it as a floating-point guard.  Proposition~\ref{prop:projection}
concerns $\bar\gamma_i$, not this redundant outer clip.
\item \emph{Look-ahead accounting.}  Methods describes the look-ahead as one of
four controller steps.  It is a full check pass, but it is the pass that the next
iteration would have performed; Proposition~\ref{prop:noextra} counts it as that
iteration's pass, and Proposition~\ref{prop:cost} counts the controller's own
edge sweep as the added work.
\item \emph{Free variables only.}  Peeled variables are fixed before the run, are
not updated by the controller, and are reported with a $\pm\infty$ posterior
marker that never enters the arithmetic.  All propositions are restricted to
free variables.
\item \emph{Floating point.}  The implementation forms the extrinsic messages
from prefix and suffix sums rather than by subtracting $m_{a\to i}$ from $R_i$.
This is equivalent in exact arithmetic and differs by rounding in floating point;
the affine response is the relation solved by the controller, not a bitwise
identity for every stored posterior.
\end{itemize}
The finite-sample identity and the overflow-continuation caveat for the research
copy are documented in Supplementary Note~3.

%% file: si/note2_theory.tex
\section*{Supplementary Note 2: Loop gain of memory min-sum}\label{si:note2}
\addcontentsline{toc}{section}{Supplementary Note 2: Loop gain of memory min-sum}
\setcounter{section}{2}
\setcounter{subsection}{0}

This note supports the predictive controller and the mechanism section of the
main text. The exact one-step response explains how memory changes a belief;
two certified families explain how memory combines with reinforcement through
the detector graph. The theorems apply under their stated hypotheses, the
certificates concern named detector error models, and the sampled diagnostics
are reported separately in Supplementary Note~3. None of these results predicts
logical failure rates or which family a decoding trajectory enters.

Notation. $H$ is the detector--fault incidence matrix after checks of degree one
have been removed, $L_i$ the prior log-likelihood ratio of fault $i$, $R_i$ its
belief, and $\mu$ and $m$ the variable-to-check and check-to-variable messages.
The memory strength is $\gamma$ (or $g$ for one variable), and $T$ counts
syndrome-extraction rounds. Growth factors are per iteration; their logarithms are growth rates.

\subsection{The exact one-step response}\label{si:response}

A memory update adds a multiple of the belief's displacement from its prior to
the next check field. Holding the current messages fixed therefore gives an
exact prediction of how the next belief changes with its own memory strength.
After degree-one checks are peeled, let $P$ copy variable values to edges,
$C=P^T$ sum check messages, and $\mathcal M_s$ be the unnormalized min-sum
check map, with syndrome $s$ and hard decision $\widehat e_i=1\{R_i\le0\}$.
The updates are
\begin{equation}
 m^+=\mathcal M_s(\mu),\qquad
 R^+=L+Cm^++\gamma(R-L),\qquad \mu^+=PR^+-m^+ .
 \label{eq:lg-update}
\end{equation}
For fixed current messages, $R_i^+(g)=F_i+gD_i$, where $F=L+C\mathcal M_s(\mu)$
and $D=R-L$. This identity is exact; it is the response used by the controller (equation~(2) of the main text). If $(F_i-U_i)D_i\ne0$, a memory strength strictly reduces the distance of the next belief from $U_i$, relative to no memory, exactly when it has the sign of $-(F_i-U_i)D_i$ and $0<|g|<2|(F_i-U_i)/D_i|$. If $D_i=0$ or $F_i=U_i$, no strict reduction is possible. A statement about this sign alone is not a stability theorem, which is why the channel analysis below is needed.

\subsection{The tied-minimum family}\label{si:c1}

When each check has at least two equally weak inputs, excluding one input does
not change the minimum magnitude it passes back. On the family below this
closes the update on two coefficients, so the effect of uniform memory on
reinforcement can be calculated exactly.

The relevant closed family arises when every check has at least two equally
weak inputs. For $w\ge0$, $w\ne0$, put $t_a=\min_{i\in \mathcal N(a)}w_i$ and
$q_i=\delta_iw_i$, $\delta_i\in\{-1,1\}$, and require
\begin{equation}
 \begin{gathered}
 \text{each minimum }t_a\text{ is attained at least twice},\qquad H^Tt=cw,
 \quad c>0,\\
 (-1)^{s_a}\prod_{i\in \mathcal N(a)}\delta_i=1\quad\text{whenever }t_a>0.
 \end{gathered}\label{eq:lg-c1}
\end{equation}
We call this condition $\mathrm{C1}(c)$. Because every minimum is attained twice, each check output equals the tied minimum with the sign fixed by the parity condition, and it does not change when the non-minimal inputs grow. Hence, for $u_0=Pq$, $m=\mathcal M_s(u_0)$ and $h=u_0-m$,
\begin{equation}
 Cm=cq,\qquad \mathcal M_s(xu_0+yh)=xm\quad(x,y\ge0).
 \label{eq:lg-flat}
\end{equation}
In particular, the additional parameter cannot be an arbitrary fitted real
number. If $w_i$ is a smallest positive entry, every adjacent active check has
$t_a=w_i$; hence $cw_i=d_i^+w_i$, with $d_i^+$ the integer active degree.
Thus $c\in\mathbb N$ and $b=c-1\in\mathbb Z_{\ge0}$. The minimum variables
need not have the same \emph{full} degree when the support is proper.

\paragraph{Proposition 2.1 (the two-pole law).}
Suppose $w\ge0$, $w\ne0$, and signs $\delta_i\in\{-1,1\}$ satisfy
condition~\eqref{eq:lg-c1}: every check minimum is attained at least twice,
$H^{\mathsf T}t=cw$ with $c>0$, and the sign parity holds on every check with
$t_a>0$. Suppose also that the priors vanish, $L=0$. On the homogeneous
family $\mu=xu_0+yh$, $R=czq$, each admissible update is
\begin{equation}
 (x^+,y^+,z^+)=(bx+c\gamma z,x,x+\gamma z),\qquad
 K_{b,\gamma}=\begin{pmatrix}b&(b+1)\gamma\\1&\gamma\end{pmatrix}.
 \label{eq:lg-core}
\end{equation}
Its nonzero characteristic factor is
\begin{equation}
 p_{b,\gamma}(\lambda)=\lambda^2-(b+\gamma)\lambda-\gamma.
 \label{eq:lg-poly}
\end{equation}
For $\gamma\ge0$ the nonnegative coefficient cone is invariant. For $\gamma<0$
the spectral statement concerns the $2\times2$ coefficient block only: the cone
need not be preserved, and Schur stability of the block is not a contraction
claim for the nonlinear iteration. On the upper real-root branch
$\gamma>\gamma_d=-b-2+2\sqrt{b+1}$, the dominant root satisfies
\begin{equation}
 \lambda_+=\frac{b+\gamma+\sqrt{(b+\gamma)^2+4\gamma}}2,
 \qquad \frac{d\lambda_+}{d\gamma}
 =\frac{\lambda_++1}{\sqrt{(b+\gamma)^2+4\gamma}}>0.
 \label{eq:lg-monotone}
\end{equation}
For $0\le b<3$ the dominant root crosses one at $\gamma_c=(1-b)/2$.
The full coefficient block is strictly Schur stable precisely when
\begin{equation}
 -1<b<3,\qquad -1<\gamma<(1-b)/2.
 \label{eq:lg-jury}
\end{equation}
At BP its roots are $b,0$, so BP is marginal in this block exactly when $b=1$.

\emph{Proof.} Substitution of \eqref{eq:lg-flat} into \eqref{eq:lg-update},
with $L=0$, gives \eqref{eq:lg-core}; its determinant gives
\eqref{eq:lg-poly}. Implicit differentiation proves \eqref{eq:lg-monotone}.
The three strict quadratic Jury inequalities are
$p(1)=1-b-2\gamma>0$, $p(-1)=1+b>0$ and $|-\gamma|<1$ (giving $1+\gamma>0$, while $\gamma<1$ is subsumed by $p(1)>0$ for $b\ge0$),
which give \eqref{eq:lg-jury}. At $\gamma_c$ the roots are $1,(b-1)/2$,
proving the crossing statement. A positive eigenray has
$y=x/\lambda$, $z=x/(\lambda-\gamma)$, with the indicated denominators
positive. For $\gamma>0$, a nonzero nonnegative $(x,z)$ excites the Perron root.
For other states, a mode grows only if its projection is nonzero.\hfill$\square$

There are essential qualifications for negative memory strength. Between
$-b-2-2\sqrt{b+1}$ and $\gamma_d$ the roots are complex and their modulus is
$\sqrt{-\gamma}$, which decreases as $\gamma$ increases. In particular,
$\gamma=-0.2,-0.3$ lie in this region for $b=1$; there is no larger real pole
there. For $b\ge3$ no memory strength makes the block Schur stable; for $b>3$ the root
crossing one at $\gamma_c$ is the smaller root. Moreover, negative memory strength need
not preserve \eqref{eq:lg-flat}'s cone. Negative-side spectral statements
require continued membership, a valid affine branch, or an independently
certified eigenray; they are not global decoder contraction claims.

The complete gross-code detector graph realizes the $b=1$ tied-minimum family explicitly. On this graph (936 detectors, 8,784 fault mechanisms,
30,672 edges; no check of degree one), we constructed a certificate with $c=2$ ($b=1$) and verified all 30,672 min-sum outputs and
condition~\eqref{eq:lg-c1} in integer arithmetic. Equation~(3) of the main text
is the general-$b$ polynomial~\eqref{eq:lg-poly}; the gross-code certificate
has $b=1$. For this block, both roots lie inside the unit circle precisely
for $-1<\gamma<0$.

The zero-amplitude variables of a tied-minimum certificate form a stopping set,
because each adjacent check has at least two such neighbors.
Stopping- and trapping-set analyses and graph-cover pseudocodewords describe
related supports and local dynamics~\cite{raveendran2021trapping,pacenti2026detectortraps,schlegel2010dynamics,vontobel2005graphcover},
whereas the result here concerns an exact invariant family and its response to memory.

\subsection{The temporal loop}\label{si:temporal}

A fault linking two successive detection events passes an extrinsic message
one step along the time direction in each iteration. At a time boundary,
a fault incident on $m+1$ detectors can return the sum of $m$ check messages,
reinforcing the signal after its passage through the temporal chain.

Throughout, $T$ counts syndrome rounds and $T+1$ counts detector slices.
We use unscaled flooding min-sum with zero priors and the unsigned map
on nonnegative edge messages,
\begin{equation}
 F(u)=(PC-I)\mathcal M_0(u),\qquad
 (\mathcal M_0(u))_{a\to i}=\min_{j\in \mathcal N(a)\setminus i}u_{j\to a},
 \label{eq:temporal-map}
\end{equation}
where $P$ copies variable values to edges and $C$ sums incoming check messages at each variable. The growth factor $\lambda_{\rm temp}$ below concerns strictly positive edge messages. This map is not the absolute-value dynamics of arbitrary signed BP; signed states or states with zero messages can have different growth factors.

\paragraph{Theorem 2.2 (temporal channel).}
Let $H$ be a finite binary detector--fault incidence matrix, with nonempty checks of degree at least two. Suppose (i) every fault has degree at least two; (ii) the degree-two faults, viewed as edges between detectors, form disjoint paths covering all detectors, each with exactly $T\ge1$ edges; (iii) all other faults have degree at least $m+1$, where $m\ge2$; and (iv) every path endpoint is incident on a degree-$(m+1)$ fault whose entire detector support consists of path endpoints. Then, for $u_{i\to a}=d_i-1>0$,
\begin{equation}
 F^{T+1}(u)=mu,\qquad \lambda_{\rm temp}=m^{1/(T+1)}.
 \label{eq:temporal-law}
\end{equation}
These are sufficient conditions, rather than a characterization of all amplifying DEMs. A graph traversal and one pass over fault supports test them in $O(M+N+\mathrm{nnz}(H))$ time, without power iteration or detector coordinates.

\emph{Proof.} Orient each path and label its detectors $r=0,\ldots,T$.
Write $u^k=F^k(u^0)$, with $u^0_{i\to a}=d_i-1$, and
$z^k=\mathcal M_0(u^k)$. Call all non-temporal faults ordinary and call
any degree-$(m+1)$ fault supported entirely on endpoints a witness.
For $0\le k\le T$ we prove simultaneously:
(a) the temporal fault joining $r$ to $r+1$ sends $m$ toward $r$ if
$k\ge T-r$ and one otherwise, and sends $m$ toward $r+1$ if
$k\ge r+1$ and one otherwise;
(b) every ordinary input is at least $m$;
(c) every witness input is exactly $m$; and
(d) $z^k_{a\to i}$ to an ordinary fault at depth $r$ is $m$ if
$k\ge\max(r,T-r)$ and one otherwise.

At $k=0$ the temporal inputs are one, the ordinary inputs are $d_i-1\ge m$,
and witness inputs are $m$. Every detector has a temporal input, so (d)
follows as well. Suppose (a)--(d) hold at $k<T$. An interior check has
exactly two temporal inputs. On excluding one of them, the other is at most
$m$ and all ordinary inputs are at least $m$, so the check selects that
other temporal value. At an endpoint, exclusion of the sole temporal input
leaves ordinary inputs with minimum exactly $m$, attained by a witness.
The degree-two variable update therefore advances each threshold in (a)
by one step. On excluding an ordinary input, at least one temporal input
remains; all such check outputs are one or $m$. Summing over the other
$d_i-1\ge m$ checks proves (b) at $k+1$, including degree-$(m+1)$ faults
that are not witnesses. At every endpoint, (d) at $k<T$ gives output one
to every ordinary fault. All checks of a witness are endpoints, so its
next outgoing sums are exactly $m$. This proves (c) at $k+1$ without
assuming that every degree-$(m+1)$ fault is a witness. Finally, at an
interior check the temporal inputs at phase $k+1$ are both $m$ exactly
when $k+1\ge r$ and $k+1\ge T-r$; at an endpoint its sole temporal
input becomes $m$ exactly at phase $T$. Together with (b), these facts
prove (d) at $k+1$ and close the induction.

At phase $T$, every temporal input is $m$. An output to an ordinary
fault has a temporal competitor equal to $m$; an output to a temporal
fault has another temporal competitor at an interior check and a witness
at an endpoint. Thus every entry of $z^T$ is exactly $m$, and
$F(u^T)_{i\to a}=m(d_i-1)$. This also covers $T=1$, when all checks are
endpoints and the single induction step precedes the final variable update.
This proves the exact identity
\begin{equation}
 F^{T+1}(u^0)=m\,u^0.
 \label{eq:temporal-seed}
\end{equation}
Now let $x$ be any strictly positive state. Since $u^0>0$, there are constants $0<c\le C$ with $cu^0\le x\le Cu^0$. Monotonicity and positive homogeneity give
\[
 c\,m^n u^0\le F^{n(T+1)}(x)\le C\,m^n u^0,
\]
and taking $(T+1)n$-th roots yields the common asymptotic growth factor $m^{1/(T+1)}$; the same squeeze applies at intermediate times. This is the cone spectral radius of the map on the strictly positive cone~\cite{lemmens2012nonlinear,gaubert2004perron}. The identity \eqref{eq:temporal-seed} is exact for the stated seed; it does not assert coordinatewise multiplication by $m$ for every positive state, only the stated asymptotic growth factor.\hfill$\square$

\paragraph{Certified growth factors.}
Across rebuilt round counts $T$, integer iteration independently certifies the exact full-cone growth factor $m^{1/(T+1)}$ for LP20 ($m=2$), Mitten-540 ($m=2$) and Coset-128 ($m=3$) at every populated entry of Supplementary Table~\ref{tab:temporal-rates}. To test scaling with $T$, we rebuilt LP20, Mitten-540 and Coset-128 at fixed per-family noise probabilities and unchanged extraction schedules. At the original round counts, circuit, DEM, matrices and priors reproduce the frozen inputs exactly. One prespecified syndrome per graph yielded seven exact signed-cycle certificates near pre-stop runaway states (LP20 at $T=16,20,24$, Mitten-540 at $T=12,18,20$, and Coset-128 at $T=12$). The other finite probes establish no absence claim. Higher-degree faults remain in all computations. They do not defeat the temporal minima under the theorem's hypotheses.

\begin{table}[ht]
\centering
\SItablefont
\caption{Certified full-positive-cone BP growth factors. No populated entry deviates from $m^{1/(T+1)}$. The separate original instances BB-144 ($T=12$), BB-72 ($T=6$) and Kasai-96 ($T=12$) have growth factors $2^{1/13}$, $2^{1/7}$ and $2^{1/13}$, respectively.}
\label{tab:temporal-rates}
\begin{tabular}{rccc}
\hline
$T$ & LP20, $m=2$ & Mitten-540, $m=2$ & Coset-128, $m=3$\\
\hline
4  & 1.148698355 & 1.148698355 & 1.245730940\\
8  & 1.080059739 & 1.080059739 & 1.129830964\\
12 & 1.054766076 & 1.054766076 & 1.088182243\\
16 & 1.041616011 & 1.041616011 & 1.066758117\\
18 & ---         & 1.037155044 & ---\\
20 & 1.033557783 & 1.033557783 & 1.053707472\\
24 & 1.028113827 & 1.028113827 & 1.044924351\\
\hline
\end{tabular}
\end{table}

All six original circuits use noiseless final data readout. Their initial detectors contain the first ancilla readout; their final detectors compare the last ancilla readout with the parity of the final data readouts. The final record counts are seven for both bivariate bicycle codes, thirteen for Kasai-96, nine for LP20, ten for Mitten-540 and nine for Coset-128. Every endpoint has a pure-boundary degree-three fault except Coset-128, where the degree is four. Explicit fault tracing locates these reflectors in preparation and last-round gate noise. Time-like DEM columns also merge measurement flips and gate faults with the same detector effect; they are not uniquely measurement-error locations.

Signed cycles require separate certificates. On the original LP20, Mitten-540
and Coset-128 graphs, signed states extracted near saved runaway trajectories
have return lengths 42, 38 and 26, respectively, with gain four. The corresponding
growth factors are $2^{1/21}$, $2^{1/19}$ and $2^{1/13}$; the last differs from Coset-128's
strictly positive growth factor $3^{1/13}$. All seven signed cycles in the round scan
have $2(T+1)$-step gain four, but exact half-period checks reject componentwise
doubling of messages and beliefs, including their absolute amplitudes.
These certificates establish homogeneous cycles near the saved trajectories,
not attraction from the real priors.

For the certified temporal class, $\lambda_{\rm temp}-1\sim\ln m/(T+1)$ as the window grows,
provided the boundary conditions are preserved. The asymptotic doubling time
is $(T+1)\ln2/\ln m$.
\IfFileExists{figures/figS6.pdf}{Supplementary Figure~\ref{fig:S6} shows this
relation and the certified growth factors.}{}

\paragraph{Uniform memory and the unit-gain state.}
For $G_\gamma(u,R)=(F(u)+\gamma PR,C\mathcal M_0(u)+\gamma R)$, set $k=\lambda-\gamma$ and $h=\gamma/\lambda$. Interior temporal check outputs obey the transfer relation
\begin{equation}
 \binom{U_{r+1}}{V_{r+1}}=A_{\rm tr}\binom{U_r}{V_r},\quad
 A_{\rm tr}=\begin{pmatrix}(1-h^2)/k&h\\-h&k\end{pmatrix},\quad
 B=\frac{m+h}{k},\quad (-B,1)A_{\rm tr}^{T}(B,1)^{\mathsf t}=0.
 \label{eq:temporal-transfer}
\end{equation}
Here $A_{\rm tr}^{T}$ denotes a matrix power; the boundary relation applies while its full-graph minimum conditions hold. At $\lambda=1$, $A_{\rm tr}=I+\gamma N$ with $N^2=0$. The first boundary branch, with the other $m$ inputs on the same slice, has $B_1=(m+h)/k$. Substitution shows that an exact state of unit gain exists at
\begin{equation}
 \gamma_0=-\frac{m-1}{(m+1)T+2}.
 \label{eq:temporal-unit}
\end{equation}
A full-graph construction uses $U_r=mT+1-(m-1)r$ and $V_r=T+1+(m-1)r$. Ordinary check outputs are $q_r=\min(U_r,V_r)$, with endpoint value $T+1$. For a non-temporal input at distance $d$ from the nearest endpoint,
\begin{equation}
 u_{i\to a}\ge mT+1+\frac{\gamma_0}{1-\gamma_0}(m-1)d
 \ge mT+1-(m-1)d.
\end{equation}
Thus every ordinary input is at least as large as the temporal competitors; endpoint witnesses attain equality. Independent integer checks certify the construction on the named DEMs. Thus $\gamma_0=-1/(3T+2)$ for $m=2$, including $-1/62$ and $-1/56$, whereas Coset-128 has $\gamma_0=-1/(2T+1)$, giving $-1/25$ at $T=12$.
The second, next-slice boundary branch is excluded at its formal unit root
on the original and rebuilt graphs by exact rational certificates, because it
requires a minimum above $m$ while a same-slice witness forces an input below $m$.
A unit-gain state alone establishes no global stability threshold, since
negative memory does not preserve monotonicity.

\paragraph{The two positive-memory growth factors quoted in the main text.}
At $\gamma=0.1$, algebraic arithmetic certifies the Mitten-540 growth factor
$\lambda=1.224080272500746\ldots$ on the next-slice boundary branch.
For LP20, exact rational bounds on the full positive augmented map give
\[
 1.2206986810<r_{0.1}<1.2207179050.
\]
The bounds follow by verifying $aX\le G_{0.1}(X)\le bX$ for a strictly
positive rational state $X$: monotonicity and homogeneity iterate the bounds,
and comparison with $X$ bounds the asymptotic growth factor between $a$ and $b$.
Both growth factors round to 1.22, compared with 1.18 from the tied-minimum law with
$b=1$. They are positive-state certificates, not certified continuations of
the signed BP cycles; a temporal BP growth factor alone does not determine the
response to memory.

\subsection{A one-step information example}\label{si:barrier}

The present local state need not determine the next incoming check field.
A two-variable example shows why observing that field can change the memory
strength needed to cancel the next belief.

\paragraph{Remark 2.3 (two variables).}
On $H=(1\ 1)$, $s=1$, $L=(1,1)$, two legally controlled, still-active histories
from standard initialization have the same local tuple
$(R_1,L_1,g_1,\rho_1)=(-1/400,1,-1/20,1)$ and the same own incoming and outgoing
messages, but next fields $F_1=\pm19/400$. Thus every shared memory strength obeys
\[
 \max\{|R_{1,A}^+(g)|,|R_{1,B}^+(g)|\}
 =\frac{19}{400}+\frac{401}{400}|g|\ge19|R_1|,
\]
whereas observing the next field permits memory strengths $\pm19/401$ that cancel the
belief separately. The histories and rational verifier are supplied with the
certificates. This is a one-step example for the stated observation, not a
long-term impossibility theorem or a claim that TEMPO-BP generates those histories.

\subsection{A certified single action}\label{si:action}

The value of a control action can lie several iterations ahead. The saved example below isolates one hold action while keeping
the common past and all other actions fixed.
\IfFileExists{figures/figS7.pdf}{Supplementary Figure~\ref{fig:S7} shows the
saved two-branch comparison described below.}{}

In one saved trajectory of the gross-code decoder, seven variables receive a hold action at one control step, that is, a raw target above $\gamma_{\max}$. We replaced each of these seven actions independently by the brake action with the same step size, keeping the common past and all other actions fixed, and replayed all $2^7=128$ combinations. The decoder reaches its first syndrome-valid state within four iterations if and only if the action on fault variable 6901 is retained: all 64 such combinations succeed at the fourth iteration with all twelve logical observables correct, and none of the other 64 satisfies the syndrome within four iterations.

Compare the branch in which all seven actions are replaced with the branch in which only that on variable 6901 is kept. The hard decisions on all 8,784 variables coincide for three iterations. At the fourth, the belief of variable 6901 is $-0.5922306265$ in the first branch and $1.2499207754$ in the second, and its exact decomposition has the following rounded values:
\begin{equation}
 \Delta R_4\approx2.0289293225-0.0677039423-0.1190739783=1.8421514019 ,
\end{equation}
where the first term is the retained difference in memory strength multiplied by the displacement of the belief from its prior, and the other two are the differences already present in the belief and in the check field. The benefit thus comes from the memory strength that is held, not from a more favorable check field.

The statement holds on intervals and not only at the two end points. Let $h\in[0,1]$ interpolate the action on variable 6901 between the brake action ($h=0$) and the hold action ($h=1$), with the other six actions set to brake. Exact rational bounds, which use the global 1-Lipschitz property of min-sum and allow signs, minimal inputs and ties to change, prove that no syndrome-valid state occurs within four iterations for $h\in[0,1/128]$ and that the first valid state occurs at the fourth iteration and is logically correct for $h\in[63/64,1]$. The interval in between is not classified. These rational certificates interpret the saved binary64 states, memory strengths and computed log-priors as exact rational inputs; they do not enclose the original transcendental log-priors or certify each native floating-point rounding operation.

This counterfactual probes one trajectory history, action sequence and four-iteration horizon from a successful run rather than estimating cohort-wide frequency: across six successful end segments examined identically, replacing all hold actions prevented success within the horizon in three, advanced it in two and left it unchanged in one.

\subsection{Certificates and verifiers}

The certificates check the specified states against every relevant check output
and variable sum on the detector graph. They distinguish an exact mechanism
from a growth rate observed on a sampled trajectory; they do not establish
how often decoding reaches that mechanism.

\begin{table}[ht]
\centering
\caption{Exact certificates retained in this note. The accompanying data and
verifiers use the arithmetic indicated.}
\label{tab:certificates}
\SItablefont
\begin{tabular}{p{67mm}p{55mm}p{30mm}}
\hline
Statement & Detector graph & Arithmetic\\
\hline
Tied-minimum channel, $b=1$ & BB-144 & integer\\
Temporal channel, $\lambda_{\rm temp}=m^{1/(T+1)}$ & 19 rebuilt graphs of LP20, Mitten-540, Coset-128; six original graphs & integer\\
Unit states at $\gamma_0$; exclusion of the second unit branch & same & integer, rational\\
Signed periodic orbits near runaway trajectories & LP20, Mitten-540, Coset-128 & integer\\
Growth factors at $\gamma=0.1$ & Mitten-540; LP20 & algebraic; rational bounds\\
One-step information example & two variables & rational\\
Single action deciding a decode & BB-144 & rational interval\\
\hline
\end{tabular}
\end{table}

The accompanying data and verification programs cover the statements in
Supplementary Table~\ref{tab:certificates}, using the arithmetic listed there.
The code and data repository supplies the certificates and their verifiers.

%% file: si/note3_mechanism.tex
\section*{Supplementary Note 3\quad Mechanism diagnostics across codes}
\addcontentsline{toc}{section}{Supplementary Note 3: Mechanism diagnostics across codes}

This note collects the mechanism experiments that support the loop-gain picture
and the controller ablations.  They are mechanism diagnostics on separately drawn samples rather than the accuracy tables of the main text.  We report negative and mixed outcomes with the same weight as positive ones, because several of them bound what the mechanism claims can be.

\subsection*{3.1 Experimental scope and definitions}

The experiments use a research copy of the decoder that exposes switches for the
controller ablations and records belief amplitudes.  With all switches off, the
research copy reproduces the released decoder on the checked output fields:
decoding bits, validity, iteration count, number of controller updates, solution
weight, warm-up iterations and main-stage iterations.  The check used 3,000
adaptive and 600 static shots of each bivariate bicycle code, with the same
global shot indices and seed.  These matched-run diagnostics confirm exact agreement across all 7,200 tested traces.  The research core also retains a diagnostic
continuation mode after non-finite arithmetic, whereas the production decoder
asserts finiteness; the identity statement therefore covers the tested finite
trajectories, not every overflow path.

For the four transfer codes the reconstruction gate compared the rebuilt circuits
with the archived originals by SHA-256 (all four identical), reused the same
$H$, $O$ and prior files as the mechanism experiment, and compared eight
center-point arms with the published rates.  The eight two-sided Fisher tests gave
$P=0.22$--$0.96$, all above the pre-specified Bonferroni threshold $0.05/8$;
this is compatibility at the achieved sample size, not an equivalence proof.
The transfer circuits and gate schedules are those of the main text.

Within every experiment all arms decode the same syndrome bank with the same
global shot indices.  Uniform-memory arms use an exactly constant $\gamma$ and
the controller is off.  Disordered arms draw $\gamma_i$ once per shot and
variable from an interval and keep the field fixed.  Arms that differ only in an
interval share the underlying uniform random stream; they do not necessarily
have the same numerical $\gamma_i$ field.  A shot is a logical failure if it does
not converge or converges to a wrong logical class.  ``Runaway'' means that the
maximum absolute posterior over free variables at the last executed iteration
exceeds $10^{12}$; it is not the same as non-convergence, and a converged shot can
be counted as a runaway.  ``Bounded non-convergence'' means non-convergence with
all recorded amplitudes below the threshold and no non-finite flag.  These are
finite-window descriptions, not certificates of a fixed point or of exponential
growth.  The mechanism runs use a 3,000-iteration cap except where stated; the
confirmation sample and the two bivariate bicycle uniform grids use 10,000.

\subsection*{3.2 Uniform memory on the six codes}

The uniform-memory curves are the cleanest test of the speed--convergence trade-off.
All six curves have all fifteen uniform strengths on the fixed grid.
The gross code, BB-72 and Coset-128 use $p=0.4\%$.
The gross-code and BB-72 scans use 10,000 common shots and a 10,000-iteration cap; the
Coset-128 scan uses 10,000 common shots and a 3,000-iteration cap.  The complete
curves all have an interior minimum and rising ends, but their shapes and
failure compositions differ, and the differences matter.

\begin{itemize}
\item \emph{Gross code.}  The grid minimum is $\gamma=0.05$ with 668 failures in
10,000 shots (6.68\%); TEMPO-BP fails on 72 of the same shots (0.72\%), a
ratio of logical failure rates per shot of 9.28.  At $\gamma=-0.30$ all 10,000 shots fail to converge,
with bounded amplitudes (median peak 43.5, 99th percentile 54.2).  At
$\gamma=+0.50$ the failure count is 2,536 (25.36\%), and 2,515 of those failures
have a final amplitude above $10^{12}$.  Most failures in this positive-memory arm therefore end in
runaway, but the minimum itself is not a zero-runaway point.
\item \emph{BB-72.}  The observed minimum is $\gamma=0.10$ with 498
failures in 10,000 (4.98\%); TEMPO-BP fails on 454 (4.54\%), a ratio of only
1.10.  The curve is shallow and locally non-monotone: $\gamma=0.05,0.075,0.10$
give 509, 519 and 498 failures.  The positive branch is dominated by converged
wrong-class outcomes, not by runaway.  At $\gamma=+0.40$ there are 923 failures,
of which 912 are wrong class and 792 are bounded wrong-class failures without a
runaway; only 131 failures have a final runaway.  The published Relay-BP interval
gives 1,223 failures, of which 1,167 are wrong class, with zero final runaway and
zero non-finite flags.  Thus a two-way classification into ``frozen'' and
``runaway'' is incomplete on this code.
\item \emph{Coset-128.}  The observed minimum is $\gamma=0.075$ with 1,717
failures in 10,000 (17.17\%); TEMPO-BP fails on 509 (5.09\%), a ratio of 3.37.
At $\gamma=-0.30$ all 10,000 shots fail, with bounded amplitudes; at
$\gamma=+0.50$ there are 6,256 failures, 6,240 of them with a final runaway.  Even
plain BP at $\gamma=0$ has 140 final runaways (65 of them failures), so exceeding
the amplitude threshold is not by itself evidence for positive memory.
\item \emph{LP20.}  All 15 grid points are complete, on 2,000 common shots at
$p=0.175\%$ with a 3,000-iteration cap.  The grid minimum is
$\gamma=+0.075$ with 290 failures (14.50\%); TEMPO-BP fails on 38 of the same
shots (1.90\%) and the published Relay-BP interval on 1,035 (51.75\%).
Both sides rise away from the minimum.  At $\gamma=-0.30$ all 2,000 shots fail
to converge with bounded amplitudes; at $\gamma=+0.50$ all 1,705 failures
have a final runaway.  Plain BP at $\gamma=0$ already has 505 final runaways,
355 of which are failures and 150 of which are successful returns.
\item \emph{Kasai-96.}  All 15 grid points are complete, from
$\gamma=-0.30$ to $+0.50$, on 2,000 common shots at
$p=0.06875\%$ with a 3,000-iteration cap.
The grid minimum is $\gamma=+0.025$ with 316 failures (15.80\%); TEMPO-BP
fails on 75 of the same shots (3.75\%) and the Relay-BP interval on 1,618 (80.90\%).  Plain BP at
$\gamma=0$ has zero final runaways here, whereas $\gamma=+0.10$ has 315; the
runaway count is therefore state- and code-dependent.  Both sides rise away
from the minimum.  At $\gamma=-0.30$ all 2,000 shots fail to converge with
bounded amplitudes; at $\gamma=+0.50$ all 1,914 failures have a final runaway.
\item \emph{Mitten-540.}  All 15 grid points are complete, on 2,000 common shots
at $p=0.1125\%$ with a 3,000-iteration cap.  The grid minimum is
$\gamma=+0.025$ with 189 failures (9.45\%); TEMPO-BP fails on 10 of the same
shots (0.50\%) and the Relay-BP interval on 755 (37.75\%).  Both sides rise
away from the minimum.  At $\gamma=-0.30$ all 2,000 shots fail to converge
with bounded amplitudes; at $\gamma=+0.50$ all 1,885 failures have a final
runaway.  At
$\gamma=0$ there are 236 final runaways (142 failures), again showing that the
threshold is crossed under plain BP.
\end{itemize}

\begin{table}[!htbp]
\centering
\SItablefont
\caption{Uniform-memory logical failures on the LP20, Kasai-96 and Mitten-540
mechanism cohorts.  Each column uses 2,000 common shots and a 3,000-iteration
limit; the physical error probabilities are $0.175\%$, $0.06875\%$ and
$0.1125\%$, respectively.  The TEMPO-BP row uses the same shots as its column.}
\label{tab:uniform-transfer}
\begin{tabular}{@{}lrrr@{}}
\toprule
Uniform memory strength & LP20 & Kasai-96 & Mitten-540 \\
\midrule
$-0.30$ & 2,000 & 2,000 & 2,000 \\
$-0.20$ & 2,000 & 2,000 & 2,000 \\
$-0.10$ & 1,926 & 1,882 & 1,645 \\
$-0.05$ & 649 & 864 & 546 \\
$0$ & 376 & 346 & 213 \\
$+0.025$ & 347 & 316 & 189 \\
$+0.05$ & 339 & 363 & 237 \\
$+0.075$ & 290 & 452 & 289 \\
$+0.10$ & 311 & 534 & 328 \\
$+0.15$ & 339 & 785 & 516 \\
$+0.20$ & 435 & 1,023 & 736 \\
$+0.25$ & 557 & 1,234 & 927 \\
$+0.30$ & 732 & 1,401 & 1,165 \\
$+0.40$ & 1,160 & 1,713 & 1,581 \\
$+0.50$ & 1,705 & 1,914 & 1,885 \\
\midrule
TEMPO-BP & 38 & 75 & 10 \\
\bottomrule
\end{tabular}
\end{table}

The complete-grid minima are at $\gamma=0.05$ (gross code), $0.10$
(BB-72), $0.075$ (Coset-128 and LP20) and $0.025$ (Kasai-96 and Mitten-540).
These are minima on the same discrete grid with different sample sizes and
iteration limits, and the paired bootstrap intervals in the source
tables are conditional on the selected cell; they must not be read as a
continuous optimum or as evidence that a single memory strength transfers
between codes.  What the six codes support is narrower: the complete curves have
an interior low-failure region, the negative side loses convergence, the
positive side eventually loses accuracy, and the failure composition on the
positive side changes with the code.  Supplementary Figure~\ref{fig:S1} shows
all fifteen strengths on each of the six curves, through $\gamma=+0.50$;
Supplementary Table~\ref{tab:uniform-transfer} gives the three complete transfer-code grids above.

\subsection*{3.3 The 42-field disordered scan}

Across a 42-field center--width grid on the gross code ($p=0.4\%$), even the top confirmed static random intervals leave 6.6\%--8.2\% logical failures---at least 9.35 times TEMPO-BP's 71/10,000 failure count. The disordered scan varies the center $c$ and width $w$ of the interval from
which a fixed per-variable field is drawn: $c\in\{-0.1,0,0.1,0.2,0.3,0.4,0.5\}$
and $w\in\{0,0.2,0.4,0.6,0.85,1.2\}$, with $\gamma_i\sim\mathcal
U[c-w/2,c+w/2)$.  All 42 cells were run on 4,000 common gross-code shots at
$p=0.4\%$ with a 3,000-iteration cap.  The lowest failure count in the discovery
sample is 284/4,000 at $c=0.10,w=0.20$; the next are 291/4,000 at
$c=0.10,w=0$ and 351/4,000 at $c=0.20,w=0$.  TEMPO-BP fails on 30/4,000
(0.75\%) on the same cohort.

The three lowest cells were selected on the discovery sample and re-evaluated on
an independent 10,000-shot sample with a 10,000-iteration cap.  The cells
$(c,w)=(0.10,0)$, $(0.10,0.20)$ and $(0.20,0)$ fail on 664, 689 and 820
shots, respectively, against 71 for TEMPO-BP.  The lowest observed logical
failure rate per shot is 9.35 times that of TEMPO-BP (paired bootstrap 95\% confidence
interval 7.6--12.0, from 200,000 resamples of the paired outcome table).
This interval is conditional on the selected cell and does not correct for
selecting the lowest count within the confirmation sample.  All three selected cells lie on non-negative centers and remain clustered near 6.6\%--8.2\% regardless of whether the 34 non-finite logical successes observed in the $c=0.10,w=0$ cell are counted as failures (changing its count from 664 to 698).

The scan also shows that removing runaways is not the same as minimizing
failures.  The best zero-runaway cell, $c=0.10,w=0.40$, fails on 358/4,000, which
is 74 more failures than the discovery minimum, while the minimum cell still has
154 final runaways (19 of them successful returns).  A control objective that
only suppresses amplitude would therefore not reproduce the observed accuracy
ordering.  Supplementary Figure~\ref{fig:S2} displays the full center--width map and the
confirmation sample.

\subsection*{3.4 Controller ablations on four codes}

Disabling or constraining individual controller elements---online updates, negative strengths, braking, holding or residual gating---raises observed logical failures on every tested code (statistically significant across all seven ablations on BB-144, Coset-128 and LP20, and six of seven on BB-72). Each ablation changes one element of the controller and keeps the timing,
stopping rule and, where applicable, the initial field.  The gross code,
BB-72 and Coset-128 use 10,000 common shots at $p=0.4\%$;
the two bivariate bicycle codes use a 10,000-iteration cap and Coset-128 a
3,000-iteration cap.  LP20 uses 2,000 common shots at $p=0.175\%$ with a
3,000-iteration cap.  The logical failure counts are:

\begin{center}
\SItablefont
\begin{tabular}{@{}lrrrr@{}}
\toprule
Arm & BB-144 & BB-72 & Coset-128 & LP20 \\
\midrule
TEMPO-BP & 75 & 454 & 509 & 38 \\
always brake (target $\gamma_{\min}$) & 2,552 & 1,132 & 6,175 & 1,261 \\
always hold (target $\gamma_{\max}$) & 2,286 & 556 & 6,335 & 1,636 \\
blind decay toward $\gamma_{\min}$ & 1,738 & 788 & 4,549 & 896 \\
plain BP ($\gamma\equiv0$) & 994 & 631 & 3,092 & 376 \\
controller off (TEMPO-static) & 1,227 & 522 & 4,385 & 715 \\
non-negative range ($\gamma_{\min}=0$) & 349 & 497 & 2,081 & 245 \\
no residual gate ($\beta=0$) & 110 & 461 & 553 & 54 \\
\bottomrule
\end{tabular}
\end{center}

The endpoint-target and blind-decay arms retain the denominator and
raw-target finiteness guards. Blind decay uses the constant step
$\operatorname{clip}_{[0,1]}(\kappa_0(1-\beta))$ when these guards permit an update.
The no-residual-gate arm also rescales $\kappa_0$ by the original factor
$(1-\beta)$, so that the step far from unsatisfied checks is unchanged.
All comparisons are paired exact McNemar tests on the same shots.  For
BB-72 the seven-arm Holm family gives $P=1.14\times10^{-171}$ (always
brake), $5.07\times10^{-11}$ (always hold), $1.12\times10^{-71}$ (blind decay),
$5.89\times10^{-22}$ (BP), $1.53\times10^{-6}$ (static), $0.0046$ (non-negative)
and $0.628$ (no residual gate).
For Coset-128 all seven Holm-adjusted comparisons reject, with the residual gate
at $P=0.0080$; the other adjusted probabilities range from $4.49\times10^{-393}$
to below $10^{-1700}$.  For the gross-code ablation cohort, the paired exact
tests computed from the saved per-shot arrays reject for all seven arms
($P<10^{-250}$ for always brake, always hold, blind decay, BP and static;
$P=6.1\times10^{-67}$ for non-negative; $P=1.27\times10^{-4}$ for the residual
gate).  On LP20 all seven Holm-adjusted comparisons reject: the residual-gate
comparison gives $P=0.0365$, and the other adjusted probabilities range from
$4.12\times10^{-53}$ to $5.04\times10^{-478}$.  All LP20 failures in the table
are failures to converge, with no wrong-class returns.  These are the seven comparisons within each code, not a single
cross-code family.

Across all four codes, these ablations show that state-dependent feedback in both directions outperforms constant-action and non-negative baselines, while residual gating provides statistically significant gains on the convergence-limited codes. On the gross code, removing the rare upper actions is worse than plain BP and worse than
switching the controller off; the same is true on BB-72, although the
margin over plain BP is smaller.  On Coset-128, ``always hold'' is catastrophic
and most of its failures are runaway.  On BB-72, by contrast,
``always hold'' has only 28 failures with a final runaway among its 556
(the total runaway count is 30, including two successful returns): 528 are
bounded wrong-class outcomes.  The non-negative arm is also not a pure sign intervention
on BB-72: it changes the initialization lower bound from
$-0.0260$ to $0$ as well as the controller lower bound, so its 497 failures
cannot be attributed to removing negative control actions alone.  The residual gate delivers statistically significant failure reductions on the convergence-limited codes (gross code, Coset-128 and LP20), whereas no statistically significant difference is detected on BB-72 ($P=0.628$), where 95.9\% of TEMPO-BP failures are already converged wrong-class corrections ($z_a=0$).

The gross-code ablation cohort and the gross-code uniform-grid cohort are
separate: the ablation table gives 75 failures
for TEMPO-BP and 994 for plain BP, while the uniform table gives 72 and 969 on
its own 10,000-shot sample.  We do not pool these cohorts; the qualitative
ordering is the same.

\subsection*{3.5 One open-loop run with the published Relay-BP interval}

A single open-loop run with the interval published for Relay-BP is not
Relay-BP-5~\cite{muller2025relay}: it has no Mem-BP warm stage, no successive legs and no collection
of five candidates.  On the six mechanism cohorts this single run fails on
1,556/10,000 (15.56\%) on the gross code, 1,223/10,000 (12.23\%) on BB-72,
1,035/2,000 (51.75\%) LP20, 1,618/2,000 (80.90\%) Kasai-96, 3,949/10,000
(39.49\%) Coset-128 and 755/2,000 (37.75\%) Mitten-540.  In every case the final
runaway count and the non-finite count are zero.  Zero final runaway counts do
not exclude converged wrong-class failures: Coset-128 has 119 such failures.
On the gross code the single run is worse than
plain BP; on BB-72 it is dominated by wrong-class returns.  This
comparison measures the failure rate per shot of one fixed random memory field;
it neither isolates the contributions of successive legs and candidate selection
to Relay-BP's accuracy nor benchmarks Relay-BP-5 itself.

\subsection*{3.6 Parameter diversity versus seed multiplicity}

On 20,000 common gross-code shots at $p=0.4\%$ (capped at 1,205 iterations per member), independent random seeds already deliver most of the 15-member ensemble reduction, while parameter diversity yields 72 against 88 selected failures (Holm $P=0.0664$, not reaching significance in this sample) alongside statistically significant exploratory gains in convergence and oracle failure count (Holm $P=8.0\times10^{-4}$ and $1.95\times10^{-3}$). The diversity experiment compares two 15-member arms.  One arm uses the
frozen 15-member parameter roster; the other uses 15 replicas of the first
member with different seeds.  Both select the correction most probable under the priors
among the converged corrections.  At $S=1$ both arms fail on 268 shots.  At $S=2$ both fail on 186:
the diverse arm trades 34 fewer non-convergences for 34 more selected wrong-class
outcomes.  At $S=15$ the diverse arm fails on 72 and the replicas on 88
(19 versus 35 non-converged; 53 selected wrong-class outcomes in both).  The
oracle failure counts are 40 and 57.  The exact McNemar test for the selected
failure count gives $P=0.0166$, and the Holm adjustment over the four
pre-specified sizes ($S=2,4,8,15$) gives $P=0.0664$; thus the primary selected-failure comparison does not reach significance here, whereas the convergence and oracle comparisons belong to separate exploratory families (Holm-adjusted $P=8.0\times10^{-4}$ and $1.95\times10^{-3}$).

The diverse arm also has a larger mean depth: 135.88 versus 117.57
iterations at $S=15$, and it has more distinct valid correction vectors per shot on average (11.585
versus 10.295).  Seed multiplicity alone already reduces the selected failures
from 268 at $S=1$ to 88 at $S=15$.  In this 20,000-shot cohort, parameter diversity reduces selected failures further to 72 and improves convergence and candidate coverage with the same directional trend confirmed at statistical significance in the larger independent 100,000-shot comparison of Supplementary Note~4, Section~4.8 (376 against 419 failures, $P=0.0087$).

\subsection*{3.7 Per-shot growth rates}
\label{sec:si-growth}

Unconverged trajectories in the finite-window records follow two certified amplification regimes: on the gross code across nine positive $\gamma$, the growing-branch median matches the $b=1$ graph-cycle rate $\ln\lambda_+(\gamma)$ to four digits, whereas on LP20 and Mitten-540 the faster temporal-chain channel dominates and sustained growth drops sharply around the certified unit-gain boundary $\gamma_0<0$.  Here, each per-shot growth rate is the late-time slope of the logarithm of the maximum absolute posterior, $\max_i|R_i|$, computed as a secant from three quarters of the recorded trace to its end; a trajectory is called growing when this rate exceeds $0.001$ per iteration. Traces stop at amplitude $10^{200}$; the final converged status is taken from a full run to the 3,000-iteration cap. The six codes use the mechanism syndrome banks at $p=0.4\%$ (gross code, BB-72, Coset-128), $p=0.175\%$ (LP20), $p=0.06875\%$ (Kasai-96) and $p=0.1125\%$ (Mitten-540), with Stim seed 20261113 and decoder seed 0. The saved cohorts contain 10,000 shots for the first three codes and 2,000 for the other three. Supplementary Table~\ref{tab:growth01} gives the examined unconverged trajectories, the growing count, the median rate of the growing ones and the candidate rate used for comparison at $\gamma=0$ and $\gamma=0.1$.

\begin{table}[!htbp]
\SItablefont
\setlength{\tabcolsep}{4pt}
\caption{Growing branch of the unconverged trajectories at $\gamma=0$ and $\gamma=0.1$. ``Examined'' counts unconverged trajectories with an estimable late-time rate; ``growing'' counts those above $0.001$ per iteration. The last column gives the candidate rate and the fraction of growing trajectories assigned to it; assignment compares numerical rates and does not certify that a trajectory belongs to an invariant family. A dash means that no examined trajectory exceeds the growth threshold.}
\label{tab:growth01}
{\centering
\begin{tabular}{@{}lrrrrl@{}}
\toprule
Code & $\gamma$ & Examined & Growing & Median rate & Candidate rate matched \\
\midrule
BB-144 & 0 & 200 & 4 & 0.002185 & none within tolerance \\
BB-144 & 0.1 & 200 & 140 & 0.169261 & $\ln\lambda_+$ ($b=1$), 138/140 \\
BB-72 & 0 & 200 & 0 & --- & --- \\
BB-72 & 0.1 & 125 & 60 & 0.169261 & $\ln\lambda_+$ ($b=1$), 59/60 \\
LP20 & 0 & 200 & 188 & 0.033010 & $2^{1/21}$, $\ln=0.033007$, 188/188 \\
LP20 & 0.1 & 200 & 200 & 0.199426 & $1.2207$, $\ln=0.199424$, 191/200 \\
Kasai-96 & 0 & 200 & 0 & --- & --- \\
Kasai-96 & 0.1 & 200 & 102 & 0.169261 & $\ln\lambda_+$ ($b=1$), 100/102 \\
Mitten-540 & 0 & 200 & 134 & 0.036484 & $2^{1/19}$, $\ln=0.036481$, 132/134 \\
Mitten-540 & 0.1 & 200 & 188 & 0.202190 & $1.2241$, $\ln=0.202206$, 120/188 \\
Coset-128 & 0 & 200 & 10 & 0.053324 & signed $2^{1/13}$, $\ln=0.053319$, 6/10 \\
Coset-128 & 0.1 & 200 & 112 & 0.169261 & $\ln\lambda_+$ ($b=1$), 103/112 \\
\bottomrule
\end{tabular}\par}
\par\vspace{2pt}\SItablenotefont
Rates are per iteration, rounded to six decimal places. The certified growth factors are the larger root of $\lambda^2-(1+\gamma)\lambda-\gamma=0$ at the stated $\gamma$, the plain-BP temporal growth factors $2^{1/21}$ (LP20) and $2^{1/19}$ (Mitten-540), the signed orbit $2^{1/13}$ (Coset-128), and the separately certified positive-memory growth factors $1.2207$ (LP20) and $1.2241$ (Mitten-540) at $\gamma=0.1$. Their logarithms are compared with the measured growth rates. Assignment uses the tolerance $\max(0.001,0.05\times\text{certified rate})$; ``none within tolerance'' is a finite-window statement, not evidence that no channel exists.
\end{table}

On the gross code, all nine positive strengths from $0.0125$ to $0.4$ have a growing branch whose median equals $\ln\lambda_+$ to four digits, while the fraction of unconverged trajectories in that branch rises from 34\% to 96\%. Supplementary Table~\ref{tab:grossgrowth} lists the values; this is the gross-code statement quoted in the main text.

\begin{table}[!htbp]
\SItablefont
\setlength{\tabcolsep}{5pt}
\caption{BB-144 growing branch at the nine positive uniform strengths. ``Growing'' is the count above $0.001$ per iteration over the examined unconverged trajectories; the median rate is compared with $\ln\lambda_+$ from $\lambda^2-(1+\gamma)\lambda-\gamma=0$.}
\label{tab:grossgrowth}
{\centering
\begin{tabular}{@{}rrrrr@{}}
\toprule
$\gamma$ & Growing/examined & Fraction growing & Median rate & $\ln\lambda_+$ \\
\midrule
0.0125 & 73/217 & 33.6\% & 0.0244 & 0.0244 \\
0.025 & 92/200 & 46.0\% & 0.0477 & 0.0477 \\
0.05 & 114/200 & 57.0\% & 0.0913 & 0.0913 \\
0.075 & 137/200 & 68.5\% & 0.1317 & 0.1317 \\
0.1 & 140/200 & 70.0\% & 0.1693 & 0.1693 \\
0.15 & 162/200 & 81.0\% & 0.2377 & 0.2377 \\
0.2 & 171/200 & 85.5\% & 0.2989 & 0.2989 \\
0.3 & 181/200 & 90.5\% & 0.4055 & 0.4055 \\
0.4 & 191/200 & 95.5\% & 0.4968 & 0.4968 \\
\bottomrule
\end{tabular}\par}
\par\vspace{2pt}\SItablenotefont
The 0.0125 point uses the larger 2,000-shot prefix; the others use the 200-trajectory stratified sample. The median is the median of the growing branch only and is printed to four decimal places. At these nine strengths the first and third quartiles of the growing branch coincide to numerical precision, so the median represents the whole interquartile range.
\end{table}

Below zero, the growing fraction falls sharply around the memory strength at which the certified temporal state has unit growth factor. LP20 goes from 26/26 at $\gamma=-0.01$ to 0/29 at $-0.02$, and Mitten-540 from 6/8 to 0/12, on either side of their exact values $\gamma_0=-1/62$ and $-1/56$. Coset-128 keeps a small growing fraction on both sides of its $\gamma_0=-1/25$; the gross code, BB-72 and Kasai-96 show no sustained growth below zero. Supplementary Table~\ref{tab:belowzero} gives these finite-window counts.

\begin{table}[!htbp]
\SItablefont
\setlength{\tabcolsep}{4pt}
\caption{Growing trajectories below zero, as a percentage with the raw count in parentheses. Columns are uniform strengths; $\gamma_0$ is the exact memory strength of the temporal state with unit growth factor from Supplementary Note~2.}
\label{tab:belowzero}
{\centering
\begin{tabular}{@{}lrrrrr@{}}
\toprule
Code & $\gamma_0$ & $\gamma=-0.01$ & $\gamma=-0.02$ & $\gamma=-0.03$ & $\gamma=-0.05$ \\
\midrule
BB-144 & $-1/38$ & 0.00\% (0/280) & 0.63\% (2/317) & 0.00\% (0/371) & 0.50\% (1/200) \\
BB-72 & $-1/20$ & 0.00\% (0/237) & 0.00\% (0/267) & 0.00\% (0/297) & 1.00\% (2/200) \\
LP20 & $-1/62$ & 100.00\% (26/26) & 0.00\% (0/29) & 0.00\% (0/29) & 0.00\% (0/17) \\
Kasai-96 & $-1/38$ & 0.00\% (0/33) & 0.00\% (0/38) & 0.00\% (0/47) & 0.00\% (0/34) \\
Mitten-540 & $-1/56$ & 75.00\% (6/8) & 0.00\% (0/12) & 0.00\% (0/21) & 0.00\% (0/8) \\
Coset-128 & $-1/25$ & 2.27\% (16/705) & 0.63\% (5/793) & 0.34\% (3/884) & 0.50\% (1/200) \\
\bottomrule
\end{tabular}\par}
\par\vspace{2pt}\SItablenotefont
The denominator is the number of unconverged trajectories with an estimable rate at that strength. The BB-144, BB-72 and Kasai-96 entries include isolated one- or two-shot positive finite-window slopes; they are not sustained growth.
\end{table}

These rate laws describe the activated growing branch on certified modes rather than every individual trajectory: bounded non-converged states can remain below the growth threshold, shot-by-shot iteration counts are non-monotone in $\gamma$, and negative-memory amplitudes do not collapse linearly with $1/|\gamma|$ (details in the companion data). Supplementary Figure~\ref{fig:S5} shows the median growing rate and the growing fraction across codes and strengths.

\subsection*{3.8 Ablation of the look-ahead}
\label{sec:si-lookahead}

Replacing the next check field $Q^{(t+1)}$ with the present field $Q^{(t)}$ increases logical failures on Coset-128 from 500 to 581 (Holm $P=8.745\times10^{-6}$), with no statistically significant difference detected on BB-144, BB-72 or LP20 (Holm $P\ge0.68$; Supplementary Tables~\ref{tab:lookahead-counts} and~\ref{tab:lookahead-tests}), isolating the effect of look-ahead prediction when only the prediction field changes. This ablation changes only the field used to predict the next belief in the
controller.  TEMPO-BP uses the next field $Q^{(t+1)}$, and the present-field arm
uses $Q^{(t)}$.  The update timing,
target, projection, step sizes, guards and random streams are unchanged.
Within each code both arms use the same shots, seeds and initial memory
fields.  The gross code and BB-72 each use 10,000 shots at
$p=0.4\%$ with an iteration limit of 10,000.  Coset-128 uses 10,000 shots at
$p=0.4\%$ with a limit of 3,000; LP20 uses 2,000 shots at $p=0.175\%$ with a
limit of 3,000.  These are the controller-ablation cohorts of Section~3.4.

\begin{table}[!htbp]
\centering
\SItablefont
\caption{Look-ahead ablation.  Failures comprise shots that do not converge
and converged shots in the wrong logical class.  Both arms were run with
the same build on the same machine.  No arm recorded non-finite arithmetic.}
\label{tab:lookahead-counts}
\begin{tabular}{@{}llrrr@{}}
\toprule
Code & Prediction field & Not converged & Wrong class & Failures \\
\midrule
BB-144 & $Q^{(t+1)}$ (TEMPO-BP) & 26 & 49 & 75 \\
 & $Q^{(t)}$ & 31 & 39 & 70 \\
\addlinespace
BB-72 & $Q^{(t+1)}$ (TEMPO-BP) & 14 & 442 & 456 \\
 & $Q^{(t)}$ & 20 & 433 & 453 \\
\addlinespace
Coset-128 & $Q^{(t+1)}$ (TEMPO-BP) & 435 & 65 & 500 \\
 & $Q^{(t)}$ & 521 & 60 & 581 \\
\addlinespace
LP20 & $Q^{(t+1)}$ (TEMPO-BP) & 38 & 0 & 38 \\
 & $Q^{(t)}$ & 48 & 0 & 48 \\
\bottomrule
\end{tabular}
\end{table}

\begin{table}[!htbp]
\centering
\SItablefont
\setlength{\tabcolsep}{4pt}
\caption{Paired two-sided exact McNemar tests against the same-machine
TEMPO-BP control.  The discordant counts are control-only and alternative-only
failures, respectively.  Holm correction covers one family of four
comparisons: the present-field arm against the TEMPO-BP control on each of four codes.}
\label{tab:lookahead-tests}
\begin{tabular}{@{}llrrrr@{}}
\toprule
Code & Field & Control only & Alternative only & Exact $P$ & Holm $P$ \\
\midrule
BB-144 & $Q^{(t)}$ & 40 & 35 & 0.6445 & 1 \\
BB-72 & $Q^{(t)}$ & 76 & 73 & 0.8699 & 1 \\
Coset-128 & $Q^{(t)}$ & 104 & 185 & $2.186\times10^{-6}$ & $8.745\times10^{-6}$ \\
LP20 & $Q^{(t)}$ & 23 & 33 & 0.2288 & 0.6864 \\
\bottomrule
\end{tabular}
\end{table}

Replacing $Q^{(t+1)}$ with $Q^{(t)}$ increases logical failures on Coset-128 after Holm correction (581 against 500 for TEMPO-BP, Holm $P=8.74\times10^{-6}$), with no significant difference detected on the gross code, BB-72 or LP20 (Holm $P\ge0.68$; whether Coset-128's sensitivity relates to its larger positive-cone multiplier $m=3$ or signed cycle structure remains to be tested).

The unmodified TEMPO-BP control was rerun alongside the present-field arm on the
same macOS/arm64 machine.  Its counts are 75 versus 75 stored failures for
the gross code, 456 versus 454 for BB-72, 500 versus 509 for
Coset-128, and 38 versus 38 for LP20.  The gross-code run reproduces the stored
outcomes exactly.  The other three stored runs used x86-64 Linux; their cohort
files are unchanged, but the local decoder build is not bit-identical to the
stored build.  The three-way outcome (success, not converged or wrong class)
differs on 94/10,000 BB-72 shots, 288/10,000 Coset-128 shots and
20/2,000 LP20 shots.  Thus equal total counts need not mean identical outcomes.
All paired tests above compare the rerun arms on the same machine, not an
alternative arm against a stored control.

%% file: si/note4_convergence.tex
\section*{Supplementary Note 4\quad Convergence-time statistics}
\addcontentsline{toc}{section}{Supplementary Note 4: Convergence-time statistics}

This note separates three things that are often conflated: a reproducible
syndrome-specific time scale, the shape of the convergence-time distribution of a
single syndrome, and the population tail obtained by mixing syndromes of
different difficulty.  Accounting for early-iteration transients, the data support a late-time approximate constant hazard per syndrome and a mixture explanation of the population power law.

\subsection*{4.1 Data and protocol}

The archive contains 40,000 gross-code X-memory shots at $p=0.4\%$, decoded with
TEMPO-BP.  A first pass identified 3,483 shots requiring more than 100
iterations; from this stratum 400 hard syndromes were sampled, together with 100
easy controls.  Each selected syndrome was decoded 128 times with independent
initial memory fields, at a 10,000-iteration cap.  ``Hard'' is a label from the
first random run, not a deterministic property of the syndrome.  The archive has
891 censored returns from 29 hard rows;
371 hard rows converged in all 128 seeds, and all 100 easy rows converged in all
seeds.  The analysis conditions on the selected syndromes and on the frozen
parameters.

Let $\tau_{s,j}$ be the convergence time for syndrome $s$ and initialization $j$.
A per-syndrome rate was fitted with the censored shifted-exponential estimator
\begin{equation}\label{eq:si-convergence-rate}
\hat\lambda_s=\frac{\#\{\text{converged seeds}\}}
{\sum_{j}\bigl(\min(\tau_{s,j},10{,}000)-\min_k \tau_{s,k}\bigr)},
\end{equation}
and the same estimator was applied separately to the first and second 64 seeds.
The inverse rate $1/\hat\lambda_s$ from equation~\eqref{eq:si-convergence-rate} defines an effective time scale.

\subsection*{4.2 A reproducible syndrome-specific time scale}

The ranking of syndromes by $\hat\lambda_s$ is nearly identical in the two
disjoint halves of the seeds: Spearman $\rho=0.986$ over the 400 hard syndromes
(Supplementary Figure~\ref{fig:S3}a).  The fitted rates span more than two decades: the 1st and
99th percentiles are $8.18\times10^{-5}$ and $5.11\times10^{-2}$ per iteration, a
factor of 624; the corresponding effective time scales $1/\hat\lambda_s$ differ by a factor
of 627 (about 20 versus 12,300 iterations).  The lower-tail Hill estimates are
0.775, 0.775 and 0.795 using the lowest 50\%, 30\% and 20\% of the rates.  All
400 hard syndromes have at least two successful seeds, so the data show that a
successful initialization exists for every tested syndrome; they do not show that
every initialization eventually succeeds, and they do not prove the absence of
an absorbing state.

\subsection*{4.3 Early transients and departures from full-range exponentials}

Although pooling shifted convergence times yields a visually near-exponential curve ($\mathrm{KS}=0.0265$), a full-pipeline parametric bootstrap rejects the apparent collapse ($P=0.001$) and conditional tests expose systematic departures from a single-exponential law within individual syndromes. A pooled collapse is formed by subtracting each syndrome's
minimum observed time, fitting one rate per syndrome, retaining only fully
converged rows, and normalizing.  The pooled shape is close to an exponential
(Kolmogorov--Smirnov distance 0.0265), but that number is not a test of the
per-syndrome law: the onset and rate are fitted on the same seeds, the selection
retains only fully converged rows, and the pooled shape can hide opposite
deviations.  A parametric bootstrap that repeats the whole fit, censoring,
selection and zero-removal steps gives a null interval $[0.0077,0.0133]$ for the
pooled KS statistic, so the observed 0.0265 is far outside it (Monte Carlo
$P=0.001$, the resolution of 999 replicates).

A conditional test removes the fitted onset and rate.  Let $a$
be the minimum return and define $z_j=\tau_j-a-1$ for each return $\tau_j>a$.
Conditional on $a$, its multiplicity and $\sum_j z_j$, a constant-hazard model
gives a uniform weak integer composition subject to $0\le z_j\le10{,}000-a-1$.
The onset and rate disappear from the conditional law, while full-convergence
selection remains as this support constraint.  Applying this test to the 371 fully converged hard rows rejects 184 rows at
a Benjamini--Hochberg false-discovery rate of 5\%; the Fisher-combined $P$ value is about $5.4\times10^{-247}$, using $\chi^2_{742}$ calibration under independence across syndrome rows.  Three individual rows
with different shapes give conditional KS statistics 0.485, 0.322 and 0.315
(all $P=10^{-4}$ at 9,999 draws).  The rejected hypothesis is the full-range
statement ``after a fixed onset, each syndrome follows a single exponential''.
No stronger per-syndrome exponential law is used in this note.

While finite iteration limits leave censoring above 10,000 steps unobserved, the empirical truncated mean is shaped jointly by the decoder-dependent tail exponent, prefactor and early transient distribution rather than by the cap alone.

\subsection*{4.4 Late-time approximate constant hazard}

The rejection of the full-range law does not exclude an approximately exponential
late tail after an early transient.  For a landmark $\ell\in\{100,500,1000,2000\}$,
only returns with $\tau>\ell$ were kept and $\tau-\ell$ was treated as a wait starting at
$\ell$, with censoring at 10,000.  A discrete constant-hazard maximum-likelihood fit
and a cap-matched geometric simulation were used to calibrate the survival KS
statistic.  No individual row was rejected after Benjamini--Hochberg correction
at 5\%; the mean-KS bootstrap probabilities were 0.121, 0.348, 0.300 and 0.747
for the four landmarks.

Out-of-sample tests on held-out seeds remain statistically compatible with an approximately constant late-time hazard after early-iteration transients ($\ell\ge 100$), yielding bootstrap KS probabilities 0.641--0.869 across all four landmarks.  In this held-out validation, the
first 64 seeds of each row were used to select 28 rows whose training minimum was
at most 100 and which had at least 10 training successes after iteration 2,000;
the last 64 seeds were then tested.  The held-out mean KS statistics were 0.088,
0.093, 0.111 and 0.151 at $\ell=100,500,1000,2000$, with bootstrap upper-tail
probabilities 0.641, 0.869, 0.749 and 0.691.  Standardizing the held-out
times by the training rate, rather than by the test data, gives survival
probabilities close to the geometric prediction: at $\ell=100$ and
$x=0.5,1,2,3,4$, observed 0.605, 0.387, 0.137, 0.049, 0.013 against predicted
0.606, 0.368, 0.135, 0.050, 0.018.  At $\ell=2{,}000$ the observed values at
$x=2,3$ are 0.181 and 0.096 against 0.135 and 0.050; a full 999-replicate
calibration of the training/selection/rate-estimation pipeline puts these
deviations within the pointwise 95\% interval.  Held-out survival thus remains compatible with an approximately constant late hazard---a statistical finding distinct from single-syndrome full-range exponentiality or a proof of a quasi-stationary measure or spectral gap; the relevant open-systems framework is the
conditionally invariant measure of~\cite{demers2006escape}.

\subsection*{4.5 The population tail}

The empirical survival of all 40,000 first-pass shots
is 1,807/40,000 (4.52\%) at 200 iterations, 468/40,000 (1.17\%) at 1,000
and 73/40,000 (0.18\%) at the 10,000-iteration limit.  The local log--log
slopes between successive endpoints 200, 500, 1,000, 2,000, 5,000 and 10,000
iterations are $-0.91$, $-0.74$, $-0.72$, $-0.76$ and $-0.95$, giving an effective exponent
of about 0.75--0.9 over the observed window.  The lower-tail
Hill estimates of the fitted rates over the 400 selected hard syndromes give
0.775--0.795 at $p=0.4\%$.

A mixture of shifted exponentials fitted to the first 64 initializations of each
of the 400 selected hard syndromes predicts survival on the last 64
initializations of those same syndromes.  At 200, 500, 1,000, 2,000, 5,000
and 10,000 iterations the training model
gives 39.93\%, 20.61\%, 11.99\%, 6.82\%, 3.11\% and 1.71\%, against held-out
values 39.59\%, 20.34\%, 11.87\%, 6.75\%, 3.13\% and 1.73\%.  These rates
condition on the selected hard syndromes; the held-out sample has 25,600
returns (400 syndromes, 64 initializations each).

Pooling all 128 initializations gives survival of 20,224/51,200 (39.50\%) at
200 iterations within the hard stratum.  Multiplying by the first-pass hard
fraction, $3{,}483/40{,}000=0.087075$, gives 3.44\%.  This is the estimated
contribution of the hard stratum to population survival, not the survival
within that stratum or a reconstruction of the full population.  Including
the 100 easy controls with the complementary weight gives 4.45\% at 200
iterations.  At the 10,000-iteration limit the hard-stratum contribution is
0.152\%; the easy controls have all converged.  These estimates use repeated
initializations of the selected syndromes and need not equal the survival
of the original 40,000 first-pass shots.

Because the multi-seed archive was collected solely at $p=0.4\%$, these exponent estimates characterize that operating point alone.

\subsection*{4.6 Why a mixture gives a power law}

The following proposition relates a distribution of per-syndrome rates to a
population power law.  It assumes a probability model; it is not derived from
the TEMPO-BP update.

\setcounter{theorem}{0}
\begin{proposition}[Bounded random transients and the population tail]\label{prop:mixture}
Suppose the convergence time of syndrome $s$ has the form
\begin{equation}\label{eq:si-transient}
\tau_s=\Delta_s+E_s,\qquad 0\le \Delta_s\le \Delta_0<\infty,\qquad
E_s\mid s\sim\operatorname{Exp}(\lambda_s),
\end{equation}
where $\Delta_s$ may depend on the syndrome and on the initialization and may be
correlated with $E_s$, but $\Delta_0$ is uniform over syndromes.  If the rate
distribution has a regular lower tail
\begin{equation}\label{eq:si-rate-tail}
\Pr_s(\lambda_s\le x)\sim C_\lambda\,x^\eta\qquad(x\downarrow0),
\end{equation}
with $C_\lambda>0$ and $\eta>0$, then
\begin{equation}\label{eq:si-pop-tail}
\Pr(\tau>t)\sim C_\lambda\,\Gamma_{\rm E}(1+\eta)\,t^{-\eta},
\end{equation}
where $\Gamma_{\rm E}$ is the Euler gamma function.  For fixed $S$ independent
initializations of the same syndrome, as $I_{\max}\to\infty$,
\begin{equation}\label{eq:si-width}
\Pr\!\left(\min_{j\le S}\tau_{s,j}>I_{\max}\right)
\sim C_\lambda\,\Gamma_{\rm E}(1+\eta)\,(S I_{\max})^{-\eta}
\sim \Pr(\tau>S I_{\max}).
\end{equation}
\end{proposition}

\begin{proof}
The additive transient gives $E_s\le \tau_s\le E_s+\Delta_0$, hence for $t\ge \Delta_0$
\begin{equation}
e^{-\lambda_s t}\le \mathcal S_s(t)=\Pr(\tau_s>t)\le e^{-\lambda_s(t-\Delta_0)}.
\end{equation}
With $\Phi(t)=\mathbb E_s e^{-t\lambda_s}$, independence across initializations
gives
\begin{equation}
\Phi(S I_{\max})\le \mathbb E_s \mathcal S_s(I_{\max})^S\le \Phi(S I_{\max}-S\Delta_0).
\end{equation}
Integration by parts and \eqref{eq:si-rate-tail} give
$\Phi(t)\sim C_\lambda\Gamma_{\rm E}(1+\eta)t^{-\eta}$ as $t\to\infty$.  Regular variation
is insensitive to the fixed shift $S\Delta_0$, so the two bounds coincide to leading
order and give \eqref{eq:si-width}; the one-member case is \eqref{eq:si-pop-tail}.
The discrete version replaces $\lambda_s$ by $-\log r_s$ for a geometric tail.
\end{proof}

The proposition explains why rejecting a full-range exponential per syndrome
does not remove the population power law: a bounded transient can change the
early shape without changing the tail exponent.  It does not prove the additive
form \eqref{eq:si-transient}, a uniform $\Delta_0$, or the regular variation
\eqref{eq:si-rate-tail} for this decoder.  On a fixed finite code there are
finitely many syndromes and, if all rates are positive, the true $t\to\infty$ tail
is eventually dominated by the slowest exponential; the power-law regime therefore characterizes intermediate iteration scales (or growing code sizes) prior to that finite-code asymptotic cutoff.

\subsection*{4.7 Depth for width at the tested budgets}

Under the assumptions of Proposition~\ref{prop:mixture},
for fixed $S$ as $I_{\max}\to\infty$, running $S$ independent members at cap
$I_{\max}$ and one member at cap $S I_{\max}$ gives the same leading tail
probability, proportional to $(S I_{\max})^{-\eta}$.  The archive allows a direct finite-budget
check.  Resampling $S$ of the 128 saved seeds for every hard syndrome, with each
member capped at $I_{\max}=1{,}205$, gives unconverged fractions 5.79\%, 3.18\% and
1.68\% for $S=2,4,8$; one run with the same total budget $S I_{\max}$ gives 5.75\%,
3.19\% and 1.79\%.  Within the 400-syndrome hard stratum, the two budget allocations track closely over $S=2,4,8$ at $I_{\max}=1{,}205$ and $p=0.4\%$. Across general finite budgets, however, a syndrome-dependent transient prefactor $a_s$ can change the ratio at order one, precluding a universal lossless depth--width law for arbitrary widths and iteration limits.  The 20,000-shot
diversity comparison in Section~3.6 is reported alongside a larger,
independent comparison using the same protocol in Section~4.8.

\subsection*{4.8 Parameter diversity and seed multiplicity at 100,000 shots}

Across 100,000 common gross-code shots at $p=0.4\%$, parameter diversity across the 15-member ensemble achieves a statistically significant failure reduction over seed multiplicity alone (376 against 419 failures; two-sided exact McNemar $P=0.0087$), while confirming that distinct seeds already capture most of the width benefit. The ensemble can be built from members with different controller parameters or
from replicas of one member with different seeds.  The two arms are compared on
100,000 common gross-code shots at $p=0.4\%$, with the same shot prefix, member
positions, member seeds and a 1,205-iteration cap per member.  The diverse arm
uses the frozen 15-member roster; the replica arm uses 15 copies of its first
member.  Both select, among the converged corrections that reproduce the
syndrome, the correction most probable under the priors, with ties resolved by member
order.  Supplementary Table~\ref{tab:diversity100k} gives the nested failure counts as members
are added.

\begin{table}[!htbp]
\SItablefont
\setlength{\tabcolsep}{7pt}
\caption{Nested failures on 100,000 common BB-144 shots at $p=0.4\%$ for the diverse and replica arms. Each row adds one member to the prefix; a failure is a shot that does not converge or converges to the wrong logical class.}
\label{tab:diversity100k}
{\centering
\begin{tabular}{@{}rrrrr@{}}
\toprule
$S$ & Diverse failures & Replica failures & Diverse not converged & Replica not converged \\
\midrule
1 & 1,360 & 1,360 & 1,002 & 1,002 \\
2 & 854 & 927 & 453 & 626 \\
3 & 679 & 754 & 305 & 476 \\
4 & 592 & 679 & 247 & 407 \\
5 & 550 & 611 & 219 & 343 \\
6 & 522 & 567 & 193 & 300 \\
7 & 504 & 539 & 156 & 276 \\
8 & 467 & 522 & 141 & 253 \\
9 & 449 & 497 & 119 & 237 \\
10 & 435 & 486 & 105 & 221 \\
11 & 413 & 468 & 103 & 208 \\
12 & 397 & 450 & 93 & 195 \\
13 & 392 & 436 & 90 & 184 \\
14 & 384 & 419 & 87 & 172 \\
15 & 376 & 419 & 83 & 166 \\
\bottomrule
\end{tabular}\par}
\par\vspace{2pt}\SItablenotefont
All counts are out of 100,000 shots. The diverse arm changes both controller parameters and the initialization interval across members.
\end{table}

At $S=15$ the diverse arm fails on 376 shots and the replica arm on 419.
The discordant counts are 107 diverse-only and 150 replica-only failures; the
two-sided exact McNemar probability is $P=0.0087$.  This is the single
pre-specified primary test.  The oracle failure counts, 193 and 287, and the
non-convergence columns are descriptive and are not given separate tests.  An
independent, smaller 20,000-shot sample of the same protocol gave 72 against 88 failures at
$S=15$ ($P=0.0166$); with Holm adjustment over the four pre-specified sizes
$S=2,4,8,15$ that sample gives $P=0.0664$ and does not reject.  The 100,000-shot
sample is independent of that 20,000-shot sample, and the two are not pooled.

\subsection*{4.9 Composition of failures at $p=0.4\%$}

On the gross code at $p=0.4\%$, a single TEMPO-BP run fails on 730 of 100,000
shots, of which 231 do not converge within 10,000 iterations and 499 converge to
the wrong logical class.  The 15-member ensemble, limited to 1,205 iterations per
member, fails on 169 of 48,000 shots, of which 39 do not converge and 130
converge to the wrong class.  Relay-BP-5, at its 18,080-iteration limit, fails on
227 of the same 48,000 shots, of which 92 do not converge and 135 converge to
the wrong class.  On their respective cohorts, the ensemble has lower rates of both
non-convergence and convergence to the wrong logical class than the single run.

\subsection*{4.10 Ensemble prefixes on the common 48,000-shot cohort}

On 48,000 common shots at $p=0.4\%$, prefixes of the frozen ensemble roster drop below five-solution Relay-BP-5's failure count from $S=8$ on the gross code (222 against 227, down from 229 at $S=7$) and from $S=2$ on BB-72 (2,059 against 2,105, down from 2,310 at $S=1$). Figure~3b of the main text plots these prefix counts across the first $S$ members (1,205 iterations per member) on the 48,000 common gross-code shots at
$p=0.4\%$ shared by the TEMPO ensemble and Relay-BP-5, while the corresponding 48,000-shot cohort of BB-72 keeps its 1,808-iteration limit.
These are descriptive crossings of failure counts, not significance thresholds.
The per-member limits stay fixed as the prefix grows. This benchmark comparison
is distinct from the 100,000-shot replicas-versus-members comparison in
Section~4.8.

%% file: si/note5_hardware.tex
\section*{Supplementary Note 5\quad Hardware latency estimate}
\addcontentsline{toc}{section}{Supplementary Note 5: Hardware latency estimate}

This note models the decoding time of one window on a fully parallel field-programmable gate array (FPGA) from the measured iteration counts.  The
primary quantity is the decoding time of one window,
\begin{equation}
 D_{\rm window}=N_{\rm iter}\,\tau_{\rm iter},
 \label{eq:si-window-time}
\end{equation}
where $N_{\rm iter}$ is the number of executed iterations and
$\tau_{\rm iter}$ is the time per iteration.  A window contains 12 syndrome
rounds for the gross code and 6 for BB-72, following the
reference implementation.

\subsection*{5.1 Reference implementation and iteration time}

The published fully parallel Relay-BP implementation for the gross code gives
each variable node and each check node its own compute unit, wired according to
$H$, and runs a flooding schedule in which the check update and the variable
update each take one clock cycle~\cite{maurer2025grossfpga}.  With a 12~ns clock
this is
\begin{equation}\label{eq:si-iteration-time}
\tau_{\rm iter}=2\ \text{cycles}=24\ \text{ns per iteration}.
\end{equation}
The reference uses 4-bit messages plus sign for split X/Z decoding, computes the
memory term with shifts and adds, uses one decoding window of 12 syndrome
rounds, and has a convergence checker that can signal the controller.  The
Relay-BP paper budgets about 20~ns per iteration from an earlier fully parallel
design; we use 24~ns consistently.

TEMPO-BP retains the BP message-passing updates.  The memory-weighted prior has
the same form as in Mem-BP and Relay-BP, and the look-ahead is the check update
that the next iteration would perform anyway (Supplementary Note~1,
Proposition~\ref{prop:noextra}).  What it adds is the controller: per variable,
the sum $Q_i$ of the new check messages, a count of at most $d_i$ unsatisfied
neighboring checks, two table lookups for $\tanh$, one reciprocal of $D_i$ by
table lookup, a few multiplications and additions, and three clippings.  The delay of this
fixed-point controller has not been measured.  We
assume it costs $c_{\rm hw}$ extra clock cycles, with $c_{\rm hw}=2$ as the nominal value and $c_{\rm hw}=4$
as a sensitivity value.

The controller runs once every $K$ iterations.  The average iteration time is
therefore
\begin{equation}\label{eq:si-tempo-time}
\tau_{\rm TEMPO}=\tau_{\rm iter}\Bigl(1+\frac{c_{\rm hw}}{2K}\Bigr),
\end{equation}
which for $c_{\rm hw}=2$ is 30~ns for $K=4$ (gross code) and 27~ns for $K=8$
(BB-72).  For the ensemble we take the smallest update interval among the
members, which is conservative: $K=4$ for the gross code (30~ns) and $K=1$ for
BB-72 (48~ns).  The factor $c_{\rm hw}/(2K)$ gives a first-order model: it assumes the
controller cycles are inserted between the two message-passing phases and that
the two phases still take one clock each.  For $c_{\rm hw}=4$ the gross-code TEMPO-BP and
ensemble iteration times become 36~ns (20\% above 30~ns); the BB-72
TEMPO-BP time becomes 30~ns (11\% above 27~ns) and its ensemble time 72~ns (50\%
above 48~ns).

\subsection*{5.2 Estimated decoding time of one window}

Under the parallel FPGA timing model, the TEMPO ensemble reduces the worst-case single-window decoding time bound to 36.15~$\mu$s on the gross code and 86.78~$\mu$s on BB-72 (12 and 5 times below Relay-BP-5's 433.92~$\mu$s) and lowers the gross-code $p=0.4\%$ mean window time from 12.83~$\mu$s to 3.94~$\mu$s; on BB-72, where member M07 uses $K=1$ ($\tau_{\rm TEMPO}=48$~ns), the ensemble trades a modestly higher nominal mean window time at $p\ge0.3\%$ for this fivefold reduction in the worst-case latency bound. Supplementary Tables~\ref{tab:S5} and~\ref{tab:S6} report the mean, the 99th
percentile and the iteration-limit (worst-case) decoding time of one window for
the gross code and BB-72.  The iteration counts are the measured
means and 99th percentiles over shots.  For the ensemble we use the ideal
parallel depth, the per-shot maximum over members; a serial sum of member times
is not an ensemble latency.  Relay-BP-5 uses $\tau_{\rm iter}=24$~ns.

The worst-case entries follow from the iteration limits.  On the gross code,
TEMPO-BP with 10,000 iterations gives $10{,}000\times30~{\rm ns}=300~\mu$s per
window; the 15-member ensemble, limited to 1,205 iterations per member, gives
$1{,}205\times30~{\rm ns}=36.15~\mu$s; and Relay-BP-5, limited to 18,080
iterations, gives $18{,}080\times24~{\rm ns}=433.92~\mu$s.  On
BB-72 the corresponding limits are 270~$\mu$s for TEMPO-BP,
$1{,}808\times48~{\rm ns}=86.78~\mu$s for the 10-member ensemble and
433.92~$\mu$s for Relay-BP-5.

Representative model estimates are obtained from the same table.  The
gross-code ensemble bound is $36.15~\mu$s $=1{,}205\times30$~ns.  The
Relay-BP-5 bound is $433.92~\mu$s $=18{,}080\times24$~ns.  At $p=0.4\%$ the
ensemble mean depth of 131.42 iterations gives $3.94~\mu$s
$=131.42\times30$~ns, and the Relay-BP-5 mean of 534.43 iterations gives
$12.83~\mu$s $=534.43\times24$~ns.  These values round to 36, 434, 3.9
and 12.8~$\mu$s.  Dividing a window time by 12 (or 6) gives an amortized
time per round; this is not the latency of a round, and we do
not use it as a deadline~\cite{tang2026hierarchical}.  At a 1~$\mu$s syndrome period a 12-round window spans
12~$\mu$s, so the 36~$\mu$s bound is three times the span of one window; it is a
bound on the decoding time of that window, not a guarantee for a stream of
overlapping windows.  Conversely, the number of TEMPO-BP iterations that fit
within the 12~$\mu$s span of one window is $12~\mu{\rm s}/30~{\rm ns}=400$; this is the
vertical line in Fig.~1e of the main text, an estimate rather than a deadline.

\subsection*{5.3 Cost of the ensemble and total work}

The bound is paid for with parallel decoders.  The gross-code ensemble uses
$S=15$ instances and the BB-72 ensemble $S=10$.  The reference
implementation uses about half of the logic of one AMD XCVU19P for one X decoder
of the gross code, so the 15-member ensemble needs about eight such devices
before the controller's extra logic, or an ASIC. The device count for the ten-member BB-72 ensemble requires a code-specific area estimate.  The per-window bound assumes that the members run concurrently.

While concurrent execution bounds per-window critical-path depth, aggregate work over all members follows a distinct scaling. Supplementary Table~\ref{tab:S7} compares the
mean total number of iterations summed over the ensemble members with the mean
iteration count of Relay-BP-5 at each of the ten operating points.  The ratio
ranges from 0.87 (gross code, $p=0.6\%$) to 2.7 (BB-72, $p=0.5\%$).  At
the iteration limit the 15 members total $15\times1{,}205=18{,}075$ iterations,
against 18,080 for Relay-BP-5.  The ensemble therefore exchanges area and, at
most points, total work for a lower per-window bound.

\subsection*{5.4 Assumptions and limits}

The hardware latency estimates in Supplementary Tables~\ref{tab:S5} and~\ref{tab:S6} rely on the following modeling scope and assumptions:
\begin{itemize}
\item The estimate counts iterations, not measured hardware time.  TEMPO-BP has
not been implemented on an FPGA, and its accuracy has been measured only in
double-precision floating point.  The effect of 4- to 6-bit fixed-point
arithmetic on the controller, and on the division by $D_i$, is untested; while 99.80\% of raw targets $\gamma_i^\star$ on the gross code are projected to the interval endpoints $\gamma_{\min}$ or $\gamma_{\max}$ before partial stepping (Supplementary Table~\ref{tab:S9}), fixed-point robustness still depends on whether quantization alters the denominator guard, target sign or projection branch.  The
division is costed as part of $c_{\rm hw}$ but has not been built.
\item A sliding window with a commit region is not modeled.  We estimate the
decoding time of one complete window; we do not simulate overlapping windows,
their commit width, or their back-pressure.  Streaming operation, in which a
new window starts before the previous one is committed, is also not modeled.
\item Convergence checking, leg transitions of Relay-BP, and the ensemble selection of the correction
most probable under the priors (a sum of prior costs followed by a comparison
over members) are not included. Their hardware costs have not been measured
and need not be the same across decoders.
\item The ensemble assumes one decoder per member.  The reference implementation
uses about half of the logic of an AMD XCVU19P for one X decoder of the gross
code, so a 15-member ensemble needs several such devices or an ASIC.  The
per-window estimate assumes that the members run concurrently; a serial API sum
is not an ensemble latency.
\item The bivariate bicycle circuits and noise model are those of IBM's public
Relay-BP benchmark (uniform circuit-level noise including idling errors).  The
FPGA study quotes per-round latencies for its own simulation settings, so its
figures and ours should not be compared digit by digit.
\item The $c_{\rm hw}=4$ sensitivity raises the gross-code TEMPO-BP and ensemble entries
by 20\%, the BB-72 TEMPO-BP entries by 11\% and the BB-72
ensemble entries by 50\%.  The gross-code comparisons with Relay-BP-5 retain their direction, whereas on BB-72---while the worst-case bound remains 3.3 times below Relay-BP-5---the ensemble mean at $p=0.2\%$ and 99th percentile at $p=0.6\%$ and the single-run mean at $p=0.6\%$ cross above Relay-BP-5.
\item Accordingly, the values in Supplementary Tables~\ref{tab:S5} and~\ref{tab:S6} represent architectural model projections for a fully parallel datapath rather than physical hardware measurements~\cite{tang2026hierarchical}.
\end{itemize}

%% file: tables/tableS1.tex
\begingroup\SItablefont
\begin{longtable}{@{}r l r r r r r r r@{}}
\caption{Full failure breakdown on BB-72 at every tested noise point.}\label{tab:S1}\\
\toprule
$p$ (\%) & Decoder & Shots & Invalid & Wrong class & Failures & Nonfinite & Mean it. & P99 it. \\
\midrule\endfirsthead
\toprule
$p$ (\%) & Decoder & Shots & Invalid & Wrong class & Failures & Nonfinite & Mean it. & P99 it. \\
\midrule\endhead
\bottomrule\endfoot
0.2 & BP & 30,000 & 34 & 62 & 96 & 0 & 25.29 & 90 \\
 & Mem-BP & 30,000 & 5 & 96 & 101 & 8 & 8.477 & 26 \\
 & DMem-BP & 30,000 & 4 & 270 & 274 & 0 & 10.57 & 39 \\
 & DMem-warm & 30,000 & 1 & 95 & 96 & 0 & 6.529 & 26 \\
 & BP+OSD+CS-10 & 30,000 & 0 & 94 & 94 & 0 & -- & -- \\
 & Relay-BP-1 & 30,000 & 0 & 94 & 94 & 0 & 6.035 & 26 \\
 & Relay-BP-5 & 30,000 & 0 & 76 & 76 & 0 & 31.67 & 132 \\
 & TEMPO-BP & 30,000 & 1 & 76 & 77 & 0 & 8.204 & 37 \\
 & TEMPO ensemble & 30,000 & 0 & 67 & 67 & 0 & 11.86 & 89 \\
0.3 & BP & 15,000 & 181 & 138 & 319 & 0 & 165.5 & 10,000 \\
 & Mem-BP & 15,000 & 14 & 258 & 272 & 20 & 27.38 & 93 \\
 & DMem-BP & 15,000 & 19 & 627 & 646 & 0 & 34.22 & 233 \\
 & DMem-warm & 15,000 & 4 & 279 & 283 & 0 & 15.37 & 98 \\
 & BP+OSD+CS-10 & 15,000 & 0 & 281 & 281 & 0 & -- & -- \\
 & Relay-BP-1 & 15,000 & 0 & 272 & 272 & 0 & 10.91 & 98 \\
 & Relay-BP-5 & 15,000 & 0 & 216 & 216 & 0 & 53.86 & 352 \\
 & TEMPO-BP & 15,000 & 6 & 226 & 232 & 0 & 22.36 & 187 \\
 & TEMPO ensemble & 15,000 & 1 & 181 & 182 & 0 & 27.82 & 276 \\
0.4 & BP & 20,000 & 869 & 474 & 1,343 & 0 & 552 & 10,000 \\
 & Mem-BP & 20,000 & 128 & 973 & 1,101 & 132 & 104.8 & 2,280 \\
 & DMem-BP & 20,000 & 131 & 2,309 & 2,440 & 0 & 125.7 & 3,140 \\
 & DMem-warm & 20,000 & 39 & 1,111 & 1,150 & 0 & 49.71 & 315 \\
 & TEMPO-static & 20,000 & 362 & 729 & 1,091 & 0 & 259.4 & 10,000 \\
 & BP+OSD+CS-10 & 20,000 & 0 & 1,257 & 1,257 & 0 & -- & -- \\
 & Relay-BP-1 & 48,000 & 1 & 2,544 & 2,545 & 0 & 21.73 & 236 \\
 & Relay-BP-5 & 48,000 & 1 & 2,104 & 2,105 & 0 & 99.83 & 828 \\
 & TEMPO-BP & 20,000 & 38 & 891 & 929 & 0 & 67.72 & 871 \\
 & TEMPO ensemble & 48,000 & 0 & 1,786 & 1,786 & 0 & 63.22 & 1,097 \\
0.5 & BP & 15,000 & 1,759 & 651 & 2,410 & 0 & 1,391 & 10,000 \\
 & Mem-BP & 15,000 & 252 & 1,658 & 1,910 & 252 & 249.6 & 10,000 \\
 & DMem-BP & 15,000 & 286 & 3,524 & 3,810 & 0 & 330.8 & 10,000 \\
 & DMem-warm & 15,000 & 116 & 1,945 & 2,061 & 0 & 148.1 & 4,566 \\
 & BP+OSD+CS-10 & 15,000 & 0 & 2,220 & 2,220 & 0 & -- & -- \\
 & Relay-BP-1 & 15,000 & 1 & 1,947 & 1,948 & 0 & 47.78 & 540 \\
 & Relay-BP-5 & 15,000 & 1 & 1,627 & 1,628 & 0 & 204.4 & 2,047 \\
 & TEMPO-BP & 15,000 & 80 & 1,622 & 1,702 & 0 & 156.8 & 3,469 \\
 & TEMPO ensemble & 15,000 & 2 & 1,368 & 1,370 & 0 & 138.6 & 1,808 \\
0.6 & BP & 10,000 & 2,450 & 661 & 3,111 & 0 & 2,741 & 10,000 \\
 & Mem-BP & 10,000 & 405 & 2,111 & 2,516 & 374 & 574.7 & 10,000 \\
 & DMem-BP & 10,000 & 425 & 3,838 & 4,263 & 0 & 693.6 & 10,000 \\
 & DMem-warm & 10,000 & 178 & 2,483 & 2,661 & 0 & 317.7 & 10,000 \\
 & BP+OSD+CS-10 & 10,000 & 0 & 2,886 & 2,886 & 0 & -- & -- \\
 & Relay-BP-1 & 10,000 & 1 & 2,516 & 2,517 & 0 & 98.12 & 1,205 \\
 & Relay-BP-5 & 10,000 & 1 & 2,194 & 2,195 & 0 & 409.3 & 4,520 \\
 & TEMPO-BP & 10,000 & 118 & 2,116 & 2,234 & 0 & 336.6 & 10,000 \\
 & TEMPO ensemble & 10,000 & 4 & 1,937 & 1,941 & 0 & 270.2 & 1,808 \\
\end{longtable}
\par\vspace{2pt}\SItablenotefont Failure is non-convergence (invalid) or convergence to the wrong logical class. Nonfinite counts shots with a non-finite returned posterior; it is reported separately and is not added to the failure total. The mean and 99th percentile are iterations; for the ensemble they are the per-shot ideal-parallel depth. At $p=0.4\%$, Relay-BP-1, Relay-BP-5 and the TEMPO ensemble use the same 48,000-shot cohort; the remaining rows use 20,000 common shots. TEMPO-static was measured only at $p=0.4\%$. All cells use the recorded accuracy cohorts.\par
\endgroup

%% file: tables/tableS2.tex
\begingroup\SItablefont\setlength{\tabcolsep}{0.6pt}
\begin{longtable}{@{}>{\raggedright\arraybackslash}p{16mm}>{\centering\arraybackslash}p{14mm}>{\centering\arraybackslash}p{24mm}>{\centering\arraybackslash}p{16mm}>{\centering\arraybackslash}p{11mm}>{\raggedright\arraybackslash}p{28mm}>{\centering\arraybackslash}p{7mm}>{\centering\arraybackslash}p{7mm}>{\centering\arraybackslash}p{16mm}>{\centering\arraybackslash}p{16mm}>{\centering\arraybackslash}p{6mm}@{}}
\caption{Paired tests in the seven analysis families. Exact two-sided McNemar probabilities; Holm adjustment within each family. The two ensemble-versus-Relay-BP-5 comparisons at $p=0.4\%$ belong only to the ten-test center family D; family A contains the eight side-point comparisons.}\label{tab:S2}\\
\toprule
Code & $p$ (\%) & $n$ (M/R) & Fails (M/R) & Pair $n$ & Comparison & M & R & $P$ & Holm & Rej. \\
\midrule\endfirsthead
\toprule
Code & $p$ (\%) & $n$ (M/R) & Fails (M/R) & Pair $n$ & Comparison & M & R & $P$ & Holm & Rej. \\
\midrule\endhead
\bottomrule\endfoot
\multicolumn{11}{@{}>{\raggedright\arraybackslash}p{168mm}}{\textit{A: TEMPO ensemble vs Relay-BP-5, eight bivariate-bicycle side points (8 tests; post hoc; $n$ = 10,000/15,000/30,000/100,000/500,000)}}\\
BB-72 & 0.2 & 30,000/30,000 & 67/76 & 30,000 & TEMPO ensemble vs Relay-BP-5 & 14 & 23 & 0.19 & 0.56 & no \\
BB-72 & 0.3 & 15,000/15,000 & 182/216 & 15,000 & TEMPO ensemble vs Relay-BP-5 & 45 & 79 & 0.0029 & 0.012 & yes \\
BB-72 & 0.5 & 15,000/15,000 & 1,370/1,628 & 15,000 & TEMPO ensemble vs Relay-BP-5 & 182 & 440 & $1.3\times10^{-25}$ & $7.9\times10^{-25}$ & yes \\
BB-72 & 0.6 & 10,000/10,000 & 1,941/2,195 & 10,000 & TEMPO ensemble vs Relay-BP-5 & 234 & 488 & $1.8\times10^{-21}$ & $9\times10^{-21}$ & yes \\
BB-144 & 0.2 & 500,000/500,000 & 13/13 & 500,000 & TEMPO ensemble vs Relay-BP-5 & 7 & 7 & 1 & 1 & no \\
BB-144 & 0.3 & 100,000/100,000 & 36/36 & 100,000 & TEMPO ensemble vs Relay-BP-5 & 17 & 17 & 1 & 1 & no \\
BB-144 & 0.5 & 15,000/15,000 & 391/643 & 15,000 & TEMPO ensemble vs Relay-BP-5 & 79 & 331 & $9.1\times10^{-38}$ & $6.4\times10^{-37}$ & yes \\
BB-144 & 0.6 & 10,000/10,000 & 1,298/2,014 & 10,000 & TEMPO ensemble vs Relay-BP-5 & 153 & 869 & $4.7\times10^{-122}$ & $3.7\times10^{-121}$ & yes \\
\addlinespace[2pt]
\multicolumn{11}{@{}>{\raggedright\arraybackslash}p{168mm}}{\textit{B: TEMPO-BP vs BP, ten bivariate-bicycle points (10 tests; post hoc; $n$ = 10,000/15,000/20,000/30,000/100,000/500,000)}}\\
BB-72 & 0.2 & 30,000/30,000 & 77/96 & 30,000 & TEMPO-BP vs BP & 22 & 41 & 0.023 & 0.023 & yes \\
BB-72 & 0.3 & 15,000/15,000 & 232/319 & 15,000 & TEMPO-BP vs BP & 50 & 137 & $1.5\times10^{-10}$ & $3\times10^{-10}$ & yes \\
BB-72 & 0.4 & 20,000/20,000 & 929/1,343 & 20,000 & TEMPO-BP vs BP & 151 & 565 & $4.4\times10^{-57}$ & $1.3\times10^{-56}$ & yes \\
BB-72 & 0.5 & 15,000/15,000 & 1,702/2,410 & 15,000 & TEMPO-BP vs BP & 232 & 940 & $2.1\times10^{-101}$ & $8.6\times10^{-101}$ & yes \\
BB-72 & 0.6 & 10,000/10,000 & 2,234/3,111 & 10,000 & TEMPO-BP vs BP & 169 & 1046 & $9.2\times10^{-155}$ & $4.6\times10^{-154}$ & yes \\
BB-144 & 0.2 & 500,000/500,000 & 31/1,419 & 500,000 & TEMPO-BP vs BP & 9 & 1397 & $<10^{-300}$ & $<10^{-300}$ & yes \\
BB-144 & 0.3 & 100,000/100,000 & 84/2,136 & 100,000 & TEMPO-BP vs BP & 20 & 2072 & $<10^{-300}$ & $<10^{-300}$ & yes \\
BB-144 & 0.4 & 100,000/100,000 & 730/9,836 & 100,000 & TEMPO-BP vs BP & 101 & 9207 & $<10^{-300}$ & $<10^{-300}$ & yes \\
BB-144 & 0.5 & 15,000/15,000 & 616/4,366 & 15,000 & TEMPO-BP vs BP & 26 & 3776 & $<10^{-300}$ & $<10^{-300}$ & yes \\
BB-144 & 0.6 & 10,000/10,000 & 1,686/5,884 & 10,000 & TEMPO-BP vs BP & 28 & 4226 & $<10^{-300}$ & $<10^{-300}$ & yes \\
\addlinespace[2pt]
\multicolumn{11}{@{}>{\raggedright\arraybackslash}p{168mm}}{\textit{C: TEMPO-BP vs Relay-BP-1: full and paired cohorts (10 tests; post hoc; full $n$ as listed (M/R); paired $n$ as listed)}}\\
BB-72 & 0.2 & 30,000/30,000 & 77/94 & 30,000 & TEMPO-BP vs Relay-BP-1 & 19 & 36 & 0.03 & 0.12 & no \\
BB-72 & 0.3 & 15,000/15,000 & 232/272 & 15,000 & TEMPO-BP vs Relay-BP-1 & 55 & 95 & 0.0014 & 0.0068 & yes \\
BB-72 & 0.4 & 20,000/48,000 & 929/2,545 & 10,000 & TEMPO-BP vs Relay-BP-1 & 99 & 172 & $1.1\times10^{-5}$ & $6.5\times10^{-5}$ & yes \\
BB-72 & 0.5 & 15,000/15,000 & 1,702/1,948 & 15,000 & TEMPO-BP vs Relay-BP-1 & 306 & 552 & $3.7\times10^{-17}$ & $3\times10^{-16}$ & yes \\
BB-72 & 0.6 & 10,000/10,000 & 2,234/2,517 & 10,000 & TEMPO-BP vs Relay-BP-1 & 332 & 615 & $2.6\times10^{-20}$ & $2.4\times10^{-19}$ & yes \\
BB-144 & 0.2 & 500,000/500,000 & 31/21 & 500,000 & TEMPO-BP vs Relay-BP-1 & 21 & 11 & 0.11 & 0.33 & no \\
BB-144 & 0.3 & 100,000/100,000 & 84/72 & 100,000 & TEMPO-BP vs Relay-BP-1 & 46 & 34 & 0.22 & 0.44 & no \\
BB-144 & 0.4 & 100,000/48,000 & 730/327 & 5,000 & TEMPO-BP vs Relay-BP-1 & 9 & 15 & 0.31 & 0.44 & no \\
BB-144 & 0.5 & 15,000/15,000 & 616/762 & 15,000 & TEMPO-BP vs Relay-BP-1 & 194 & 340 & $2.7\times10^{-10}$ & $1.9\times10^{-9}$ & yes \\
BB-144 & 0.6 & 10,000/10,000 & 1,686/2,136 & 10,000 & TEMPO-BP vs Relay-BP-1 & 298 & 748 & $3.3\times10^{-45}$ & $3.3\times10^{-44}$ & yes \\
\addlinespace[2pt]
\multicolumn{11}{@{}>{\raggedright\arraybackslash}p{168mm}}{\textit{D: Bivariate bicycle center family (10 tests; post hoc; $n$ = 20,000/48,000/100,000)}}\\
BB-72 & 0.4 & 20,000/20,000 & 929/1,091 & 20,000 & TEMPO-BP vs TEMPO-static & 119 & 281 & $3.2\times10^{-16}$ & $9.7\times10^{-16}$ & yes \\
BB-72 & 0.4 & 20,000/20,000 & 929/1,101 & 20,000 & TEMPO-BP vs Mem-BP & 197 & 369 & $4.4\times10^{-13}$ & $8.8\times10^{-13}$ & yes \\
BB-72 & 0.4 & 20,000/20,000 & 929/1,150 & 20,000 & TEMPO-BP vs DMem-warm & 183 & 404 & $3.9\times10^{-20}$ & $1.6\times10^{-19}$ & yes \\
BB-72 & 0.4 & 20,000/20,000 & 929/1,257 & 20,000 & TEMPO-BP vs BP+OSD+CS-10 & 198 & 526 & $3.6\times10^{-35}$ & $2.1\times10^{-34}$ & yes \\
BB-72 & 0.4 & 48,000/48,000 & 1,786/2,105 & 48,000 & TEMPO ensemble vs Relay-BP-5 & 292 & 611 & $1.1\times10^{-26}$ & $5.3\times10^{-26}$ & yes \\
BB-144 & 0.4 & 100,000/100,000 & 730/12,337 & 100,000 & TEMPO-BP vs TEMPO-static & 58 & 11665 & $<10^{-300}$ & $<10^{-300}$ & yes \\
BB-144 & 0.4 & 100,000/100,000 & 730/7,486 & 100,000 & TEMPO-BP vs Mem-BP & 118 & 6874 & $<10^{-300}$ & $<10^{-300}$ & yes \\
BB-144 & 0.4 & 100,000/100,000 & 730/4,818 & 100,000 & TEMPO-BP vs DMem-warm & 170 & 4258 & $<10^{-300}$ & $<10^{-300}$ & yes \\
BB-144 & 0.4 & 100,000/100,000 & 730/2,510 & 100,000 & TEMPO-BP vs BP+OSD+CS-10 & 304 & 2084 & $<10^{-300}$ & $<10^{-300}$ & yes \\
BB-144 & 0.4 & 48,000/48,000 & 169/227 & 48,000 & TEMPO ensemble vs Relay-BP-5 & 65 & 123 & $2.8\times10^{-5}$ & $2.8\times10^{-5}$ & yes \\
\addlinespace[2pt]
\multicolumn{11}{@{}>{\raggedright\arraybackslash}p{168mm}}{\textit{E: LP20 and Kasai-96 new points (12 tests; pre-declared; $n$ = 2,000)}}\\
LP20 & 0.15 & 2,000/2,000 & 3/160 & 2,000 & TEMPO-BP vs BP & 1 & 158 & $4.4\times10^{-46}$ & $2.6\times10^{-45}$ & yes \\
LP20 & 0.15 & 2,000/2,000 & 3/18 & 2,000 & TEMPO-BP vs Relay-BP-5 & 3 & 18 & 0.0015 & 0.0015 & yes \\
LP20 & 0.15 & 2,000/2,000 & 0/18 & 2,000 & TEMPO ensemble vs Relay-BP-5 & 0 & 18 & $7.6\times10^{-6}$ & $3.1\times10^{-5}$ & yes \\
LP20 & 0.2 & 2,000/2,000 & 66/609 & 2,000 & TEMPO-BP vs BP & 19 & 562 & $5.3\times10^{-140}$ & $5.3\times10^{-139}$ & yes \\
LP20 & 0.2 & 2,000/2,000 & 66/778 & 2,000 & TEMPO-BP vs Relay-BP-5 & 17 & 729 & $8.9\times10^{-191}$ & $9.8\times10^{-190}$ & yes \\
LP20 & 0.2 & 2,000/2,000 & 4/778 & 2,000 & TEMPO ensemble vs Relay-BP-5 & 0 & 774 & $2\times10^{-233}$ & $2.4\times10^{-232}$ & yes \\
Kasai-96 & 0.0625 & 2,000/2,000 & 6/125 & 2,000 & TEMPO-BP vs BP & 0 & 119 & $3\times10^{-36}$ & $1.5\times10^{-35}$ & yes \\
Kasai-96 & 0.0625 & 2,000/2,000 & 6/24 & 2,000 & TEMPO-BP vs Relay-BP-5 & 0 & 18 & $7.6\times10^{-6}$ & $3.1\times10^{-5}$ & yes \\
Kasai-96 & 0.0625 & 2,000/2,000 & 6/24 & 2,000 & TEMPO ensemble vs Relay-BP-5 & 0 & 18 & $7.6\times10^{-6}$ & $3.1\times10^{-5}$ & yes \\
Kasai-96 & 0.075 & 2,000/2,000 & 123/543 & 2,000 & TEMPO-BP vs BP & 5 & 425 & $8.7\times10^{-119}$ & $7.9\times10^{-118}$ & yes \\
Kasai-96 & 0.075 & 2,000/2,000 & 123/392 & 2,000 & TEMPO-BP vs Relay-BP-5 & 4 & 273 & $2\times10^{-75}$ & $1.4\times10^{-74}$ & yes \\
Kasai-96 & 0.075 & 2,000/2,000 & 116/392 & 2,000 & TEMPO ensemble vs Relay-BP-5 & 2 & 278 & $4.1\times10^{-80}$ & $3.2\times10^{-79}$ & yes \\
\addlinespace[2pt]
\multicolumn{11}{@{}>{\raggedright\arraybackslash}p{168mm}}{\textit{F: Mitten-540: center selected using Relay-BP-5 only (9 tests; pre-declared; $n$ = 2,000)}}\\
Mitten-540 & 0.0984375 & 2,000/2,000 & 3/72 & 2,000 & TEMPO-BP vs BP & 1 & 70 & $6.1\times10^{-20}$ & $3\times10^{-19}$ & yes \\
Mitten-540 & 0.0984375 & 2,000/2,000 & 3/6 & 2,000 & TEMPO-BP vs Relay-BP-5 & 2 & 5 & 0.45 & 0.45 & no \\
Mitten-540 & 0.0984375 & 2,000/2,000 & 1/6 & 2,000 & TEMPO ensemble vs Relay-BP-5 & 0 & 5 & 0.063 & 0.13 & no \\
Mitten-540 & 0.1125 & 2,000/2,000 & 8/170 & 2,000 & TEMPO-BP vs BP & 0 & 162 & $3.4\times10^{-49}$ & $2.1\times10^{-48}$ & yes \\
Mitten-540 & 0.1125 & 2,000/2,000 & 8/63 & 2,000 & TEMPO-BP vs Relay-BP-5 & 3 & 58 & $3.3\times10^{-14}$ & $9.9\times10^{-14}$ & yes \\
Mitten-540 & 0.1125 & 2,000/2,000 & 2/63 & 2,000 & TEMPO ensemble vs Relay-BP-5 & 0 & 61 & $8.7\times10^{-19}$ & $3.5\times10^{-18}$ & yes \\
Mitten-540 & 0.1265625 & 2,000/2,000 & 21/391 & 2,000 & TEMPO-BP vs BP & 1 & 371 & $7.8\times10^{-110}$ & $7\times10^{-109}$ & yes \\
Mitten-540 & 0.1265625 & 2,000/2,000 & 21/351 & 2,000 & TEMPO-BP vs Relay-BP-5 & 4 & 334 & $1.9\times10^{-93}$ & $1.4\times10^{-92}$ & yes \\
Mitten-540 & 0.1265625 & 2,000/2,000 & 8/351 & 2,000 & TEMPO ensemble vs Relay-BP-5 & 0 & 343 & $1.1\times10^{-103}$ & $8.9\times10^{-103}$ & yes \\
\addlinespace[2pt]
\multicolumn{11}{@{}>{\raggedright\arraybackslash}p{168mm}}{\textit{G: Coset-128 new 8,000 shots per point (9 tests; post hoc, post-observation extension; $n$ = 8,000)}}\\
Coset-128 & 0.35 & 8,000/8,000 & 84/1,005 & 8,000 & TEMPO-BP vs BP & 4 & 925 & $1.4\times10^{-269}$ & $9.6\times10^{-269}$ & yes \\
Coset-128 & 0.35 & 8,000/8,000 & 84/109 & 8,000 & TEMPO-BP vs Relay-BP-5 & 22 & 47 & 0.0035 & 0.0035 & yes \\
Coset-128 & 0.35 & 8,000/8,000 & 61/109 & 8,000 & TEMPO ensemble vs Relay-BP-5 & 14 & 62 & $2.3\times10^{-8}$ & $4.6\times10^{-8}$ & yes \\
Coset-128 & 0.4 & 8,000/8,000 & 277/2,138 & 8,000 & TEMPO-BP vs BP & 9 & 1870 & $<10^{-300}$ & $<10^{-300}$ & yes \\
Coset-128 & 0.4 & 8,000/8,000 & 277/406 & 8,000 & TEMPO-BP vs Relay-BP-5 & 63 & 192 & $2.4\times10^{-16}$ & $7.3\times10^{-16}$ & yes \\
Coset-128 & 0.4 & 8,000/8,000 & 215/406 & 8,000 & TEMPO ensemble vs Relay-BP-5 & 30 & 221 & $3.9\times10^{-37}$ & $1.5\times10^{-36}$ & yes \\
Coset-128 & 0.45 & 8,000/8,000 & 718/3,727 & 8,000 & TEMPO-BP vs BP & 9 & 3018 & $<10^{-300}$ & $<10^{-300}$ & yes \\
Coset-128 & 0.45 & 8,000/8,000 & 718/1,163 & 8,000 & TEMPO-BP vs Relay-BP-5 & 100 & 545 & $4.9\times10^{-75}$ & $2.5\times10^{-74}$ & yes \\
Coset-128 & 0.45 & 8,000/8,000 & 618/1,163 & 8,000 & TEMPO ensemble vs Relay-BP-5 & 46 & 591 & $1.3\times10^{-121}$ & $7.7\times10^{-121}$ & yes \\
\end{longtable}
\par\vspace{2pt}\SItablenotefont M and R denote method and reference. Their separate columns give the two discordant counts on the paired subset; ``Fails (M/R)'' gives failures on each decoder's full cohort. Pre-declared families were fixed in a dated plan before the corresponding data were collected. Families A--D and G are post hoc; E and F are pre-declared. The Mitten-540 center in F, $p=0.1125\%$, was selected using a Relay-BP-5 pilot scan alone. Family G uses the 8,000 shots collected after an initial 2,000-shot sample had been examined. In family C the eight side rows pair the same saved accuracy cohort; the two $p=0.4\%$ rows use the fixed matched timing subsets (10,000 inputs on BB-72, 5,000 on BB-144) because the full accuracy cohorts differ (TEMPO-BP 20,000 and 100,000 shots, respectively; Relay-BP-1 48,000 shots on each code), so only those fixed inputs are pairable. Probabilities below $10^{-300}$ are displayed as $<10^{-300}$. Non-rejection is not evidence of equivalence.\par
\endgroup

%% file: tables/tableS3.tex
\begin{table}[!htbp]\SItablefont
\caption{Per-seed ensemble results at $p=0.4\%$. Rates are logical failure rates per shot.}\label{tab:S3}
{\centering
\begin{tabular}{@{}l r l r r r r r@{}}
\toprule
Code & Seed & Decoder & Shots & Invalid & Wrong class & Failures & Rate (\%) \\
\midrule
BB-144 & 1 & TEMPO ensemble & 16,000 & 8 & 41 & 49 & 0.306 \\
BB-144 & 1 & Relay-BP-5 & 16,000 & 28 & 48 & 76 & 0.475 \\
BB-144 & 1 & Relay-BP-1 & 16,000 & 28 & 75 & 103 & 0.644 \\
\addlinespace[1pt]
BB-144 & 2 & TEMPO ensemble & 16,000 & 18 & 43 & 61 & 0.381 \\
BB-144 & 2 & Relay-BP-5 & 16,000 & 34 & 48 & 82 & 0.513 \\
BB-144 & 2 & Relay-BP-1 & 16,000 & 34 & 85 & 119 & 0.744 \\
\addlinespace[1pt]
BB-144 & 3 & TEMPO ensemble & 16,000 & 13 & 46 & 59 & 0.369 \\
BB-144 & 3 & Relay-BP-5 & 16,000 & 30 & 39 & 69 & 0.431 \\
BB-144 & 3 & Relay-BP-1 & 16,000 & 30 & 75 & 105 & 0.656 \\
\addlinespace[1pt]
BB-72 & 1 & TEMPO ensemble & 16,000 & 0 & 598 & 598 & 3.737 \\
BB-72 & 1 & Relay-BP-5 & 16,000 & 0 & 733 & 733 & 4.581 \\
BB-72 & 1 & Relay-BP-1 & 16,000 & 0 & 893 & 893 & 5.581 \\
\addlinespace[1pt]
BB-72 & 2 & TEMPO ensemble & 16,000 & 0 & 594 & 594 & 3.712 \\
BB-72 & 2 & Relay-BP-5 & 16,000 & 0 & 680 & 680 & 4.250 \\
BB-72 & 2 & Relay-BP-1 & 16,000 & 0 & 828 & 828 & 5.175 \\
\addlinespace[1pt]
BB-72 & 3 & TEMPO ensemble & 16,000 & 0 & 594 & 594 & 3.712 \\
BB-72 & 3 & Relay-BP-5 & 16,000 & 1 & 691 & 692 & 4.325 \\
BB-72 & 3 & Relay-BP-1 & 16,000 & 1 & 823 & 824 & 5.150 \\
\bottomrule
\end{tabular}\par}
\par\vspace{2pt}\SItablenotefont Each seed contains 16,000 common shots. The ensemble has 15 members on BB-144 and 10 on BB-72; the per-member iteration limits are 1,205 and 1,808, respectively. The three independent sample sets are descriptive replicates; the pooled tests are in Supplementary Table~2 and the main text. Relay-BP-1 rows are shown for context but are not part of the ensemble comparison. All three decoders share the same 48,000 shots per code, distinct from the single-path cohorts of 100,000 BB-144 shots and 20,000 BB-72 shots.
\end{table}

%% file: tables/tableS4.tex
\begingroup\SItablefont
\begin{longtable}{@{}l r l r r r r r l@{}}
\caption{Isolated single-worker timing on all codes.}\label{tab:S4}\\
\toprule
Code & $p$ (\%) & Decoder & Inputs & Mean (s) & Median (s) & Tail (s) & Mean it. & Latency? \\
\midrule\endfirsthead
\toprule
Code & $p$ (\%) & Decoder & Inputs & Mean (s) & Median (s) & Tail (s) & Mean it. & Latency? \\
\midrule\endhead
\bottomrule\endfoot
BB-144 & 0.4 & BP & 5,000 & 0.1754 & -- & 1.593 (P99) & 1200 & yes \\
 & 0.4 & Mem-BP & 5,000 & 0.1005 & -- & 1.358 (P99) & 811.6 & yes \\
 & 0.4 & DMem-BP & 5,000 & 0.2695 & -- & 1.681 (P99) & 1637 & yes \\
 & 0.4 & DMem-warm & 5,000 & 0.0862 & -- & 1.687 (P99) & 517.4 & yes \\
 & 0.4 & TEMPO-static & 5,000 & 0.1504 & -- & 1.283 (P99) & 1244 & yes \\
 & 0.4 & BP+OSD+CS-10 & 5,000 & 0.0621 & -- & 0.2104 (P99) & -- & yes \\
 & 0.4 & Relay-BP-1 & 5,000 & 0.02641 & -- & 0.3745 (P99) & 159 & yes \\
 & 0.4 & Relay-BP-5 & 5,000 & 0.08716 & -- & 1.471 (P99) & 517.8 & yes \\
 & 0.4 & TEMPO-BP & 5,000 & 0.01919 & -- & 0.2561 (P99) & 90.59 & yes \\
 & 0.4 & TEMPO ensemble & 5,000 & 0.1592 & -- & 2.138 (P99) & 807.9 & no (serial sum) \\
BB-72 & 0.4 & BP & 10,000 & 0.02225 & -- & 0.4279 (P99) & 546.3 & yes \\
 & 0.4 & Mem-BP & 10,000 & 0.004051 & -- & 0.106 (P99) & 111.1 & yes \\
 & 0.4 & DMem-BP & 10,000 & 0.005302 & -- & 0.1246 (P99) & 125.7 & yes \\
 & 0.4 & DMem-warm & 10,000 & 0.00214 & -- & 0.01308 (P99) & 51.27 & yes \\
 & 0.4 & TEMPO-static & 10,000 & 0.01061 & -- & 0.3995 (P99) & 266.1 & yes \\
 & 0.4 & BP+OSD+CS-10 & 10,000 & 0.00209 & -- & 0.01256 (P99) & -- & yes \\
 & 0.4 & Relay-BP-1 & 10,000 & 0.0009261 & -- & 0.01017 (P99) & 21.47 & yes \\
 & 0.4 & Relay-BP-5 & 10,000 & 0.004406 & -- & 0.03948 (P99) & 102.4 & yes \\
 & 0.4 & TEMPO-BP & 10,000 & 0.003528 & -- & 0.03969 (P99) & 69.43 & yes \\
 & 0.4 & TEMPO ensemble & 10,000 & 0.01427 & -- & 0.1811 (P99) & 269.3 & no (serial sum) \\
LP20 & 0.175 & BP & 100 & 5.575 & 0.5208 & 29.76 (P95) & -- & yes \\
 & 0.175 & Relay-BP-5 & 100 & 46.01 & 58.52 & 60.26 (P95) & -- & yes \\
 & 0.175 & TEMPO-BP & 100 & 0.3894 & 0.1643 & 0.7359 (P95) & -- & yes \\
 & 0.175 & TEMPO ensemble & 100 & 5.218 & 3.471 & 14.18 (P95) & -- & no (serial sum) \\
Kasai-96 & 0.06875 & BP & 100 & 2.955 & 0.1604 & 15.64 (P95) & -- & yes \\
 & 0.06875 & Relay-BP-5 & 100 & 5.144 & 1.102 & 29.07 (P95) & -- & yes \\
 & 0.06875 & TEMPO-BP & 100 & 0.4382 & 0.1148 & 0.9191 (P95) & -- & yes \\
 & 0.06875 & TEMPO ensemble & 100 & 5.56 & 2.699 & 21.48 (P95) & -- & no (serial sum) \\
Coset-128 & 0.4 & BP & 100 & 0.8585 & 0.08177 & 2.559 (P95) & -- & yes \\
 & 0.4 & Relay-BP-5 & 100 & 0.7071 & 0.1097 & 4.826 (P95) & -- & yes \\
 & 0.4 & TEMPO-BP & 100 & 0.1617 & 0.01403 & 1.044 (P95) & -- & yes \\
 & 0.4 & TEMPO ensemble & 100 & 1.047 & 0.4402 & 4.666 (P95) & -- & no (serial sum) \\
Mitten-540 & 0.1125 & BP & 100 & 2.069 & 0.2167 & 15.44 (P95) & -- & yes \\
 & 0.1125 & Relay-BP-5 & 100 & 9.98 & 5.795 & 31.21 (P95) & -- & yes \\
 & 0.1125 & TEMPO-BP & 100 & 0.2901 & 0.1112 & 0.6954 (P95) & -- & yes \\
 & 0.1125 & TEMPO ensemble & 100 & 4.159 & 2.586 & 13.63 (P95) & -- & no (serial sum) \\
\end{longtable}
\par\vspace{2pt}\SItablenotefont All times are over the stated fixed inputs, measured with one worker and one thread. Transfer timings use 100 reused inputs per code and decoder. ``Tail'' is the 99th percentile for the bivariate bicycle codes and the 95th percentile for the transfer codes. Ensemble times sum the serial member calls (10 or 15 members) and are not parallel ensemble latencies. Medians are unavailable for the bivariate bicycle records. On the gross code (BB-144), TEMPO-BP is faster in isolated single-worker runtime than BP+OSD+CS-10; on BB-72 the order reverses. BP+OSD+CS-10 was not run in the transfer timing comparison.\par
\endgroup

%% file: tables/tableS5.tex
\begin{table}[!htbp]\SItablefont
\caption{Estimated decoding time of one 12-round window on BB-144.}\label{tab:S5}
{\centering
\setlength{\tabcolsep}{4pt}
\begin{tabular}{@{}r rrr rrr rrr@{}}
\toprule
 & \multicolumn{3}{c}{TEMPO-BP} & \multicolumn{3}{c}{TEMPO ensemble} & \multicolumn{3}{c}{Relay-BP-5} \\
\cmidrule(lr){2-4}\cmidrule(lr){5-7}\cmidrule(lr){8-10}
$p$ (\%) & mean & P99 & limit & mean & P99 & limit & mean & P99 & limit \\
\midrule
0.2 & 0.31 & 0.96 & 300 & 0.55 & 2.76 & 36.15 & 1.53 & 5.71 & 433.92 \\
0.3 & 0.71 & 4.50 & 300 & 1.32 & 15.33 & 36.15 & 3.42 & 25.85 & 433.92 \\
0.4 & 2.75 & 38.34 & 300 & 3.94 & 36.15 & 36.15 & 12.83 & 254.69 & 433.92 \\
0.5 & 12.21 & 300 & 300 & 10.67 & 36.15 & 36.15 & 57.37 & 433.92 & 433.92 \\
0.6 & 50.28 & 300 & 300 & 21.32 & 36.15 & 36.15 & 174.98 & 433.92 & 433.92 \\
\bottomrule
\end{tabular}\par}
\par\vspace{2pt}\SItablenotefont
Decoding time of one window in $\mu$s, not per syndrome round. The window is 12 rounds. The mean and 99th percentile are over shots; ``limit'' is the iteration cap times the per-iteration time. Per-iteration times are 30~ns for TEMPO-BP and the ensemble ($c_{\rm hw}=2$, $K=4$), and 24~ns for Relay-BP-5. The ensemble uses the ideal parallel depth, the per-shot maximum over members, not a serial sum. Dividing a window time by 12 gives an amortized time per round, not a latency. With $c_{\rm hw}=4$ the TEMPO-BP and ensemble entries rise by 20\%. These are estimates, not measurements of a TEMPO-BP implementation; assumptions are in Supplementary Note~5.
\end{table}

%% file: tables/tableS6.tex
\begin{table}[!htbp]\SItablefont
\caption{Estimated decoding time of one 6-round window on BB-72.}\label{tab:S6}
{\centering
\setlength{\tabcolsep}{4pt}
\begin{tabular}{@{}r rrr rrr rrr@{}}
\toprule
 & \multicolumn{3}{c}{TEMPO-BP} & \multicolumn{3}{c}{TEMPO ensemble} & \multicolumn{3}{c}{Relay-BP-5} \\
\cmidrule(lr){2-4}\cmidrule(lr){5-7}\cmidrule(lr){8-10}
$p$ (\%) & mean & P99 & limit & mean & P99 & limit & mean & P99 & limit \\
\midrule
0.2 & 0.22 & 1.00 & 270 & 0.57 & 4.27 & 86.78 & 0.76 & 3.17 & 433.92 \\
0.3 & 0.60 & 5.05 & 270 & 1.34 & 13.25 & 86.78 & 1.29 & 8.45 & 433.92 \\
0.4 & 1.83 & 23.52 & 270 & 3.03 & 52.66 & 86.78 & 2.40 & 19.87 & 433.92 \\
0.5 & 4.23 & 93.67 & 270 & 6.65 & 86.78 & 86.78 & 4.91 & 49.13 & 433.92 \\
0.6 & 9.09 & 270 & 270 & 12.97 & 86.78 & 86.78 & 9.82 & 108.49 & 433.92 \\
\bottomrule
\end{tabular}\par}
\par\vspace{2pt}\SItablenotefont
Decoding time of one window in $\mu$s, not per syndrome round. The window is 6 rounds. The mean and 99th percentile are over shots; ``limit'' is the iteration cap times the per-iteration time. Per-iteration times are 27~ns for TEMPO-BP ($c_{\rm hw}=2$, $K=8$), 48~ns for the ensemble ($K=1$), and 24~ns for Relay-BP-5. The ensemble uses the ideal parallel depth, the per-shot maximum over members, not a serial sum. Dividing a window time by 6 gives an amortized time per round, not a latency. With $c_{\rm hw}=4$ the TEMPO-BP entries rise by 11\% and the ensemble entries by 50\%. These are estimates, not measurements of a TEMPO-BP implementation; assumptions are in Supplementary Note~5.
\end{table}

%% file: tables/tableS7_work.tex
\begingroup\SItablefont\setlength{\tabcolsep}{4pt}
{\centering
\begin{tabular}{@{}lrrrrrrrrrr@{}}
\toprule
Code & $p$ (\%) & $S$ & $I_{\max}$ & Ens. mean & Ens. P99 & Ens. total & Relay mean & Relay P99 & Depth ratio & Work ratio \\
\midrule
BB-72 & 0.2 & 10 & 1,808 & 11.856 & 89 & 64.551 & 31.674 & 132 & 2.67 & 2.04 \\
BB-72 & 0.3 & 10 & 1,808 & 27.82 & 276 & 127.34 & 53.858 & 352 & 1.94 & 2.36 \\
BB-72 & 0.4 & 10 & 1,808 & 63.222 & 1097 & 257.53 & 99.835 & 828.01 & 1.58 & 2.58 \\
BB-72 & 0.5 & 10 & 1,808 & 138.61 & 1808 & 548.74 & 204.4 & 2047.1 & 1.47 & 2.68 \\
BB-72 & 0.6 & 10 & 1,808 & 270.24 & 1808 & 1080.2 & 409.29 & 4520.3 & 1.51 & 2.64 \\
BB-144 & 0.2 & 15 & 1,205 & 18.201 & 92 & 155.67 & 63.568 & 238 & 3.49 & 2.45 \\
BB-144 & 0.3 & 15 & 1,205 & 44.165 & 511 & 305.85 & 142.32 & 1077 & 3.22 & 2.15 \\
BB-144 & 0.4 & 15 & 1,205 & 131.42 & 1205 & 823.18 & 534.43 & 10612 & 4.07 & 1.54 \\
BB-144 & 0.5 & 15 & 1,205 & 355.65 & 1205 & 2483.8 & 2390.4 & 18080 & 6.72 & 1.04 \\
BB-144 & 0.6 & 15 & 1,205 & 710.54 & 1205 & 6308.6 & 7290.8 & 18080 & 10.26 & 0.87 \\
\bottomrule
\end{tabular}\par}
\par\vspace{2pt}\SItablenotefont Ensemble depth is the per-shot ideal-parallel depth: the maximum number of iterations over members on each shot; the mean and 99th percentile are over shots. ``Ens. total'' is the recorded mean total iterations summed over all members per shot and $I_{\max}$ is the per-member iteration limit. ``Relay'' is Relay-BP-5. Depth ratio $=$ Relay mean / ensemble mean depth ($>1$ means the ensemble is shallower). Work ratio $=$ ensemble mean total work / Relay mean iterations ($>1$ means more total work). All rows use the same common cohort for the two decoders.\par
\endgroup

%% file: tables/tableS8_parameters.tex
\begingroup\SItablefont\setlength{\tabcolsep}{3pt}
\begin{longtable}{@{}lrrrrrrrrrrr@{}}
\caption{TEMPO-BP parameters for the two single-run TEMPO-BP profiles and every ensemble member, to four significant digits. The min-sum normalization is $\alpha=1$ for every configuration.}\label{tab:S8}\\
\toprule
Traj. & $K$ & $W$ & $\Gamma$ & $[\gamma^-,\gamma^+)$ & $\kappa_0$ & $T_{\rm conf}$ & $\beta$ & $\zeta$ & $T_{\rm tgt}$ & $\epsilon$ & $I_{\max}$ \\
\midrule\endfirsthead
\toprule
Traj. & $K$ & $W$ & $\Gamma$ & $[\gamma^-,\gamma^+)$ & $\kappa_0$ & $T_{\rm conf}$ & $\beta$ & $\zeta$ & $T_{\rm tgt}$ & $\epsilon$ & $I_{\max}$ \\
\midrule\endhead
\bottomrule\endfoot
\multicolumn{12}{@{}>{\raggedright\arraybackslash}p{150mm}}{\textit{BB-144: single-run TEMPO-BP}}\\
single-run TEMPO-BP & 4 & 0 & $[-0.08358,0.8615]$ & $[0.2442,0.3513)$ & 0.6482 & 2.945 & 0.8014 & 0.04282 & 2.114 & 0.0001 & 10,000 \\
\multicolumn{12}{@{}>{\raggedright\arraybackslash}p{150mm}}{\textit{BB-144: members}}\\
M01 & 4 & 0 & $[-0.08358,0.8615]$ & $[0.2442,0.3513)$ & 0.6482 & 2.945 & 0.8014 & 0.04282 & 2.114 & 0.0001 & 1,205 \\
M02 & 8 & 0 & $[-0.03926,1.223]$ & $[0.1595,0.3622)$ & 0.517 & 3.046 & 0.5741 & 0.04959 & 3.19 & 0.0001 & 1,205 \\
M03 & 4 & 0 & $[-0.1056,1.026]$ & $[0.2767,0.2945)$ & 0.6276 & 2.527 & 0.8434 & 0.05922 & 1.678 & 0.0001 & 1,205 \\
M04 & 4 & 0 & $[-0.1031,1.008]$ & $[0.287,0.3099)$ & 0.6362 & 2.682 & 0.8568 & 0.05855 & 2.073 & 0.0001 & 1,205 \\
M05 & 4 & 0 & $[-0.09162,0.967]$ & $[0.0601,0.3208)$ & 0.6305 & 2.492 & 0.8455 & 0.0381 & 1.709 & 0.0001 & 1,205 \\
M06 & 4 & 0 & $[-0.03085,0.9704]$ & $[0.312,0.3709)$ & 0.5992 & 2.902 & 0.8278 & 0.0517 & 2.016 & 0.0001 & 1,205 \\
M07 & 8 & 0 & $[-0.04946,1.187]$ & $[-0.001815,0.4011)$ & 0.5286 & 3.797 & 0.554 & 0.04226 & 4.388 & 0.0001 & 1,205 \\
M08 & 8 & 0 & $[-0.07944,0.8991]$ & $[0.05709,0.2425)$ & 0.5214 & 1.608 & 0.6244 & 0.06855 & 3.432 & 0.0001 & 1,205 \\
M09 & 4 & 0 & $[-0.1116,1.034]$ & $[0.2229,0.3902)$ & 0.6151 & 2.345 & 0.8924 & 0.06635 & 1.329 & 0.0001 & 1,205 \\
M10 & 8 & 0 & $[-0.06621,1.3]$ & $[0.04772,0.436)$ & 0.5678 & 2.353 & 0.6089 & 0.04404 & 3.017 & 0.0001 & 1,205 \\
M11 & 8 & 0 & $[-0.08361,0.8231]$ & $[0.1489,0.2594)$ & 0.5213 & 1.744 & 0.5964 & 0.07472 & 3.053 & 0.0001 & 1,205 \\
M12 & 4 & 0 & $[-0.1061,1.038]$ & $[0.2925,0.3113)$ & 0.6149 & 2.605 & 0.8778 & 0.05838 & 1.323 & 0.0001 & 1,205 \\
M13 & 8 & 0 & $[-0.03917,1.299]$ & $[-0.03552,0.3159)$ & 0.5324 & 2.99 & 0.5807 & 0.05161 & 3.588 & 0.0001 & 1,205 \\
M14 & 8 & 0 & $[-0.05386,0.7999]$ & $[0.1847,0.2286)$ & 0.5563 & 1.605 & 0.6453 & 0.05112 & 2.77 & 0.0001 & 1,205 \\
M15 & 4 & 0 & $[-0.08191,0.9734]$ & $[0.06562,0.3455)$ & 0.6484 & 2.815 & 0.8064 & 0.03845 & 1.637 & 0.0001 & 1,205 \\
\midrule
\multicolumn{12}{@{}>{\raggedright\arraybackslash}p{150mm}}{\textit{BB-72: single-run TEMPO-BP}}\\
single-run TEMPO-BP & 8 & 0 & $[-0.1484,0.5276]$ & $[-0.02601,0.1553)$ & 0.1558 & 2.096 & 0.9406 & 0.0864 & 3.216 & 0.0001 & 10,000 \\
\multicolumn{12}{@{}>{\raggedright\arraybackslash}p{150mm}}{\textit{BB-72: members}}\\
M01 & 8 & 0 & $[-0.1484,0.5276]$ & $[-0.02601,0.1553)$ & 0.1558 & 2.096 & 0.9406 & 0.0864 & 3.216 & 0.0001 & 1,808 \\
M02 & 16 & 0 & $[-0.0319,0.641]$ & $[0.0334,0.08849)$ & 0.5932 & 4.874 & 0.5599 & 0.06832 & 2.213 & 0.0001 & 1,808 \\
M03 & 8 & 0 & $[-0.08689,0.5812]$ & $[0.06995,0.2988)$ & 0.6294 & 3.449 & 0.7994 & 0.08858 & 4.43 & 0.0001 & 1,808 \\
M04 & 8 & 0 & $[-0.06809,0.5351]$ & $[0.1727,0.1736)$ & 0.6263 & 3.67 & 0.9299 & 0.06429 & 4.611 & 0.0001 & 1,808 \\
M05 & 4 & 0 & $[-0.0269,0.8084]$ & $[0.2318,0.2803)$ & 0.5461 & 2.289 & 0.7433 & 0.01002 & 1.172 & 0.0001 & 1,808 \\
M06 & 4 & 0 & $[-0.01582,0.9837]$ & $[0.009395,0.3086)$ & 0.1529 & 4.341 & 0 & 0.06338 & 2.252 & 0.0001 & 1,808 \\
M07 & 1 & 0 & $[-0.0782,0.6265]$ & $[0.104,0.2995)$ & 0.3418 & 2.219 & 0.8702 & 0.01161 & 4.146 & 0.0001 & 1,808 \\
M08 & 80 & 0 & $[-0.07055,1.198]$ & $[0.1306,0.1588)$ & 0.4487 & 5.43 & 0.6736 & 0.006031 & 3.613 & 0.0001 & 1,808 \\
M09 & 16 & 80 & $[-0.1371,0.7877]$ & $[0.1598,0.2002)$ & 0.5829 & 4.384 & 0.7043 & 0.0316 & 1.707 & 0.0001 & 1,808 \\
M10 & 4 & 0 & $[-0.1611,0.8808]$ & $[0.1991,0.2668)$ & 0.4732 & 2.476 & 0.7693 & 0.009779 & 1.02 & 0.0001 & 1,808 \\
\end{longtable}
\par\vspace{2pt}\SItablenotefont $K$ is the update interval and $W$ the number of warm-up iterations. $\Gamma=[\gamma_{\min},\gamma_{\max}]$ is the admissible interval of memory strengths and $[\gamma^-,\gamma^+)$ the initial sampling interval. $\kappa_0$ is the step scale, $T_{\rm conf}$ the confidence scale, $\beta$ the residual-gate weight, $\zeta$ the target magnitude bound, $T_{\rm tgt}$ the target scale, $\epsilon$ the denominator guard and $I_{\max}$ the iteration limit, including warm-up. The 15 BB-144 members and the 10 members of BB-72 are listed in their fixed roster order.\par
\endgroup

%% file: tables/tableS9_actions.tex
\begingroup\SItablefont\setlength{\tabcolsep}{3pt}
{\centering
\begin{tabular}{@{}lrrrr@{}}
\toprule
Requested memory strength & Control events & Share of events & $\sum|\Delta\gamma|$ & Share of $|\Delta\gamma|$ \\
\midrule
below $\gamma_{\min}$ & 2,030,083,755 & 98.97\% & 19,467,459.22 & 86.58\% \\
above $\gamma_{\max}$ & 17,079,994 & 0.83\% & 2,760,282.83 & 12.28\% \\
inside $[\gamma_{\min},\gamma_{\max}]$ & 4,109,520 & 0.20\% & 258,081.52 & 1.15\% \\
\midrule
Total & 2,051,273,269 & 100.00\% & 22,485,823.57 & 100.00\% \\
Rare actions (above $\gamma_{\max}$ + inside) & 21,189,514 & 1.03\% & 3,018,364.35 & 13.42\% \\
\bottomrule
\end{tabular}\par}
\par\vspace{2pt}\SItablenotefont TEMPO-BP on 10,000 shots of BB-144 at $p=0.4\%$, iteration limit $I_{\max}=10{,}000$. Denominators: $2,051,273,269$ control events and $\sum|\Delta\gamma|=22,485,823.57$ over the ten thousand shots. The quoted 99.0\%/0.83\%/0.20\% are the event shares (98.967/0.8327/0.2003); the 13\% is the combined above-plus-inside share of the total absolute change (13.42\%).\par
\endgroup

%% file: main.bbl
\begin{thebibliography}{10}
\expandafter\ifx\csname url\endcsname\relax
  \def\url#1{\burl{#1}}\fi
\expandafter\ifx\csname urlprefix\endcsname\relax\def\urlprefix{URL }\fi
\providecommand{\bibinfo}[2]{#2}
\providecommand{\eprint}[2][]{\url{#2}}
\providecommand{\doi}[1]{\url{https://doi.org/#1}}
\bibcommenthead

\bibitem{terhal2015qec}
\bibinfo{author}{Terhal, B.~M.}
\newblock \bibinfo{title}{Quantum error correction for quantum memories}.
\newblock \emph{\bibinfo{journal}{Reviews of Modern Physics}}
  \textbf{\bibinfo{volume}{87}}, \bibinfo{pages}{307--346}
  (\bibinfo{year}{2015}).
\newblock \urlprefix\url{https://doi.org/10.1103/revmodphys.87.307}.

\bibitem{breuckmann2021qldpc}
\bibinfo{author}{Breuckmann, N.~P.} \& \bibinfo{author}{Eberhardt, J.~N.}
\newblock \bibinfo{title}{Quantum low-density parity-check codes}.
\newblock \emph{\bibinfo{journal}{PRX Quantum}} \textbf{\bibinfo{volume}{2}},
  \bibinfo{pages}{040101} (\bibinfo{year}{2021}).
\newblock \urlprefix\url{https://doi.org/10.1103/prxquantum.2.040101}.

\bibitem{bravyi2024high}
\bibinfo{author}{Bravyi, S.} \emph{et~al.}
\newblock \bibinfo{title}{High-threshold and low-overhead fault-tolerant
  quantum memory}.
\newblock \emph{\bibinfo{journal}{Nature}} \textbf{\bibinfo{volume}{627}},
  \bibinfo{pages}{778--782} (\bibinfo{year}{2024}).
\newblock \urlprefix\url{https://doi.org/10.1038/s41586-024-07107-7}.

\bibitem{xu2024constant}
\bibinfo{author}{Xu, Q.} \emph{et~al.}
\newblock \bibinfo{title}{Constant-overhead fault-tolerant quantum computation
  with reconfigurable atom arrays}.
\newblock \emph{\bibinfo{journal}{Nature Physics}}
  \textbf{\bibinfo{volume}{20}}, \bibinfo{pages}{1084--1090}
  (\bibinfo{year}{2024}).
\newblock \urlprefix\url{https://doi.org/10.1038/s41567-024-02479-z}.

\bibitem{tang2026hierarchical}
\bibinfo{author}{Tang, X.} \emph{et~al.}
\newblock \bibinfo{title}{Hierarchical decoding for quantum error correction}
  (\bibinfo{year}{2026}).
\newblock
  \bibinfo{eprint}{{\href{https://arxiv.org/abs/2611.16271}{{arXiv:2611.16271}}}}.

\bibitem{skoric2023parallel}
\bibinfo{author}{Skoric, L.}, \bibinfo{author}{Browne, D.~E.},
  \bibinfo{author}{Barnes, K.~M.}, \bibinfo{author}{Gillespie, N.~I.} \&
  \bibinfo{author}{Campbell, E.~T.}
\newblock \bibinfo{title}{Parallel window decoding enables scalable fault
  tolerant quantum computation}.
\newblock \emph{\bibinfo{journal}{Nature Communications}}
  \textbf{\bibinfo{volume}{14}}, \bibinfo{pages}{7040} (\bibinfo{year}{2023}).
\newblock \urlprefix\url{https://doi.org/10.1038/s41467-023-42482-1}.

\bibitem{poulin2008iterative}
\bibinfo{author}{Poulin, D.} \& \bibinfo{author}{Chung, Y.}
\newblock \bibinfo{title}{On the iterative decoding of sparse quantum codes}.
\newblock \emph{\bibinfo{journal}{Quantum Information and Computation}}
  \textbf{\bibinfo{volume}{8}}, \bibinfo{pages}{987--1000}
  (\bibinfo{year}{2008}).
\newblock \urlprefix\url{https://doi.org/10.26421/qic8.10-8}.

\bibitem{bausch2024learning}
\bibinfo{author}{Bausch, J.} \emph{et~al.}
\newblock \bibinfo{title}{Learning high-accuracy error decoding for quantum
  processors}.
\newblock \emph{\bibinfo{journal}{Nature}} \textbf{\bibinfo{volume}{635}},
  \bibinfo{pages}{834--840} (\bibinfo{year}{2024}).
\newblock \urlprefix\url{https://doi.org/10.1038/s41586-024-08148-8}.

\bibitem{gong2024graphneural}
\bibinfo{author}{Gong, A.}, \bibinfo{author}{Cammerer, S.} \&
  \bibinfo{author}{Renes, J.~M.} \bibinfo{title}{Graph neural networks for
  enhanced decoding of quantum {LDPC} codes}.
\newblock \emph{\bibinfo{booktitle}{2024 IEEE International Symposium on
  Information Theory (ISIT)}}, \bibinfo{pages}{2700--2705}
  (\bibinfo{publisher}{IEEE}, \bibinfo{year}{2024}).
\newblock \urlprefix\url{https://doi.org/10.1109/isit57864.2024.10619589}.
\newblock
  \bibinfo{eprint}{{\href{https://arxiv.org/abs/2310.17758}{{arXiv:2310.17758}}}}.

\bibitem{panteleev2021degenerate}
\bibinfo{author}{Panteleev, P.} \& \bibinfo{author}{Kalachev, G.}
\newblock \bibinfo{title}{Degenerate quantum {LDPC} codes with good finite
  length performance}.
\newblock \emph{\bibinfo{journal}{Quantum}} \textbf{\bibinfo{volume}{5}},
  \bibinfo{pages}{585} (\bibinfo{year}{2021}).
\newblock \urlprefix\url{https://doi.org/10.22331/q-2021-11-22-585}.

\bibitem{roffe2020decoding}
\bibinfo{author}{Roffe, J.}, \bibinfo{author}{White, D.~R.},
  \bibinfo{author}{Burton, S.} \& \bibinfo{author}{Campbell, E.}
\newblock \bibinfo{title}{Decoding across the quantum low-density parity-check
  code landscape}.
\newblock \emph{\bibinfo{journal}{Physical Review Research}}
  \textbf{\bibinfo{volume}{2}}, \bibinfo{pages}{043423} (\bibinfo{year}{2020}).
\newblock \urlprefix\url{https://doi.org/10.1103/physrevresearch.2.043423}.

\bibitem{kuo2022exploiting}
\bibinfo{author}{Kuo, K.-Y.} \& \bibinfo{author}{Lai, C.-Y.}
\newblock \bibinfo{title}{Exploiting degeneracy in belief propagation decoding
  of quantum codes}.
\newblock \emph{\bibinfo{journal}{npj Quantum Information}}
  \textbf{\bibinfo{volume}{8}}, \bibinfo{pages}{111} (\bibinfo{year}{2022}).
\newblock \urlprefix\url{https://doi.org/10.1038/s41534-022-00623-2}.

\bibitem{muller2025relay}
\bibinfo{author}{Müller, T.} \emph{et~al.}
\newblock \bibinfo{title}{Improved belief propagation is sufficient for
  real-time decoding of quantum memory} (\bibinfo{year}{2025}).
\newblock
  \bibinfo{eprint}{{\href{https://arxiv.org/abs/2506.01779}{{arXiv:2506.01779}}}}.

\bibitem{chen2025improved}
\bibinfo{author}{Chen, J.}, \bibinfo{author}{Yi, Z.}, \bibinfo{author}{Liang,
  Z.} \& \bibinfo{author}{Wang, X.}
\newblock \bibinfo{title}{Improved belief propagation decoding algorithms for
  surface codes}.
\newblock \emph{\bibinfo{journal}{IEEE Transactions on Quantum Engineering}}
  \textbf{\bibinfo{volume}{6}}, \bibinfo{pages}{1--16} (\bibinfo{year}{2025}).
\newblock \urlprefix\url{https://doi.org/10.1109/tqe.2025.3577769}.

\bibitem{cordova2026syndromegain}
\bibinfo{author}{Cordova, H.}, \bibinfo{author}{Balatsoukas-Stimming, A.},
  \bibinfo{author}{Gültekin, Y.~C.}, \bibinfo{author}{Liga, G.} \&
  \bibinfo{author}{Alvarado, A.}
\newblock \bibinfo{title}{Syndrome adaptive gain control for min-sum decoding
  of quantum {LDPC} codes} (\bibinfo{year}{2026}).
\newblock
  \bibinfo{eprint}{{\href{https://arxiv.org/abs/2605.10433}{{arXiv:2605.10433}}}}.

\bibitem{wang2012feedback}
\bibinfo{author}{Wang, Y.-J.}, \bibinfo{author}{Sanders, B.~C.},
  \bibinfo{author}{Bai, B.-M.} \& \bibinfo{author}{Wang, X.-M.}
\newblock \bibinfo{title}{Enhanced feedback iterative decoding of sparse
  quantum codes}.
\newblock \emph{\bibinfo{journal}{IEEE Transactions on Information Theory}}
  \textbf{\bibinfo{volume}{58}}, \bibinfo{pages}{1231--1241}
  (\bibinfo{year}{2012}).
\newblock \urlprefix\url{https://doi.org/10.1109/tit.2011.2169534}.

\bibitem{taghipour2026rewinding}
\bibinfo{author}{Taghipour, M.}, \bibinfo{author}{Chytas, D.} \&
  \bibinfo{author}{Vasić, B.}
\newblock \bibinfo{title}{Multistage rewinding decoder for {QLDPC} codes}
  (\bibinfo{year}{2026}).
\newblock
  \bibinfo{eprint}{{\href{https://arxiv.org/abs/2608.07783}{{arXiv:2608.07783}}}}.

\bibitem{relaybprepo}
\bibinfo{author}{Müller, T.} \emph{et~al.}
\newblock \bibinfo{title}{{Relay-BP}: open-source implementation and bivariate
  bicycle benchmark}.
\newblock \bibinfo{howpublished}{\url{https://github.com/trmue/relay}}
  (\bibinfo{year}{2025}).

\bibitem{luby1993restart}
\bibinfo{author}{Luby, M.}, \bibinfo{author}{Sinclair, A.} \&
  \bibinfo{author}{Zuckerman, D.}
\newblock \bibinfo{title}{Optimal speedup of {Las} {Vegas} algorithms}.
\newblock \emph{\bibinfo{journal}{Information Processing Letters}}
  \textbf{\bibinfo{volume}{47}}, \bibinfo{pages}{173--180}
  (\bibinfo{year}{1993}).
\newblock \urlprefix\url{https://doi.org/10.1016/0020-0190(93)90029-9}.

\bibitem{gomes2000heavytails}
\bibinfo{author}{Gomes, C.~P.}, \bibinfo{author}{Selman, B.},
  \bibinfo{author}{Crato, N.} \& \bibinfo{author}{Kautz, H.}
\newblock \bibinfo{title}{Heavy-tailed phenomena in satisfiability and
  constraint satisfaction problems}.
\newblock \emph{\bibinfo{journal}{Journal of Automated Reasoning}}
  \textbf{\bibinfo{volume}{24}}, \bibinfo{pages}{67--100}
  (\bibinfo{year}{2000}).
\newblock \urlprefix\url{https://doi.org/10.1023/a:1006314320276}.

\bibitem{cain2026shor}
\bibinfo{author}{Cain, M.} \emph{et~al.}
\newblock \bibinfo{title}{{Shor's} algorithm is possible with as few as 10,000
  reconfigurable atomic qubits} (\bibinfo{year}{2026}).
\newblock
  \bibinfo{eprint}{{\href{https://arxiv.org/abs/2603.28627}{{arXiv:2603.28627}}}}.

\bibitem{kasai2026orthogonality}
\bibinfo{author}{Kasai, K.}
\newblock \bibinfo{title}{Breaking the orthogonality barrier in quantum {LDPC}
  codes}.
\newblock \emph{\bibinfo{journal}{Quantum}} \textbf{\bibinfo{volume}{10}},
  \bibinfo{pages}{2205} (\bibinfo{year}{2026}).
\newblock \urlprefix\url{https://doi.org/10.22331/q-2026-09-09-2205}.

\bibitem{zhao2026ultrahigh}
\bibinfo{author}{Zhao, C.}, \bibinfo{author}{Duckering, C.},
  \bibinfo{author}{Gu, A.}, \bibinfo{author}{Maskara, N.} \&
  \bibinfo{author}{Zhou, H.}
\newblock \bibinfo{title}{Towards ultra-high-rate quantum error correction with
  reconfigurable atom arrays} (\bibinfo{year}{2026}).
\newblock
  \bibinfo{eprint}{{\href{https://arxiv.org/abs/2604.16209}{{arXiv:2604.16209}}}}.

\bibitem{aydin2026coset}
\bibinfo{author}{Aydin, A.}, \bibinfo{author}{Tamo, I.} \&
  \bibinfo{author}{Barg, A.}
\newblock \bibinfo{title}{Breaking the bicycle frame: coset-based quantum
  {LDPC} codes} (\bibinfo{year}{2026}).
\newblock
  \bibinfo{eprint}{{\href{https://arxiv.org/abs/2606.17268}{{arXiv:2606.17268}}}}.

\bibitem{bhardwaj2026highrate}
\bibinfo{author}{Bhardwaj, A.} \emph{et~al.}
\newblock \bibinfo{title}{High-rate {qLDPC} processors} (\bibinfo{year}{2026}).
\newblock
  \bibinfo{eprint}{{\href{https://arxiv.org/abs/2607.28795}{{arXiv:2607.28795}}}}.

\bibitem{gidney2021stim}
\bibinfo{author}{Gidney, C.}
\newblock \bibinfo{title}{Stim: a fast stabilizer circuit simulator}.
\newblock \emph{\bibinfo{journal}{Quantum}} \textbf{\bibinfo{volume}{5}},
  \bibinfo{pages}{497} (\bibinfo{year}{2021}).
\newblock \urlprefix\url{https://doi.org/10.22331/q-2021-07-06-497}.

\bibitem{tempocode}
\bibinfo{author}{You, Z.}, \bibinfo{author}{Li, Y.}, \bibinfo{author}{Yuan, X.}
  \& \bibinfo{author}{Yao, Y.}
\newblock \bibinfo{title}{{TEMPO-BP}: open-source implementation and experiment
  repository}.
\newblock \bibinfo{howpublished}{\url{https://github.com/Yomure/tempo-bp}}
  (\bibinfo{year}{2026}).

\bibitem{berthusen2025local}
\bibinfo{author}{Berthusen, N.} \emph{et~al.}
\newblock \bibinfo{title}{Toward a {2D} local implementation of quantum
  low-density parity-check codes}.
\newblock \emph{\bibinfo{journal}{PRX Quantum}} \textbf{\bibinfo{volume}{6}},
  \bibinfo{pages}{010306} (\bibinfo{year}{2025}).
\newblock \urlprefix\url{https://doi.org/10.1103/prxquantum.6.010306}.

\bibitem{sun2026layer}
\bibinfo{author}{Sun, Z.} \& \bibinfo{author}{Cai, Z.}
\newblock \bibinfo{title}{Layer codes as quantum memories: syndrome extraction,
  thresholds and idle robustness} (\bibinfo{year}{2026}).
\newblock
  \bibinfo{eprint}{{\href{https://arxiv.org/abs/2609.39522}{{arXiv:2609.39522}}}}.

\bibitem{pecorari2025highrate}
\bibinfo{author}{Pecorari, L.}, \bibinfo{author}{Jandura, S.},
  \bibinfo{author}{Brennen, G.~K.} \& \bibinfo{author}{Pupillo, G.}
\newblock \bibinfo{title}{High-rate quantum {LDPC} codes for
  long-range-connected neutral atom registers}.
\newblock \emph{\bibinfo{journal}{Nature Communications}}
  \textbf{\bibinfo{volume}{16}}, \bibinfo{pages}{1111} (\bibinfo{year}{2025}).
\newblock \urlprefix\url{https://doi.org/10.1038/s41467-025-56255-5}.

\bibitem{tempodata}
\bibinfo{author}{You, Z.}, \bibinfo{author}{Li, Y.}, \bibinfo{author}{Yuan, X.}
  \& \bibinfo{author}{Yao, Y.}
\newblock \bibinfo{title}{{TEMPO-BP}: benchmark datasets, detector error models
  and experiment provenance}.
\newblock \bibinfo{howpublished}{Zenodo} (\bibinfo{year}{2026}).
\newblock \urlprefix\url{https://doi.org/10.5281/zenodo.24280168}.

\end{thebibliography}
